\documentclass[twocolumn,aps,prl,showpacs,superscriptaddress,tightenlines]{revtex4-1}
\usepackage{amsmath}
\usepackage{amsfonts}
\usepackage{graphicx}
\usepackage{epsfig}
\usepackage{color}
\usepackage{diagbox}
\usepackage[colorlinks,citecolor=blue]{hyperref}

\begin{document}

\title{Microscopic Theory of Heat Transfer across a Vacuum}
\author{Yue-Hui Zhou}
\affiliation{Key Laboratory of Low-Dimensional Quantum Structures and Quantum Control of Ministry of Education, Key Laboratory for Matter Microstructure and Function of Hunan Province, Department of Physics and Synergetic Innovation Center for Quantum Effects and Applications, Hunan Normal University, Changsha 410081, China}
\author{Jie-Qiao Liao}
\email{Contact author: jqliao@hunnu.edu.cn}
\affiliation{Key Laboratory of Low-Dimensional Quantum Structures and Quantum Control of Ministry of Education, Key Laboratory for Matter Microstructure and Function of Hunan Province, Department of Physics and Synergetic Innovation Center for Quantum Effects and Applications, Hunan Normal University, Changsha 410081, China}
\affiliation{Hunan Research Center of the Basic Discipline for Quantum Effects and Quantum Technologies,  Hunan Normal University, Changsha 410081, China}
\affiliation{Institute of Interdisciplinary Studies, Hunan Normal University, Changsha, 410081, China}

\begin{abstract}
Heat transfer is a fundamental concept in physics, and how to characterize the microscopical physical mechanism of heat transfer is a significant topic. Recently, a new mechanism for heat transfer, phonon heat transfer across a vacuum through quantum fluctuations, has been experimentally demonstrated in a two-vibrating-membrane system. However, a microscopic quantum electrodynamics theory behind this phenomenon remains unexplored. Here, we establish the nonrelativistic microscopic theory of phonon heat transfer across a vacuum confined by two movable end mirrors of a one-dimensional optomechanical cavity. Under the multimode-cavity-field framework, we obtain a second-order effective Hamiltonian governing the heat transfer effect. This Hamiltonian describes a phonon-exchange interaction between the two mirrors with a strength proportional to a multimode factor and $l_{0}^{-3}$, where $l_{0}$ is the rest cavity length. We also confirm the dependence of the mode temperatures and heat flux on the rest cavity length, corresponding to the thermal steady state of the mirrors. This work initiates the microscopic investigation of thermodynamics based on the optomechanical cavity platform, and will have profound implications for both thermal management in nanoscale devices and cavity optomechanics.
\end{abstract}

\date{\today}
\maketitle

\emph{Introduction.}---The quantization of the electromagnetic field confined by moving boundaries~\cite{MooreJMPs1970,DeWitt1975,LawPRA1995} is not only an important fundamental research topic in quantum field theory~\cite{Birrell1984} and quantum electrodynamics (QED)~\cite{Milonni1994}, but also has wide applications in the frontiers of quantum physics, especially in cavity optomechanics~\cite{Kippenberg2008,AspelmeyerRMP2014}. Since the early days of the development of QED~\cite{Schwinger2003}, people have realized that the vacuum is not empty, but is full of quantum fluctuations~\cite{Mostepanenkobook1997,Bordagbook2009,Milonni2019}. In fact, the vacuum underlies many of observable physical phenomena predicted by the QED, such as the Lamb shift~\cite{Lamb1947} and the Casimir effect~\cite{Mostepanenkobook1997,Bordagbook2009,Casimirpknaw1948}. The Casimir effect physics was widely studied based on the optomechanical platforms~\cite{WilsonNature2011,MacriPRX2018,Xunn2022}. Recently, a novel and important physical effect, phonon heat transfer across a vacuum through quantum fluctuations, has been demonstrated in a two-vibrating-membrane system~\cite{FongNature2019}. Though this phenomenon has been explained based on the Casimir force~\cite{FongNature2019}, a systematical microscopic quantum electrodynamics theory underlying this physical effect remains unexplored.

In this Letter, we establish a nonrelativistic microscopic theory of phonon heat transfer across a vacuum through quantum fluctuations in an optomechanical cavity formed by two movable end mirrors. By treating both the cavity fields and the moving end mirrors quantum mechanically, we obtain the microscopic Hamiltonian describing the physical processes within the linear expansion and multimode-cavity-field framework. When the cavity field is in a vacuum, we obtain a second-order effective Hamiltonian to describe a phonon-exchange interaction between the two mirrors, which can explain the phonon heat transfer induced by the virtual process associated with quantum fluctuations. Both the frequency shifts and coupling strength are proportional to $1/l_{0}^{3}$ ($l_{0}$ being the rest cavity length) and the multimode factors determined by the cutoff frequency. The heat transfer phenomenon is confirmed via examining both the mode temperatures of the two mirrors and the heat flux.

\emph{Physical model and equations of motion.}---Consider a one-dimensional Fabry-P\'{e}rot-type optomechanical cavity composed of two perfectly reflective and movable end mirrors (with masses $m_{1}$ and $m_{2}$). The two mirrors move in individual potential wells $V_{1}(q_{1}(t))$ and $V_{2}(q_{2}(t))$, where $q_{1}(t)$ and $q_{2}(t)$ denote, respectively, the transient positions of the right and left movable mirrors with $q_{1}(t)-q_{2}(t)>0$. Note that the two-mechanical-resonator optomechanical system has been widely studied from the viewpoint of quantum optomechanics, including ground-state cooling~\cite{Genes2008,LaiPRA2020R,LiuPRA2022,HuangPRA2022}, back-action-evading measurements~\cite{WoolleyPRA2013,OckeloenPRL2016}, mechanical entanglement~\cite{HartmannPRL2008,LiaoPRA2014,FlayacPRL2014,RiedingerNature2018,OckeloenNature2018,KotlerScience2021,LepinayScience2021,LaiPRL2022}, and phonon transfer~\cite{StefanoPRL2019,YangNature2020,Nian2021PRA}. In the absence of free charge and current, the vector potential $\vec{A}(x,t)$ in the cavity region [i.e., $q_{2}(t) \leq x\leq q_{1}(t)$] obeys the wave equation $(c=1)$
\begin{equation}
\partial^{2}_{x}\vec{A}(x,t)=\partial^{2}_{t}\vec{A}(x,t).  \label{vpequationmps}
\end{equation}
In this work, we consider a linear polarized light such that the vector potential $\vec{A}(x,t)$ can be treated as a scalar quantity $A(x,t)$ and we adopt the time-dependent boundary condition $\vec{A}(q_{1}(t),t)=\vec{A}(q_{2}(t),t)=0$~\cite{MooreJMPs1970,Akopyan2021PRD,seeSM}. Below, we mark $q_{n=1,2}(t)$ as $q_{n}$ for conciseness.

The motion of the two moving mirrors is governed by the nonrelativistic equations of motion
\begin{equation}
m_{n}\ddot{q}_{n}=-\partial_{q_{n}}V_{n}(q_{n})+(-1)^{n+1}\frac{1}{2}\left. \left[\partial_{x}A(x,t)\right]^{2}\right\vert_{x=q_{n}}   \label{mirrorsequationsmps}
\end{equation}
for $n=1$ and $2$ corresponding to the right and left end mirrors, respectively. The first and second terms in Eq.~(\ref{mirrorsequationsmps}) describe the conservative force and  the radiation-pressure force~\cite{LawPRA1995,seeSM}, respectively.

Under the above-mentioned boundary conditions, we make a Fourier series expansion of the scalar quantity $A(x,t) =\sum_{k=1}^{\infty}\varphi_{k}Q_{k}$, where $\varphi_{k}=[2/(q_{1}-q_{2})]^{1/2}\sin[\omega_{k}(x-q_{2})]$ for positive integer $k$ is the $k$th instantaneous mode function, with $\omega_{k}=k\pi/(q_{1}-q_{2})$ being the position-dependent cavity frequencies, and $Q_{k}= \int_{q_{2}}^{q_{1}}\varphi_{k} A(x,t) dx$ is the generalized coordinates of the $k$th cavity mode~\cite{JiPRA1998,JPADodonov1998,Butera2022PRD}. After a lengthy calculation, the equations of motion of $Q_{k}$ and $q_{n}$ can be obtained as~\cite{seeSM}
\begin{subequations}
\label{gcood}
\begin{align}
\ddot{Q}_{k}=&-\omega_{k}^{2}Q_{k}-\sum_{n=1,2}\sum_{j=1}^{\infty}\frac{2g_{jk}^{(n)}\dot{q}_{n}\dot{Q}_{j}}{q_{1}-q_{2}}+\sum_{l=1}^{\infty }\frac{2\dot{q}_{1}\dot{q}_{2}g_{lk}^{(2)}Q_{l}}{(q_{1}-q_{2})^{2}}                                              \notag \\
&+\sum_{j,l=1}^{\infty}\frac{2\dot{q}_{1}\dot{q}_{2}g_{kj}^{(1)}g_{lj}^{(2)}Q_{l}}{(q_{1}-q_{2})^{2}}+\sum_{n=1,2}%
\sum_{j,l=1}^{\infty}\frac{\dot{q}_{n}^{2}g_{kj}^{(n)}g_{lj}^{(n)}Q_{l}}{(q_{1}-q_{2})^{2}}                      \notag \\
&-\sum_{n=1,2}\sum_{j=1}^{\infty}g_{jk}^{(n)}\frac{(q_{1}-q_{2})\ddot{q}_{n}+(-1)^{n}\dot{q}_{n}^{2}}{(q_{1}-q_{2})^{2}}Q_{j},\label{gcooda}   \\
\ddot{q}_{n=1,2}=&-\frac{1}{m_{n}}\frac{\partial V_{n}(q_{n})}{\partial q_{n}}+\sum_{k,j=1}^{\infty}[(-1)^{j+k}\delta_{n,1}-\delta_{n,2}] \notag \\
&\times\frac{\omega_{k}\omega_{j}Q_{k}Q_{j}}{m_{n}(q_{1}-q_{2})},   \label{gcoodb}
\end{align}
\end{subequations}
where we introduce $g_{kj}^{(n)}=2kj[(-1)^{j+k}\delta_{n,1}-\delta_{n,2}]/(j^{2}-k^{2})$ when $j\neq k$ and $g_{kk}^{(n)}=0$.

\emph{Hamiltonian and canonical quantization.}---Following the canonical quantization, we can construct a Lagrangian function for the system as~\cite{seeSM}
\begin{eqnarray}
L &=&\frac{1}{2}\sum_{k=1}^{\infty}(\dot{Q}_{k}^{2}-\omega_{k}^{2}Q_{k}^{2})+\sum_{n=1,2}\sum_{j,k,l=1}^{\infty}\frac{%
g_{jl}^{(n)}g_{kl}^{(n)}\dot{q}_{n}^{2}Q_{k}Q_{j}}{2(q_{1}-q_{2})^{2}}                       \notag \\
&&+\sum_{j,k,l=1}^{\infty}\frac{g_{kl}^{(1)}g_{jl}^{(2)}\dot{q}_{1}\dot{q}%
_{2}Q_{k}Q_{j}}{(q_{1}-q_{2})^{2}}+\frac{1}{2}(m_{1}\dot{q}_{1}^{2}+m_{2}\dot{q}_{2}^{2})     \notag \\
&&+\sum_{n=1,2}\sum_{j,k=1}^{\infty}\frac{g_{jk}^{(n)}\dot{q}_{n}\dot{Q}_{k}Q_{j}}{q_{1}-q_{2}}-V_{1}(q_{1})-V_{2}(q_{2}).   \label{palagrangian}
\end{eqnarray}
We have checked the validity of the Lagrangian function $L$ by confirming that Eqs.~(\ref{gcooda}) and ~(\ref{gcoodb}) can be recovered with the Euler-Lagrangian equations.

To be consistent, we also consider the radiation-pressure effect of the electromagnetic fields outside the cavity on the moving mirrors~\cite{seeSM} by constructing a three-cavity system in which two moving and perfectly reflective mirrors are placed within a fixed cavity. When taking the lengths of the two external cavities to infinity, the external fields can be considered as the modes of the rest part of the 1D universe according to the quasi-normal mode theory~\cite{LangPRA1973}. The radiation-pressure force of the external field on the movable end mirrors can be approximatively neglected when the length of the two external cavities approaches infinity~\cite{seeSM}.

By the Legendre transformation, the Hamiltonian of the system can be obtained as
\begin{equation}
H=\frac{1}{2}\sum_{k=1}^{\infty}(P_{k}^{2}+\omega_{k}^{2}Q_{k}^{2})+\sum_{n=1,2}\left[\frac{1}{2m_{n}}\left(p_{n}+\Gamma
_{n}\right)^{2}+V_{n}(q_{n})\right],  \label{Hamiltonainexpress}
\end{equation}
where $P_{k}$ ($Q_{k}$) and $p_{n}$ ($q_{n}$) are the canonical momenta (coordinates) of the $k$th cavity mode and the $n$th moving mirror ($n=1,2$), respectively, which can be expressed as $P_{k}=\dot{Q}_{k}+\sum_{n=1,2}\sum_{j=1}^{\infty }g_{jk}^{(n)}\dot{q}_{n}Q_{j}/(q_{1}-q_{2})$ and $p_{n}=m_{n}\dot{q}_{n}-\Gamma_{n}$, with $\Gamma_{n}=\sum_{j,k=1}^{\infty}g_{kj}^{(n)}Q_{j}P_{k}/(q_{1}-q_{2})$ describing the difference between the mechanical and canonical momenta of the $n$th moving mirror.

To perform the canonical quantization, we replace the variables $p_{1}$, $p_{2}$, $P_{k}$, $q_{1}$, $q_{2}$, and $Q_{k}$ with the corresponding operator $\hat{p}_{1}$, $\hat{p}_{2}$, $\hat{P}_{k}$, $\hat{q}_{1} $, $\hat{q}_{2}$, and $\hat{Q}_{k}$, which hold the nonzero commutation relations $[\hat{q}_{1},\hat{p}_{1}]=i\hbar$, $[\hat{q}_{2},\hat{p}_{2}]=i\hbar$, and $[\hat{Q}_{j},\hat{P}_{k}] =i\hbar\delta_{jk}$. Since the cavity field frequency depends on the transient distance between the two mirrors, we can define the annihilation operator $\hat{a}_{k}(\hat{q}_{r})=[2\hbar\omega_{k}(\hat{q}_{r})]^{-1/2}[\omega_{k}(\hat{q}_{r})\hat{Q}_{k}+i\hat{P}_{k}]$
of the $k$th cavity mode to describe the quantum state of the cavity field in the Fock space, where $\hat{q}_{r}=\hat{q}_{1}-\hat{q}_{2}$ is the relative coordinate operator. The relative momentum operator corresponding to $\hat{q}_{r}$ is given by $\hat{p}_{r}=\mu\hat{p}_{1}/m_{1}-\mu\hat{p}_{2}/m_{2}$ with the reduced mass $\mu =m_{1}m_{2}/m_{t}$ and the total mass $m_{t}=m_{1}+m_{2}$.

In the center-of-mass system, the quantized Hamiltonian~\cite{seeSM} of the whole system can be written as
\begin{eqnarray}
\hat{H} &=&\sum_{k=1}^{\infty}\hbar\omega_{k}(\hat{q}_{r})\hat{a}_{k}^{\dagger}(\hat{q}_{r})\hat{a}_{k}(\hat{q}_{r})+\frac{1}{2}%
\sum_{k=1}^{\infty}\hbar\omega_{k}(\hat{q}_{r})+V(\hat{q}_{c},\hat{q}_{r})   \notag \\
&&+\sum_{l=1,2}\frac{1}{2m_{l}}\left[\frac{m_{l}}{m}\hat{p}_{c}+(-1)^{l+1}\hat{p}_{r}+\hat{\Gamma}_{l}\right]^{2},   \label{Hamiltonianinzhan}
\end{eqnarray}
where $V(\hat{q}_{c},\hat{q}_{r})=V_{1}(\hat{q}_{c}+\mu\hat{q}_{r}/m_{1})+V_{2}(\hat{q}_{c}-\mu\hat{q}_{r}/m_{2})$ describes the potential energy of the two moving mirrors with the center-of-mass coordinate operator $\hat{q}_{c}=\mu\hat{q}_{1}/m_{2}+\mu\hat{q}_{2}/m_{1}$ and momentum operator $\hat{p}_{c}=\hat{p}_{1}+\hat{p}_{2}$. We see from Eq.~(\ref{Hamiltonianinzhan}) that, due to the interaction term $\hat{p}_{c}(\hat{\Gamma}_{1}+\hat{\Gamma}_{2})/m$, the center-of-mass mode cannot become a dark mode~\cite{seeSM} when the two moving mirrors are degenerate. However, under the single-mode approximation, the center-of-mass mode is decoupled from the cavity field and becomes a dark mode~\cite{Genes2008,YangNatcomm2012,seeSM,ShkarinPRL2014,SommerPRL2019,LaiPRA2020R,HuangarXive2023,ChegnizadehScience2024,CaoPRL2025} for two degenerate moving mirrors. In addition, the zero-point energy $\sum_{k=1}^{\infty}\hbar\omega_{k}\left(\hat{q}_{r}\right)/2$ can be replaced by $-\hbar\pi c/(24\hat{q}_{r})$~\cite{LawPRA1995,Butera2022PRD,Schwartz2014}, which contributes to the phonon transfer between two moving mirrors~\cite{FongNature2019}.

\emph{Linear expansion approximation.}---We assume that the two mirrors are bounded by the harmonic potential wells $V_{1}(q_{1})$ and $V_{2}(q_{2})$, with the equilibrium positions $x=0$ and $x=l_{0}$ for the left and right mirrors, respectively. Consider small vibrations around the equilibrium position, i.e., $x_{1}\equiv q_{1}-l_{0}\ll l_{0}$ and $x_{2}\equiv q_{2}\ll l_{0}$, then $\omega_{k}(\hat{q}_{r})$ and $\hat{a}_{k}(\hat{q}_{r})$ can be approximated as $\omega_{k}(\hat{q}_{r})\simeq\omega _{k0}(1-\hat{x}_{r}/l_{0})$ and $\hat{a}_{k}(\hat{q}_{r})\simeq\hat{a}_{k0}-\hat{x}_{r}\hat{a}_{k0}^{\dagger}/(2l_{0})$, where $\hat{x}_{r}=\hat{x}_{1}-\hat{x}_{2}$ is relative displacement operator, $\omega_{k0}=k\pi/l_{0}$ is the resonance frequency of the $k$th cavity mode, and $\hat{a}_{k0}=\hat{a}_{k}(l_{0})$ [$\hat{a}_{k0}^{\dagger}=\hat{a}_{k}^{\dagger}(l_{0})$] is the annihilation (creation) operator associated with the equilibrium position. Under the linear expansion, the quantized Hamiltonian of the system in the original coordinate system can be expressed as~\cite{seeSM}
\begin{eqnarray}
\hat{H}^{\prime} &=&\frac{\hbar}{2}\sum_{k,n=1}^{\infty }[x_{\text{2,zpf}}(\hat{b}_{2}^{\dagger}+\hat{b}_{2})-(-1)^{n+k}x_{\text{1,zpf}}(\hat{b}%
_{1}^{\dagger }+\hat{b}_{1})]\hat{F}_{n,k}                                                \notag \\
&&+\hbar\omega_{\text{M},1}\hat{b}_{1}^{\dagger}\hat{b}_{1}+\hbar\omega_{\text{M},2}\hat{b}_{2}^{\dagger}\hat{b}_{2}+\hbar \sum_{n=1}^{\infty
}\omega_{n0}\hat{a}_{n0}^{\dagger}\hat{a}_{n0},                                    \label{Hamiltonianlast1}
\end{eqnarray}
with $\hat{F}_{n,k}\equiv\sqrt{\omega_{n0}\omega_{k0}}(\hat{a}_{k0}^{\dagger}\hat{a}_{n0}^{\dagger}+\hat{a}_{k0}^{\dagger}\hat{a}_{n0}+H.c.)/l_{0}$. In Eq.~(\ref{Hamiltonianlast1}), we have discarded these terms $\hbar\sum_{n=1}^{\infty}\omega_{n0}/2$ and $\sum_{n=1}^{\infty}\hbar\omega_{n0}[x_{\text{2,zpf}}(\hat{b}_{2}^{\dagger}+\hat{b}_{2})-x_{\text{1,zpf}}(\hat{b}_{1}^{\dagger}+\hat{b}_{1})]/2l_{0}$ associated with the zero-point energy of the cavity fields.

\emph{Effective phonon-exchange interaction and heat transfer.}--- We see from Eq.~(\ref{Hamiltonianlast1}) that the two end mirrors are coupled to these cavity modes via the photon-phonon parametric processes, which are far-off-resonant. Physically, an effective interaction between the two mirrors can be induced via the second-order physical process under proper condition. When the cavity fields are in a vacuum and the two moving mirrors are degenerate ($\omega_{\text{M},1}$=$\omega_{\text{M},2}=\omega_{M}$). Using the time-average method~\cite{Gamel2010PRA}, we can obtain the effective Hamiltonian~\cite{seeSM}
\begin{equation}
\label{efftiveHamitonian}
\hat{H}_{\text{eff}}=\hbar\Delta_{b,1}\hat{b}_{1}^{\dagger}\hat{b}_{1}+\hbar\Delta_{b,2}\hat{b}_{2}^{\dagger}\hat{b}_{2}
+\hbar\xi(\hat{b}_{2}\hat{b}_{1}^{\dagger}+\hat{b}_{1}\hat{b}_{2}^{\dagger}).
\end{equation}
Here, the frequency shift and the effective coupling strength in Eq.~(\ref{efftiveHamitonian}) are defined by
\begin{equation}
\Delta_{b,n=1,2}=-\pi c\frac{x_{n,\text{zpf}}^{2}}{l_{0}^{3}}\Sigma_{1},%
\hspace{0.5cm}\xi=\pi c\frac{x_{1,\text{zpf}}x_{2,\text{zpf}}}{l_{0}^{3}}\Sigma_{2},  \label{Hamiparameters}
\end{equation}
where we introduce the multimode factors
\begin{equation}
\Sigma_{n=1,2}=\sum_{k,j=1}^{\infty}\frac{[\delta_{n,1}+\delta_{n,2}(-1)^{j+k}]jk(j+k)}{(j+k)^{2}-\beta^{2}}, \label{dimfrshifandcoupm}
\end{equation}
with $\beta=\omega_{M}/\Delta\omega_{c}$ and $\Delta\omega_{c}=\pi c/l_{0}$. Equation~(\ref{efftiveHamitonian}) gives an effective phonon-exchange coupling between the two mirrors, which provides the physical mechanism for phonon transfer. The factors $\Sigma_{n=1,2}$ can be used to characterize the multimode-cavity field enhancement, and we have $\Sigma_{n=1,2}\simeq\lbrack\delta_{n,1}+\delta_{n,2}(-1)^{j+k}]jk/(j+k)$ when $\beta\ll 1$.
\begin{figure}[tbp]
\center
\includegraphics[bb=5 83 409 263, width=0.47 \textwidth]{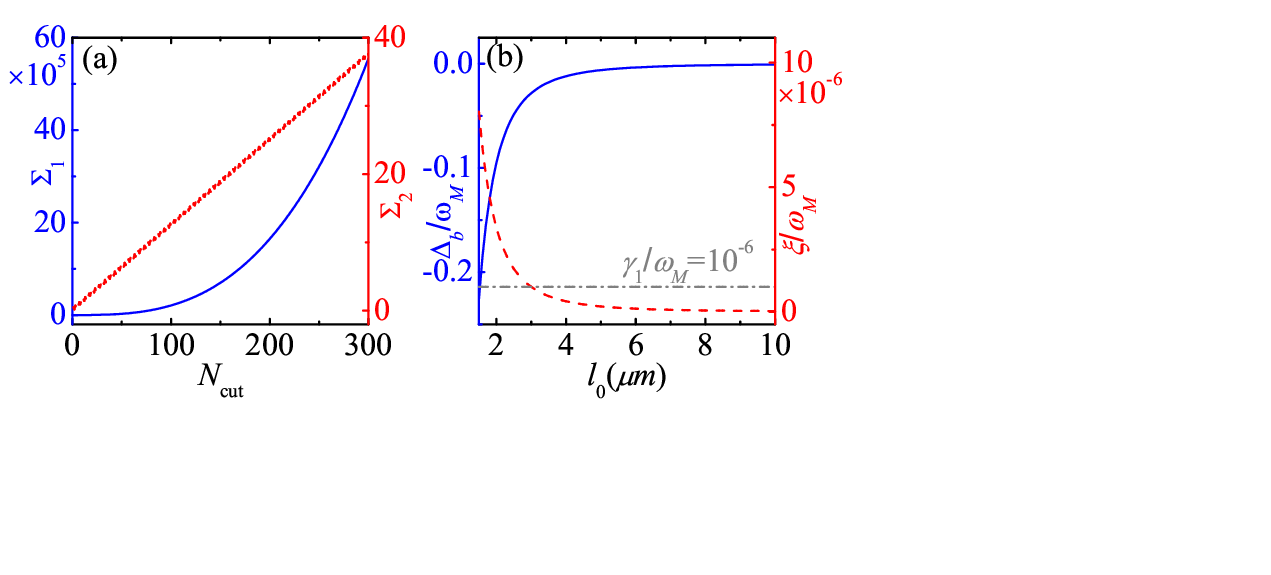}
\caption{(Color online) (a) The multimode factors $\Sigma_{1}$ (blue solid line) and $\Sigma_{2}$ (red dashed line) as functions of the cut-off mode number $N_{\text{cut}}$. (b) The scaled frequency shift $\Delta_{b}/\omega_{M}$ (blue solid line) and the scaled coupling strength $\xi/\omega_{M}$ (red dashed line) as functions of the bare cavity length $l_{0}$ when the cut-off mode number $N_{\text{cut}}=130$. Other parameters used are $m_{1}=m_{2}=0.3\times 10^{-18}$ kg, $\omega_{M}=2\pi\times50\times 10^{6}$ Hz, and $\gamma_{1}/\omega_{M}=\gamma_{2}/\omega_{M}=10^{-6}$ (as shown by the gray dash-dotted line for reference, $\xi>\gamma_{1,2}$ and $\xi<\gamma_{1,2}$ denote the strong and weak coupling regimes of two coupled mirrors, respectively).}\label{fequencyandcoupling}
\end{figure}

In Eq.~(\ref{efftiveHamitonian}), both the interaction between the two mirrors and the frequency shift are proportional to $l_{0}^{-3}$. We point out that a phonon-exchange interaction with strength $\propto l_{0}^{-3}$ can be obtained by phenomenologically expanding the Casimir force $-\hbar\pi c/(24\hat{q}_{r})^{2}$~\cite{seeSM}. However, the ratio of the phonon-exchange magnitude of the second-order effective coupling over the effective Casimir-interaction magnitude is given by $12\Sigma_{2}$~\cite{seeSM}, which indicates that the induced second-order coupling strength is much stronger than the Casimir coupling because of $\Sigma_{2}\gg1$.

We see from Eq.~(\ref{Hamiparameters}) that the interference constructiveness and cancellation occur in $\Delta_{b,n}$ and $\xi$, respectively. In the specific numerical calculation of Eq.~(\ref{Hamiparameters}), we need to introduce the cut-off frequency $\omega_{\text{cut}}=\Delta\omega_{c}N_{\text{cut}}$, with $N_{\text{cut}}$ being the cut-off-mode number~\cite{Butera2013PRL}. We consider the case $m_{1}=m_{2}=m_{M}$, then $\Delta_{b,1}=\Delta_{b,2}=\Delta_{b}$. In this case, we have $\Delta _{b}/\xi =-\Sigma _{1}/\Sigma _{2}$, which is only determined by the cut-off mode number. Figure~\ref{fequencyandcoupling}(a) shows the presence of multimode enhancement effect. As the cut-off-mode number $N_{\text{cut}}$ increases, the multimode factors $\Sigma_{1}$ becomes much larger than $\Sigma_{2}$. Figure~\ref{fequencyandcoupling}(b) shows the dependence of the frequency shift $\Delta_{b}$ and coupling strength $\xi$ on the bare cavity length $l_{0}$. We see from Fig.~\ref{fequencyandcoupling}(b) that the system can enter the weak coupling region ($\xi\ll \gamma_{1}, \gamma_{2}$) and the strong coupling region ($\xi\gg \gamma_{1}, \gamma_{2}$) by adjusting the bare cavity length $l_{0}$.

Including the dissipations, the dynamic evolution of the two mirrors is governed by the quantum master equation
\begin{equation}
\frac{\partial\hat{\rho}}{\partial t}=\frac{1}{i\hbar}[\hat{H}_{\text{eff}},\hat{\rho}]-\sum_{n=1,2}\gamma_{n}[(\bar{n}_{\text{th},n}+1) \mathcal{\hat{D}}(\hat{b}_{n})+\bar{n}_{\text{th},n}\mathcal{\hat{D}}(\hat{b}_{n}^{\dagger})]\hat{\rho}, \label{masterequation}
\end{equation}
where $\hat{H}_{\text{eff}}$ is given by Eq.~(\ref{efftiveHamitonian}), $\mathcal{\hat{D}}(\hat{o})\hat{\rho}=(\hat{o}^{\dagger}\hat{o}\hat{\rho}+\hat{\rho}\hat{o}^{\dagger}\hat{o})/2-\hat{o}\hat{\rho}\hat{o}^{\dagger}$ is the Lindbland supperoperators~\cite{SCZubairy1997}, $\gamma_{n=1,2}$ is the decay rate of the $n$th moving mirror, and $\bar{n}_{\text{th},n=1,2}=1/(e^{\hbar\omega_{M}/k_{B}T_{n}}-1)$ is the mean thermal phonons associated with the bath temperature $T_{n}$. By solving Eq.~(\ref{masterequation}), we can analytically prove that the second-order correlation function $g^{(2)}(0)=2$ corresponding to the steady state of the two mirrors~\cite{seeSM,SCZubairy1997}, which indicate that the thermal steady state of the two mirrors. The corresponding mode temperatures~\cite{seeSM,Biehs2020zn} of the two mirrors in the high-temperature limit can be further calculated as
\begin{subequations}
\begin{align}
T_{m1}=&T_{1}+\frac{4\xi^{2}\gamma_{2}(T_{2}-T_{1})}{(\gamma_{1}+\gamma_{2})(4\xi^{2}+\gamma_{1}\gamma_{2})}, \\
T_{m2}=&T_{2}+\frac{4\xi^{2}\gamma_{1}(T_{1}-T_{2})}{(\gamma_{1}+\gamma_{2})(4\xi^{2}+\gamma_{1}\gamma_{2})}.
\end{align}%
\end{subequations}
Since the coupling strength depends on the static cavity length, the mode temperatures are related to the coupling strength. Moreover, the coupling between the two mirrors will cause a leveling effect for the two mode temperatures. For example, when $T_{1}$ is greater than $T_{2}$, the mode temperature $T_{m2}$ is greater than $T_{2}$, and $T_{m1}$ is less than $T_{1}$.

\begin{figure}[tbp]
\center
\includegraphics[bb=26 84 408 272, width=0.47 \textwidth]{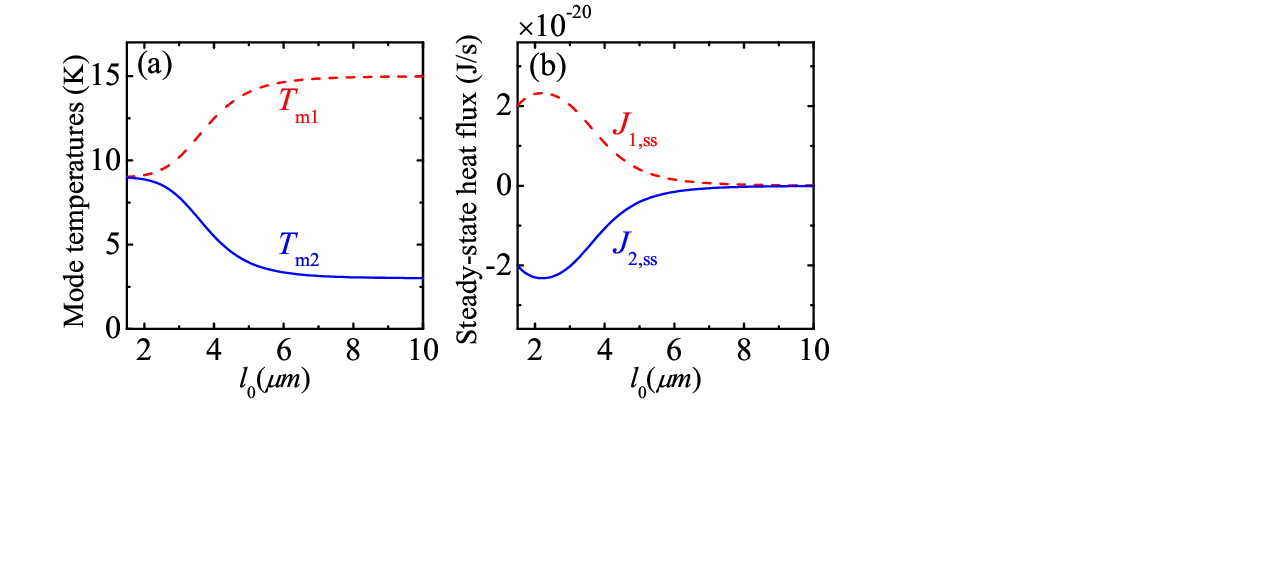}
\caption{(Color online) (a) The mode temperatures $T_{m1}$ (red dashed line) and $T_{m2}$ (blue solid line), and (b) the heat fluxes $J_{1}$ (red dashed line) and $J_{2}$ (blue solid line) as functions of the bare cavity length $l_{0}$ when the cut-off mode number $N_{\text{cut}}=130$. $T_{1}=15$K, $T_{2}=3$K, and other parameters used are the same as those in Fig.~\ref{modetemandheatflux}.}\label{modetemandheatflux}
\end{figure}

In Fig.~\ref{modetemandheatflux}(a), we plot the mode temperatures $T_{m1}$ and $T_{m2}$ as functions of the cavity length $l_{0}$. When $l_{0}$ is small, the two mirrors are working in the strong-coupling regime ($\xi^{2}\gg \gamma_{1}\gamma_{2}$), then the two mode temperatures will be the same, i.e., $T_{m1}=T_{m2}=(\gamma_{1}T_{1}+\gamma_{2}T_{2})/(\gamma_{1}+\gamma_{2})$, due to the existed phonon exchange. With the increase of $l_{0}$, the effective coupling between the two mirrors decreases, then the mode temperatures will approach to their individual environment temperature $T_{mn}=T_{n}$ for $n=1,2$ due to the thermalization effect~\cite{LiaoPRA2011}.

To further expound the heat transfer, we examine the heat flux in this system. According to Eq.~(\ref{efftiveHamitonian}), we can obtain the Hamiltonian $\hat{H}_{S}=\hbar\omega_{b,1}\hat{b}_{1}^{\dagger}\hat{b}_{1}+\hbar\omega_{b,2}\hat{b}_{2}^{\dagger}\hat{b}_{2}
+\hbar\xi(\hat{b}_{2}\hat{b}_{1}^{\dagger}+\hat{b}_{1}\hat{b}_{2}^{\dagger})$ in the Schr\"{o}dinger picture, where the frequency $\omega_{b,n}$ is defined by $\omega_{b,n}=\omega_{M}+\Delta_{b,n}$. According the expression $J_{\text{tot}}(t)=$Tr$[\dot{\rho}\hat{H}_{S}]$ of the total heat flux from the baths to the system, we obtain the steady-state heat flux in the degenerate resonator case,
\begin{equation}
J_{n,\text{ss}}=(-1)^{n+1}4\hbar\omega_{b}\frac{(\bar{n}_{\text{th},1}-\bar{n}_{\text{th},2})
\gamma_{1}\gamma_{2}\xi^{2}}{(\gamma_{1}+\gamma_{2})(\gamma_{1}\gamma_{2}+4\xi^{2})}, n=1,2
\end{equation}
with $\omega_{b}=\omega_{M}+\Delta_{b}$. In Fig.~\ref{modetemandheatflux}(b), we show the variation of the steady-state heat flux with the cavity length $l_{0}$. Due to the frequency shift of the two moving mirrors, we see that the flux disappears in the large $l_{0}$ case. This is reasonable because the heat flux becomes small with the decrease of the effective phonon exchange coupling.

\emph{Discussions and conclusion.}---Finally, we present some discussions concerning this work. (i) The physical model under consideration is a 1D model, which is different from the 3D experimental system considered in Ref.~\cite{FongNature2019}. However, the 1D model can be used to clarify the physical mechanism for heat transfer across a vacuum, and our method can be straightforwardly generalized to study the 3D model. (ii) The multimode description of the cavity field is critical for the microscopic theory. It not only provides the microscopic picture for describing the physical processes, but also gives the correct results of the physical quantities. In addition, the multimode-field description can lead to correct results for some physical effects such as the dark mode effect and the energy-spectrum lower-bound problem. (iii) Though the frequency shift and the coupling strength introduced in Eqs.~(\ref{Hamiparameters}) depend on the cut-off frequency, which is mostly determined by the materials of the mirrors in realistic cases, the microscopic mechanism for the vacuum-induced heat transfer established in this work is general, namely, the vacuum fluctuation can act as an intermediary to induce an effective phonon-exchange interaction between the two moving mirrors via virtual processes. (iv) Our results indicate that the vacuum induced second-order phonon-exchange interaction is much stronger than the phenomenologically-estimated Casimir-induced-interaction between the two moving mirrors. This means that the vacuum induced virtual process should be the main mechanism for the heat transfer in our model.

In conclusion, we have established a nonrelativistic microscopic theory for explaining the phonon heat transfer between two mirrors across a vacuum. This is achieved by considering both the cavity field and two moving mirrors quantum mechanically. Under the multimode-vacuum-field and linear-expansion framework, we have obtained a second-order effective phonon-exchange interaction to govern the heat transfer between the two mirrors, where both the frequency shift and coupling strength are proportional to the negative third power of the static cavity length, and the ratio between them is a constant determined by the cutoff mode number of the cavity fields. The heat transfer has been confirmed by examining both the mode temperatures and heat flux.

\emph{Acknowledgments.}---J.-Q.L. would like to thank Prof. Chang-Pu Sun, Prof. Le-Man Kuang, and Prof. Chi-Kwong Law for helpful discussions. J.-Q.L. was supported in part by National Natural Science Foundation of China (Grants No. 12175061, No. 12247105, No.11935006, and No. 12421005), National Key Research and Development Program of China (Grant No. 2024YFE0102400), and Hunan Provincial Major Sci-Tech Program (Grant No. 2023ZJ1010).

\newpage
\onecolumngrid
\newpage
\begin{center}
\textbf{\large Supplementary materials for ``Microscopic Theory of Heat Transfer across a Vacuum"}
\end{center}
\setcounter{equation}{0}
\setcounter{figure}{0}
\setcounter{table}{0}
\setcounter{page}{1}
\makeatletter
\renewcommand{\theequation}{S\arabic{equation}}
\renewcommand{\thefigure}{S\arabic{figure}}

\begin{center}
Yue-Hui Zhou$^{1}$ and Jie-Qiao Liao$^{1,2,3,^*}$
\end{center}

\begin{minipage}[]{18cm}
\small{\it
\centering $^{1}$Key Laboratory of Low-Dimensional Quantum Structures and Quantum Control of Ministry of Education, Key Laboratory for Matter Microstructure and Function of Hunan Province, Department of Physics and Synergetic Innovation Center for Quantum Effects and Applications, Hunan Normal University, Changsha 410081, China \\
\centering $^{2}$Hunan Research Center of the Basic Discipline for Quantum Effects and Quantum Technologies,  Hunan Normal University, Changsha 410081, China \\
\centering $^{3}$Institute of Interdisciplinary Studies, Hunan Normal University, Changsha, 410081, China \\}

\end{minipage}

\vspace{8mm}

In the supplementary materials, we present the detailed calculations and derivations, which are helpful to the understanding and following of the content of this paper. This document consists of nine parts: (I) The equations of motion for the two-moving-end-mirror optomechanical cavity; (II) The Lagrange function of the system; (III) The quantized Hamiltonian of the system described in the center-of-mass system; (IV) The quantized Hamiltonian under the linear-expansion approximation; (V) The effective Hamiltonian describing the dynamics of the two moving mirrors; (VI) The mode temperatures of the two moving end mirrors; (VII) Analyses of the system parameters; (VIII) Comparison of the phonon-exchange coupling strengths between the virtual-process mechanism and the Casimir-coupling mechanism; (IX) The quantized Hamiltonian of the three-cavity optomechanical system.

\section{I. The equations of motion for the two-moving-end-mirror optomechanical cavity}  \label{dq}

In this section, we derive the equations of motion for a one-dimensional cavity optomechanical system with two moving end mirrors under the  framework of nonrelativistic microscopic theory.
\subsection{A. The equations of motion and boundary condition of the cavity fields}

\begin{figure}
\center
\includegraphics[bb=23 648 276 708, width=0.6 \textwidth]{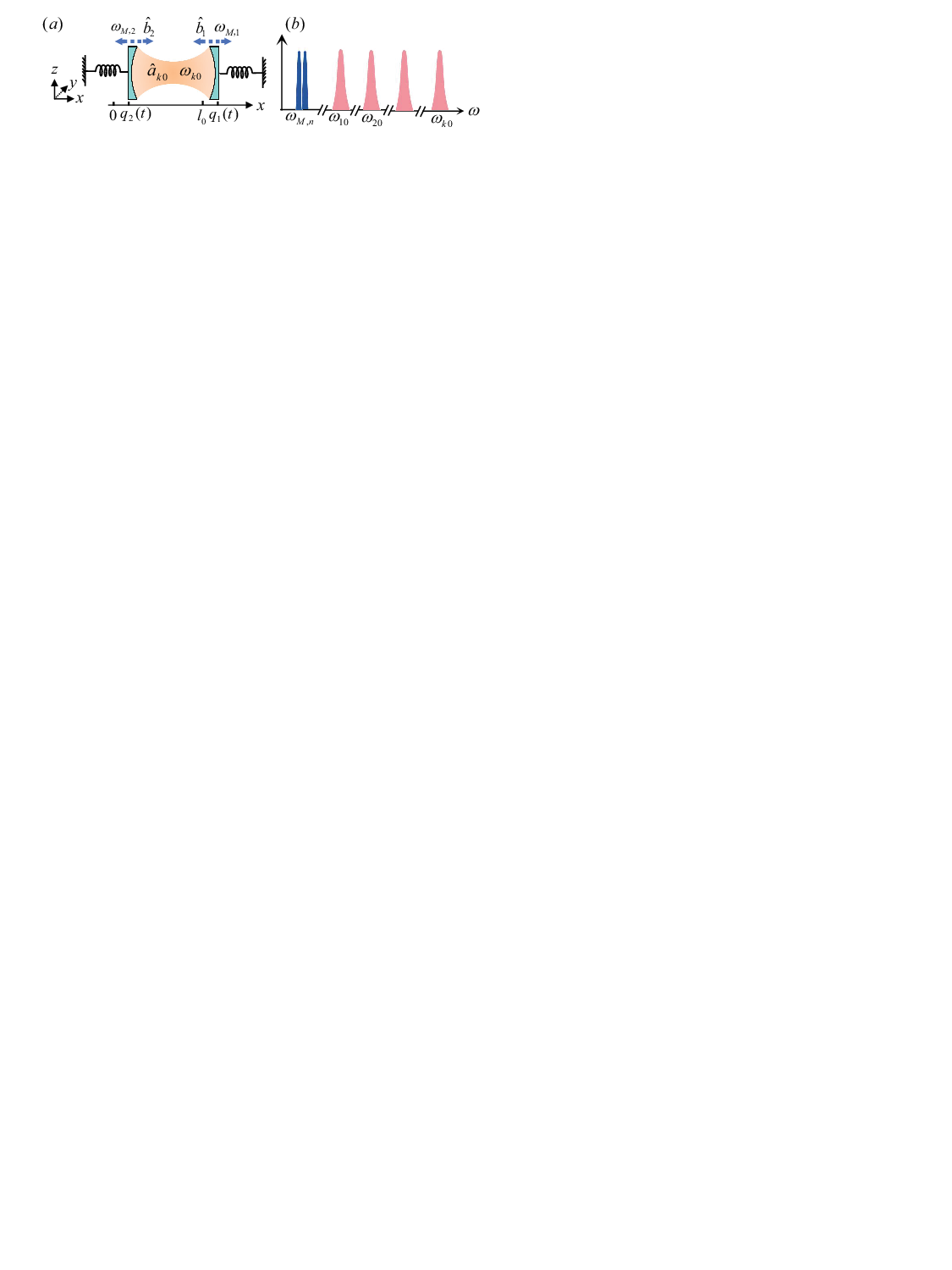}
\caption{(Color online) (a) Schematic of the one-dimensional optomechanical cavity formed by two movable end mirrors trapped in individual harmonic oscillator potential wells $V_{1}(q_{1})$ and $V_{2}(q_{2})$. The $\hat{b}_{1}$ ($\hat{b}_{2}$), $\hat{b}_{1}^{\dagger}$ ($\hat{b}_{2}^{\dagger}$), and $\omega_{\text{M},1}$ ($\omega_{\text{M},2}$) are the annihilation operator, creation operator, and resonance frequency of the right (left) moving mirror, respectively. The rest length of the cavity is $l_{0}$. The cavity-field modes corresponding to the rest cavity are denoted by $a_{k0}$ with the resonance frequency $\omega_{k0}$ ($k=1$, $2$, $3$, $\cdots$). (b) The resonance frequency distribution of the two moving mirrors and the rest-cavity field modes. Here, $\omega_{k0}=k\cdot\omega_{10}=k\pi c/l_{0}$ ($k=1$, $2$, $\cdots$) are the resonance frequency of the $k$th cavity-field mode for the rest cavity [$q_{1}(t)-q_{2}(t)=l_{0}$]. The resonance frequencies $\omega_{\text{M},1}$ and $\omega_{\text{M},2}$ of the two moving mirrors are much smaller than the basic mode frequency $\omega_{10}$ of the cavity field.}\label{mode}
\end{figure}

As introduced in the main text, the system under consideration is a one-dimensional cavity optomechanical system formed by two moving end mirrors [see Fig.~\ref{mode}(a)]. We denote the position (mass) of the two moving mirrors as $q_{1}(t)$ and $q_{2}(t)$ ($m_{1}$ and $m_{2}$), where the subscripts $1$ and $2$ denote the right and left end mirrors, respectively. In the absence of the free charge and free current, the vector potential $\vec{A}(x,t)$ in the region $q_{2}(t)\leq x\leq q_{1}(t)$ obeys the wave equation $(c=1)$
\begin{equation}
\partial^{2}_{x}\vec{A}(x,t)=\partial^{2}_{t}\vec{A}(x,t),  \label{vpequation}
\end{equation}
where the notations $\partial_{t}\equiv\frac{\partial}{\partial t}$ and $\partial_{x}\equiv\frac{\partial}{\partial x}$ denote the partial derivative with respect to $t$ and $x$, respectively. In the derivation of Eq.~(\ref{vpequation}), we have chosen $\nabla\cdot\vec{A}=0$ and scalar potential $\varphi=0$. Considering the gauge invariance of the vector potential $\vec{A}(x,t)$, the general forms of the vector potential $\vec{A}(x,t)$, electric field intensity $\vec{E}(x,t)$, and magnetic induction intensity $\vec{B}(x,t)$ can be written as
\begin{subequations}
\label{ydirequation}
    \begin{align}
    \vec{A}(x,t)&=A_{y}(x,t)\vec{e}_{y}+A_{z}(x,t)\vec{e}_{z}, \label{ydirequation:1a} \\
    \vec{E}(x,t)&=-\partial_{t}A_{y}(x,t)\vec{e}_{y}-\partial_{t}A_{z}(x,t)\vec{e}_{z}, \label{ydirequation:1b} \\
    \vec{B}(x,t)&=\partial_{x}A_{y}(x,t) \vec{e}_{z}-\partial_{x}A_{z}(x,t)\vec{e}_{y}, \label{ydirequation:1c}
    \end{align}
\end{subequations}
where $\vec{e}_{y}$ and $\vec{e}_{z}$ are the unit vectors along the $y$- and $z$-axis directions, respectively. In addition, $A_{y}$ and $A_{z}$ are the $y$- and $z$-axis components of the vector potential, respectively. Here, we point out that the quantities $\vec{A}(x,t)$, $\vec{E}(x,t)$, and $\vec{B}(x,t)$ are defined in the laboratory frame.

For finding the boundary condition of the cavity field, we introduce the reference frame $S_{j}$, in which the $j$th ($j=1,2$) moving mirror is instantaneously static. In particular, here we approximatively treat the transformation from the laboratory reference frame to the reference frame $S_{j}$ as an inertial reference frame transformation. Then, by the Lorentz transformation
\begin{subequations}
\label{eq}
    \begin{align}
\vec{E}_{j=1,2}&=\frac{1}{\sqrt{1-\dot{q}_{j}^{2}(t)}}[\vec{E}+\dot{q}_{j}(t)\vec{e}_{x}\times\vec{B}],  \label{eq:1a} \\
\vec{B}_{j=1,2}&=\frac{1}{\sqrt{1-\dot{q}_{j}^{2}(t)}}[\vec{B}-\dot{q}_{j}(t)\vec{e}_{x}\times\vec{E}],  \label{eq:1b}
\end{align}
\end{subequations}
the electric field intensity $\vec{E}_{j}(x,t)$ and the magnetic induction intensity $\vec{B}_{j}(x,t)$ in the reference frame $S_{j}$ can be written as
\begin{subequations}
\label{stationreelecfield}
    \begin{align}
\vec{E}_{j=1,2}(x,t)&=-\frac{1}{\sqrt{1-\dot{q}_{j}^{2}(t)}}\left\{\left[\frac{\partial A_{y}(x,t)}{\partial t}+\dot{q}_{j}(t)\frac{\partial
A_{y}(x,t)}{\partial x}\right]\vec{e}_{y}+\left[\frac{\partial A_{z}(x,t)}{\partial t}+\dot{q}_{j}(t)\frac{\partial A_{z}(x,t)}{\partial x}\right] \vec{e}_{z}\right\},                      \label{stationreelecfield:1a} \\
\vec{B}_{j=1,2}(x,t)&=-\frac{1}{\sqrt{1-\dot{q}_{j}^{2}(t)}}\left\{\left[\frac{\partial A_{z}(x,t)}{\partial x}+\dot{q}_{j}(t)\frac{\partial
A_{z}(x,t)}{\partial t}\right] \vec{e}_{y}-\left[\frac{\partial A_{y}(x,t)}{%
\partial x}+\dot{q}_{j}(t)\frac{\partial A_{y}(x,t)}{\partial t}\right]\vec{e}_{z}\right\}. \label{stationreelecfield:1b}
\end{align}
\end{subequations}
Here, $\dot{q}_{j=1,2}(t)$ is the instantaneous velocity of the $j$th moving mirror. Further, the electric field intensity on the surface of the $j$th ($j=1,2$) moving mirror can be expressed in the reference frame $S_{j}$ as
\begin{equation}
\vec{E}_{j=1,2}(q_{j}(t),t) =-\frac{1}{\sqrt{1-\dot{q}_{j}^{2}(t)}}\left[\frac{dA_{y}(q_{j}(t),t)}{dt}\vec{e}_{y}+\frac{dA_{z}(q_{j}(t),t)}{dt}\vec{e}_{z}\right].
\end{equation}
In order to confirm $\vec{E}_{1}(q_{1}(t),t)=0$ and $\vec{E}_{2}(q_{2}(t),t) =0$, we can choose
\begin{equation}
\vec{A}(q_{1}(t),t)=0,\hspace{1.0 cm} \vec{A}(q_{2}(t),t)=0,  \label{bocondition}
\end{equation}
which are the boundary conditions of the cavity fields in this system.

For simplicity, we consider the linear polarized light fields in this work such that the vector potential $\vec{A}(x,t)$ can be treated as a scalar quantity $A(x,t)$. Then Eq.~(\ref{vpequation}) is reduced to
\begin{equation}
\partial^{2}_{x}A(x,t)=\partial^{2}_{t}A(x,t),  \label{scequation}
\end{equation}
where the boundary conditions given in Eq.~(\ref{bocondition}) can be written as $A(q_{1}(t),t)=A(q_{2}(t),t)=0$~\cite{MooreJMPs1970,Akopyan2021PRD}. Therefore, the solution of Eq.~(\ref{scequation}) depends on the transient positions of the two moving end mirrors.

\subsection{B. The equations of motion of the two moving end mirrors}

For the present cavity optomechanical system, the coordinators $q_{1}(t)$ and $q_{2}(t)$ of the two end mirrors should be considered as dynamic variables rather than parameters. Without causing confusion, below we mark $q_{1}(t)$ and $q_{2}(t)$ as $q_{1}$ and $q_{2}$, respectively, for keeping the expressions concise. In the reference frame $S_{j}$ for $j=1,2$, the momentum density $\vec{g}_{j}$ of the electromagnetic fields is defined by
\begin{equation}
\vec{g}_{j=1,2}=(-1)^{j+1}w_{j}\vec{e}_{x},
\end{equation}
where $\vec{e}_{x}$ is the unit vector in the $x$-axis direction, $w_{j}=\vec{B}_{j,i}\cdot\vec{B}_{j,i}$ is the energy flow density with $\vec{B}_{j,i}$ being the magnetic induction intensity vertically incident on the surface of the $j$th ($j=1,2$) end mirror, with the subscript ``$i$'' representing the meaning of incidence. We have assumed that the incident light is a plane wave. Then within
the time interval $\Delta t$, the momentum of the electromagnetic wave incident vertically on the surface of the $j$th end mirror with the area $\Delta s_{j}$ can be written as
\begin{equation}
\vec{P}_{j=1,2}=\int\vec{g}_{j,i}dV=(-1)^{j+1}\Delta t\cdot \Delta s_{j}\cdot w_{j,i}\cdot\vec{e}_{x}.
\end{equation}

In the process of complete reflection, the increment of momentum can be written as $-2\vec{P}_{j}$, then the momentum theorem can be expressed as
\begin{equation}
\vec{F}_{j=1,2}\cdot\Delta t=-2\vec{P}_{j},
\end{equation}
where $\vec{F}_{j}$ is the force exerted by the $j$th end mirror with an area $\Delta s_{j}$ on the electromagnetic fields. According to Newton's third law, the force exerted by the electromagnetic fields on the mirror is $-\vec{F}_{j}$. For the one-dimensional cavity, there is no medium in the cavity, then the radiation pressure $\vec{f}_{j}$ experienced by the mirror is
\begin{equation}
\vec{f}_{j=1,2}=-\frac{\vec{F}_{j}}{\Delta s_{j}}=2(-1)^{j+1}(\vec{B}_{j,i}\cdot\vec{B}_{j,i})|_{x=q_{j}}\vec{e}_{x}.
\end{equation}
According to the boundary conditions of the electromagnetic fields, the magnetic induction intensity $\vec{B}_{j}$ is equal to $2\vec{B}_{j,i}$, then the radiation-pressure force $\vec{f}_{j}$ of the cavity field acting on the $j$th moving end mirror can be written as~\cite{LawPRA1995}
\begin{equation}
\vec{f}_{j=1,2}=\frac{(-1)^{j+1}}{2}[\vec{B}_{j}(x,t)\cdot\vec{B}_{j}(x,t)]|_{x=q_{j}}\vec{e}_{x}. \label{rpfsz}
\end{equation}%

According to Eq.~(\ref{stationreelecfield:1a}) and the conditions $\vec{E}_{j=1,2}(q_{j},t)=0$, we have the relations,
\begin{subequations}
\label{eqqbe}
    \begin{align}
\left.\frac{\partial A_{y}(x,t)}{\partial t}\right|_{x=q_{j}}&=-\dot{q}_{j}\left.\frac{\partial A_{y}(x,t)}{\partial x}\right|_{x=q_{j}},   \\
\left.\frac{\partial A_{z}(x,t)}{\partial t}\right|_{x=q_{j}}&=-\dot{q}_{j}\left.\frac{\partial A_{z}(x,t)}{\partial x}\right|_{x=q_{j}}.
\end{align}
\end{subequations}
Using Eqs.~(\ref{stationreelecfield:1b}) and~(\ref{eqqbe}), we have
\begin{equation}
\left.\vec{B}_{j}(x,t)|_{x=q_{j}}=-(1-\dot{q}_{j}^{2})^{1/2}\left[\frac{\partial
A_{z}(x,t)}{\partial x}\vec{e}_{y}-\frac{\partial A_{y}(x,t)}{\partial x}\vec{e}_{z}\right]\right|_{x=q_{j}}. \label{bjcicb}
\end{equation}
Substituting the above equation into Eq.~(\ref{rpfsz}), the radiation-pressure force $\vec{f}_{j}$ in the laboratory frame can be expressed as
\begin{equation}
\vec{f}_{j=1,2}=\frac{(-1)^{j+1}}{2}(1-\dot{q}_{j}^{2})\left.\left\{\left[\frac{%
\partial A_{z}(x,t)}{\partial x}\right]^{2}+\left[\frac{\partial A_{y}(x,t)}{\partial x}\right]^{2}\right\}\right|_{x=q_{j}}\vec{e}_{x}.\label{rforcesp}
\end{equation}
Based on Eq.~(\ref{rforcesp}), we find that it is sufficient to treat $A(x,t)$ as a scalar field in our system. In this work, we consider the case of $\dot{q}_{1}\ll1$ and $\dot{q}_{2}\ll1$, then the expression of the radiation-pressure force $\vec{f}_{j}$ can be approximately reduced as
\begin{equation}
\vec{f}_{j}=\frac{(-1)^{j+1}}{2}\left.\left[\frac{\partial A(x,t)}{\partial x}\right]\right|_{x=q_{j}}^{2}\vec{e}_{x}.
\end{equation}

According to Newton's second law, the equations of motion of the two moving mirrors can be expressed as
\begin{equation}
m_{j}\ddot{q}_{j}=-\partial_{q_{j}}V_{j}(q_{j})+(-1)^{j+1}\frac{1}{2}\left. \left[\partial_{x}A(x,t)\right]^{2}\right\vert_{x=q_{j}},  \hspace{1 cm}j=1,2,   \label{mirrorsequations}
\end{equation}
where the first term on the right-hand side of Eq.~(\ref{mirrorsequations}) is called the conservative force. We assume that the two mirrors cannot touch each other ($q_{1}-q_{2}>0$), which can be achieved by properly choosing the potential wells $V_{1}(q_{1})$ and $V_{2}(q_{2})$.

\subsection{C. The equations of motion of the cavity-field modes and the two moving mirrors expressed with the generalized coordinates}

In this section, we show how to derive the equations of motion of the system expressed with the generalized coordinates of the optomechanical cavity including the cavity fields and the vibrating mirrors. For the cavity fields, we can define a set of complete generalized coordinates as
\begin{equation}
Q_{k}\left( t\right) \equiv \sqrt{\frac{2}{q_{1}-q_{2}}}\int_{q_{2}}^{q_{1}}\sin \left[ \omega _{k}\left( x-q_{2}\right) \right] A\left( x,t\right) dx,  \label{genecoord}
\end{equation}
where $k$ is a positive integer ($k=1$, $2$, $3$, $\cdots$) and $\omega_{k}$ denotes the position-dependent cavity frequency defined by
\begin{equation}
\omega_{k}=\frac{k\pi}{q_{1}-q_{2}}.  \label{cavityfreq}
\end{equation}
Using the orthogonality of the sinusoidal functions, the scalar quantity $A(x,t)$ can be expressed as
\begin{equation}
A(x,t) =\sum_{k=1}^{\infty}Q_{k}(t)\varphi_{k}(x),   \label{vector potential}
\end{equation}
where the mode function $\varphi_{k}(x)$ is defined by
\begin{equation}
\varphi_{k}(x)=\sqrt{\frac{2}{q_{1}-q_{2}}}\sin\left[\omega_{k}(x-q_{2})\right], \hspace{1 cm} q_{2}\leq x\leq q_{1}.           \label{function}
\end{equation}
Considering the boundary condition $A(q_{1}(t),t)=A(q_{2}(t),t)=0$, we can also obtain Eq.~(\ref{vector potential}) by using the Fourier series expansion, where the generalized coordinates $Q_{k}(t)$ can be regarded as the coefficients of the Fourier series expansion. Without causing confusion, below we denote $Q_{k}(t)$ and $\varphi_{k}(x)$ as $Q_{k}$ and $\varphi_{k}$, respectively, for keeping the equations concise.

According to Eqs.~(\ref{vector potential}) and (\ref{function}), we can obtain the first and second derivatives of $A(x,t)$ with respect to the position $x$ and time $t$, and these results are given by
\begin{subequations}
\label{vpdos}
    \begin{align}
\frac{\partial A(x,t)}{\partial x} &=\sum_{k=1}^{\infty}Q_{k}\frac{\partial\varphi_{k}}{\partial x},  \label{vpdos:1a} \\
\frac{\partial^{2}A(x,t)}{\partial x^{2}} &=-\sum_{k=1}^{\infty}\omega_{k}^{2}Q_{k}\varphi_{k},       \label{vpdos:1b} \\
\frac{\partial A(x,t)}{\partial t} &=\sum_{k=1}^{\infty}\dot{Q}_{k}\varphi_{k}+\sum_{k=1}^{\infty}\sum_{n=1,2}Q_{k}\frac{\partial
\varphi_{k}}{\partial q_{n}}\dot{q}_{n},                                                                \label{vpdos:1c} \\
\frac{\partial^{2}A(x,t)}{\partial t^{2}} &=\sum_{k=1}^{\infty}\left[\ddot{Q}_{k}\varphi_{k}+\sum_{n=1,2}Q_{k}\left(\frac{\partial^{2}\varphi
_{k}}{\partial q_{n}^{2}}\dot{q}_{n}^{2}+\frac{\partial\varphi_{k}}{\partial q_{n}}\ddot{q}_{n}\right)+2\sum_{n=1,2}\dot{Q}_{k}\frac{\partial
\varphi_{k}}{\partial q_{n}}\dot{q}_{n}+2Q_{k}\frac{\partial^{2}\varphi_{k}}{\partial q_{1}\partial q_{2}}\dot{q}_{1}\dot{q}_{2}\right].\label{vpdos:1d}
\end{align}
\end{subequations}
Below, we will derive the equations of the systems using the generalized coordinates of the cavity fields. On one hand, based on Eqs.~(\ref{scequation}) and (\ref{vpdos}), we can obtain the equation of motion for the cavity fields as
\begin{eqnarray}
-\sum_{k=1}^{\infty}\omega_{k}^{2}Q_{k}\varphi_{k} &=&\sum_{k=1}^{\infty}\ddot{Q}_{k}\varphi_{k}+2\sum_{k=1}^{\infty}\sum_{n=1,2}\dot{Q}_{k}
\frac{\partial\varphi_{k}}{\partial q_{n}}\dot{q}_{n}+\sum_{k=1}^{\infty}\sum_{n=1,2}Q_{k}\frac{\partial\varphi_{k}}{\partial q_{n}}\ddot{q}_{n}                                       \nonumber \\
&&+2\sum_{k=1}^{\infty}Q_{k}\frac{\partial^{2}\varphi_{k}}{\partial q_{1}\partial q_{2}}\dot{q}_{1}\dot{q}_{2}+\sum_{k=1}^{\infty
}\sum_{n=1,2}Q_{k}\frac{\partial^{2}\varphi_{k}}{\partial q_{n}^{2}}\dot{q}_{n}^{2}.         \label{qk}
\end{eqnarray}
Multiply both sides of Eq.~(\ref{qk}) by the function $\varphi_{s}$ and perform the integral with respect to $x$, we can further obtain the equations of motion of the cavity fields as
\begin{eqnarray}
\ddot{Q}_{k} &=&-\omega_{k}^{2}Q_{k}-\sum_{n=1,2}\sum_{j=1}^{\infty}\frac{%
2g_{jk}^{(n)}\dot{q}_{n}\dot{Q}_{j}}{q_{1}-q_{2}}+\sum_{l=1}^{\infty}\frac{2\dot{q}_{1}\dot{q}_{2}g_{lk}^{(2)}Q_{l}}{(q_{1}-q_{2})^{2}}%
+\sum_{j,l=1}^{\infty}\frac{2\dot{q}_{1}\dot{q}_{2}g_{kj}^{(1)}g_{lj}^{(2)}Q_{l}}{(q_{1}-q_{2})^{2}}                    \nonumber \\
&&+\sum_{n=1,2}\sum_{j,l=1}^{\infty}\frac{\dot{q}_{n}^{2}g_{kj}^{(n)}g_{lj}^{(n)}Q_{l}}{(q_{1}-q_{2})^{2}}-\sum_{n=1,2}%
\sum_{j=1}^{\infty}g_{jk}^{(n)}\frac{(q_{1}-q_{2})\ddot{q}_{n}+(-1)^{n}\dot{q}_{n}^{2}}{(q_{1}-q_{2})^{2}}Q_{j},         \label{dynaequationsa}
\end{eqnarray}
where the dimensionless coefficients $g_{kj}^{(n)}$ are introduced as
\begin{equation}
g_{kj}^{(n)}=\left\{
\begin{array}{c}
\frac{2kj}{j^{2}-k^{2}}[(-1)^{j+k}\delta_{n,1}-\delta_{n,2}], \hspace{0.25 cm} j\neq k, \\
0,\hspace{0.25 cm} j=k,
\end{array}%
\right.  \label{dimensionlesscoeff}
\end{equation}
with $g_{jk}^{(n)}=-g_{kj}^{(n)}$. Note that in the derivation of Eq.~(\ref{dynaequationsa}), we have used the following relations:
\begin{subequations}
\label{matherela}
    \begin{align}
\int_{q_{2}}^{q_{1}}\varphi_{j}\varphi_{k}dx &=\delta_{j,k},            \label{matherela:1a} \\
\int_{q_{2}}^{q_{1}}\varphi_{j}\frac{\partial\varphi_{k}}{\partial q_{n}}dx &=\frac{g_{kj}^{(n)}}{q_{1}-q_{2}},\hspace{0.25 cm} n=1,2, \label{matherela:1b} \\
\int_{q_{2}}^{q_{1}}\frac{\partial\varphi_{j}}{\partial q_{n}}\frac{\partial\varphi_{l}}{\partial q_{n^{\prime}}}dx &=\frac{%
\sum_{k=1}^{\infty}g_{jk}^{(n)}g_{lk}^{(n^{\prime})}}{(q_{1}-q_{2})^{2}},\hspace{0.25 cm} n,n^{\prime}=1,2,   \label{matherela:1c} \\
\int_{q_{2}}^{q_{1}}\varphi_{j}\frac{\partial ^{2}\varphi_{k}}{\partial q_{n}^{2}}dx &=\frac{(-1)^{n}g_{kj}^{(n)}}{(q_{1}-q_{2})^{2}}-\frac{%
\sum_{s=1}^{\infty}g_{js}^{(n)}g_{ks}^{(n)}}{(q_{1}-q_{2})^{2}},\hspace{0.25 cm} n=1,2,   \label{matherela:1d} \\
\int_{q_{2}}^{q_{1}}\varphi_{j}\frac{\partial ^{2}\varphi_{k}}{\partial q_{1}\partial q_{2}}dx &=-\frac{g_{kj}^{(2)}}{(q_{1}-q_{2})^{2}}-\frac{%
\sum_{s=1}^{\infty}g_{ks}^{(2)}g_{js}^{(1)}}{(q_{1}-q_{2})^{2}},   \label{matherela:1e} \\
g_{kj}^{(2)}+\sum_{s=0}^{\infty }g_{ks}^{(2)}g_{js}^{(1)} &=0,\hspace{0.25 cm} j+k=\text{odd},  \label{matherela:1f}
\end{align}
\end{subequations}
where $j+k=\text{odd}$ means that the sum of $j$ and $k$ is an odd number.

On the other hand, for the moving mirrors, Eq.~(\ref{mirrorsequations}) can be rewritten, in terms of Eqs.~(\ref{vector potential})-(\ref{vpdos}),  as
\begin{equation}
\ddot{q}_{n}=-\frac{1}{m_{n}}\frac{\partial V_{n}(q_{n})}{\partial q_{n}}+\sum_{k,j=1}^{\infty}[(-1)^{j+k}\delta_{n,1}-\delta_{n,2}]\frac{\omega
_{k}\omega_{j}Q_{k}Q_{j}}{m_{n}(q_{1}-q_{2})}, \hspace{0.25 cm} n=1,2.   \label{dynaequationsb}
\end{equation}
The dynamics of the coupled system is governed by Eqs.~(\ref{dynaequationsa}) and~(\ref{dynaequationsb}), which are typically difficult to be solved. It can be seen from Eq.~(\ref{dynaequationsb}) that, no matter how to combine $q_{1}$ and $q_{2}$ as a new variable, the first derivative of this new variable over the time is related to the generalized coordinates $Q_{k}$ of the cavity fields in the absence of the single-mode approximation, which implies that there is no mechanical dark mode in the present Fabry-P\'{e}rot-type two-moving-mirror cavity optomechanical system after the canonical quantization under the multimode-cavity-field framework.

\subsection{D. Derivation of the relations in Eqs.~(\ref{matherela:1a})-(\ref{matherela:1f})}

In this section, we present the detailed derivation of the relations given in Eqs.~(\ref{matherela:1a})-(\ref{matherela:1f}). Based on Eq.~(\ref{function}), we know that
\begin{eqnarray}
\int_{q_{2}}^{q_{1}}\varphi_{j}\varphi_{k}dx &=&\frac{2}{q_{1}-q_{2}}\int_{q_{2}}^{q_{1}}\sin [\omega _{j}(x-q_{2})]\sin [\omega _{k}(x-q_{2})]dx   \nonumber\\
&=&\frac{2}{q_{1}-q_{2}}\int_{q_{2}}^{q_{1}}\left\{\cos[(\omega_{j}-\omega_{k})(x-q_{2})]-\cos[(\omega_{j}+\omega_{k})(x-q_{2})]\right\} dx \nonumber\\
&=&\delta_{j,k},   \label{ssjihc}
\end{eqnarray}
where we used the integral formulas
\begin{subequations}
\label{cjifen}
    \begin{align}
    \int_{q_{2}}^{q_{1}}\cos[(\omega_{j}-\omega_{k})(x-q_{2})]dx&=(q_{1}-q_{2})\delta_{j,k}, \label{cjifen:1a} \\
    \int_{q_{2}}^{q_{1}}\cos[(\omega _{j}+\omega_{k})(x-q_{2})] dx&=0. \label{cjifen:1b}
    \end{align}
\end{subequations}
For derivation of other relations, we need to calculate the first-order partial derivative of the function $\varphi_{k}$ with respect to $q_{n=1,2}$,
\begin{equation}
\frac{\partial\varphi_{k}}{\partial q_{n}}=\frac{1}{2}\frac{(-1)^{n}}{q_{1}-q_{2}}\varphi_{k}+\sqrt{2}\frac{(-1)^{n+1}\omega_{k}(q_{(n\:\text{mod}   \:2)+1}-x)}{(q_{1}-q_{2})^{\frac{3}{2}}}\cos[\omega_{k}(x-q_{2})],   \label{yijiedao}
\end{equation}
where ``$n$\:mod\:2'' represents the remainder after dividing the divisor $n$ by 2. Namely, $q_{(1\:\text{mod}\:2)+1}=q_{2}$ and $q_{(2\:\text{mod}\:2)+1}=q_{1}$. Multiply both sides of Eq.~(\ref{yijiedao}) by the function $\varphi_{j}$ and perform the integral with respect to $x$, we can obtain
\begin{eqnarray}
\int_{q_{2}}^{q_{1}}\varphi_{j}\frac{\partial\varphi_{k}}{\partial q_{n}}dx &=&\frac{1}{2}\frac{(-1)^{n}}{q_{1}-q_{2}}\delta_{j,k}+\frac{%
(-1)^{n+1}\omega_{k}q_{(n\:\text{mod}\:2)+1}}{(q_{1}-q_{2})^{2}}\int_{q_{2}}^{q_{1}}2\sin[\omega_{j}(x-q_{2})]\cos[\omega_{k}(x-q_{2})]dx                \nonumber\\
&&-\frac{(-1)^{n+1}\omega_{k}}{(q_{1}-q_{2})^{2}}\int_{q_{2}}^{q_{1}}2x\sin[\omega_{j}(x-q_{2})]\cos[\omega_{k}(x-q_{2})]dx, \label{24bzhenm}
\end{eqnarray}
where we used the result in Eq.~(\ref{matherela:1a}). Using the product-to-sum formula, we have
\begin{equation}
2\sin[\omega_{j}(x-q_{2})]\cos[\omega_{k}(x-q_{2})] =\sin[(\omega_{j}+\omega_{k})(x-q_{2})]+\sin[(\omega_{j}-\omega_{k})(x-q_{2})] \label{scjihc}.
\end{equation}
Integrating $x$ on both sides of Eq.~(\ref{scjihc}) yields
\begin{equation}
\label{scjihcjifen}
2\int_{q_{2}}^{q_{1}}\sin[\omega_{j}(x-q_{2})]\cos[\omega_{k}(x-q_{2})]dx=\left\{
\begin{array}{c}
-\frac{2j(q_{1}-q_{2})}{\pi (j^{2}-k^{2})}[(-1)^{j+k}-1],\hspace{1cm}j\neq k,  \\
0,\hspace{1cm}j=k.
\end{array}
\right.
\end{equation}
By multiplying both sides of Eq.~(\ref{scjihc}) by $x$ and performing the integration with respect to $x$, we have
\begin{equation}
\label{xscjihcjifen}
2\int_{q_{2}}^{q_{1}}x\sin[\omega_{j}(x-q_{2})]\cos [\omega_{k}(x-q_{2})]dx=\left\{
\begin{array}{c}
-\frac{2j(q_{1}-q_{2})}{\pi(j^{2}-k^{2})}[(-1)^{j+k}q_{1}-q_{2}],\hspace{1 cm}     j\neq k, \\
-\frac{(q_{1}-q_{2})^{2}}{\pi(j+k)},\hspace{1 cm} j=k.
\end{array}%
\right.
\end{equation}
In terms of Eqs.~(\ref{scjihcjifen}) and~(\ref{xscjihcjifen}), Eq.~(\ref{24bzhenm}) can be reduced to Eq.~(\ref{matherela:1b}).

According to Eq.~(\ref{matherela:1b}), we have the relations,
\begin{equation}
g_{jk}^{(n)}=(q_{1}-q_{2})\int_{q_{2}}^{q_{1}}\varphi_{k}(x)\frac{\partial\varphi_{j}(x)}{\partial q_{n}}dx,\hspace{1cm}g_{lk}^{(n^{\prime
})}=(q_{1}-q_{2})\int_{q_{2}}^{q_{1}}\varphi_{k}(x^{\prime})\frac{\partial\varphi_{l}(x^{\prime})}{\partial q_{n^{\prime}}}dx^{\prime},
\end{equation}
and we can then obtain the double integral
\begin{equation}
\sum_{k=1}^{\infty}g_{jk}^{(n)}g_{lk}^{(n^{\prime})}=(q_{1}-q_{2})^{2}\int_{q_{2}}^{q_{1}}\int_{q_{2}}^{q_{1}}\left[
\sum_{k=1}^{\infty}\varphi_{k}(x)\varphi_{k}(x^{\prime})\right]\frac{\partial\varphi_{j}(x)}{\partial q_{n}}\frac{\partial\varphi_{l}(x^{\prime})}{\partial
q_{n^{\prime}}}dxdx^{\prime}. \label{doubleinte}
\end{equation}
Utilize the completeness relation
\begin{equation}
\sum_{k=1}^{\infty}\varphi_{k}(x)\varphi_{k}(x^{\prime})=\delta(x-x^{\prime}),
\end{equation}
Eq.~(\ref{doubleinte}) can be reduced to Eq.~(\ref{matherela:1c}) accordingly.

To derive the relation in Eq.~(\ref{matherela:1d}), we need to calculate the second-order partial derivative of the function $\varphi_{k}$ with respect to $q_{n=1,2}$. After a detailed derivation, we have
\begin{equation}
\frac{\partial^{2}\varphi_{k}}{\partial q_{n}^{2}}=\frac{3}{4}\frac{1}{(q_{1}-q_{2})^{2}}\varphi_{k}-3\sqrt{2}\frac{\omega_{k}(
q_{(n\:\text{mod}\:2)+1}-x)}{(q_{1}-q_{2})^{\frac{5}{2}}}\cos[\omega_{k}(x-q_{2})]
-\frac{\omega_{k}^{2}(q_{(n\:\text{mod}\:2)+1}-x)^{2}}{(q_{1}-q_{2})^{2}}\varphi_{k}.  \label{ejiedao}
\end{equation}
Multiply both sides of Eq.~(\ref{ejiedao}) by the function $\varphi_{j}$ and perform the integral with respect to $x$, we can obtain the following equation
\begin{eqnarray}
\int_{q_{2}}^{q_{1}}\varphi_{j}\frac{\partial^{2}\varphi_{k}}{\partial q_{n}^{2}}dx &=&\frac{1}{4(q_{1}-q_{2})^{2}}(3-4\omega
_{k}^{2}q_{(n\:\text{mod}\:2)+1}^{2})\delta_{j,k}-\frac{6\omega_{k}q_{(n\:\text{mod}\:2)+1}}{(q_{1}-q_{2})^{3}}\int_{q_{2}}^{q_{1}}\sin [\omega_{j}(x-q_{2})]\cos[\omega_{k}(x-q_{2})]dx                                                                       \nonumber\\
&&+\frac{6\omega_{k}}{(q_{1}-q_{2})^{3}}\int_{q_{2}}^{q_{1}}x\sin[\omega_{j}(x-q_{2})]\cos[\omega_{k}(x-q_{2})]dx              \nonumber\\
&&-\frac{2\omega_{k}^{2}}{(q_{1}-q_{2})^{3}}\int_{q_{2}}^{q_{1}}x^{2}\sin[\omega_{j}(x-q_{2})]\sin[\omega_{k}(x-q_{2})]dx       \nonumber\\
&&+\frac{4\omega_{k}^{2}q_{(n\:\text{mod}\:2)+1}}{(q_{1}-q_{2})^{3}}\int_{q_{2}}^{q_{1}}x\sin[\omega_{j}(x-q_{2})]\sin[\omega_{k}(x-q_{2})]dx,   \label{24dzhenm}
\end{eqnarray}
where we used the relation in Eq.~(\ref{matherela:1a}). According to Eqs.~(\ref{scjihcjifen}) and~(\ref{xscjihcjifen}), as well as the following integration formulas
\begin{subequations}
\label{xandxxzzjifen}
    \begin{align}
\int_{q_{2}}^{q_{1}}x\sin[\omega_{j}(x-q_{2})]\sin[\omega_{k}(x-q_{2})] dx&=\left\{
\begin{array}{c}
\frac{2jk}{\pi^{2}(j+k)^{2}(j-k)^{2}}(q_{1}-q_{2})^{2}[(-1)^{j+k}-1],\hspace{0.5 cm} j\neq k,   \\
\frac{q_{1}^{2}-q_{2}^{2}}{4},\hspace{0.5 cm} j=k,
\end{array}%
\right.            \label{xzzjifen:1a} \\
\int_{q_{2}}^{q_{1}}x^{2}\sin[\omega_{j}(x-q_{2})]\sin[\omega_{k}(x-q_{2})] dx&=\left\{
\begin{array}{c}
\frac{4jk}{\pi^{2}(j-k)^{2}(j+k)^{2}}(q_{1}-q_{2})^{2}[(-1)^{j+k}q_{1}-q_{2}],\hspace{0.5 cm}    j\neq k, \\
\frac{q_{1}^{3}-q_{2}^{3}}{6}-\frac{(q_{1}-q_{2})^{3}}{\pi^{2}(j+k)^{2}},\hspace{0.5 cm} j=k,
\end{array}%
\right. \label{xxzzjifen:1b}
\end{align}
\end{subequations}
Eq.~(\ref{24dzhenm}) can be reduced to
\begin{equation}
\label{erjiejifjieg}
\int_{q_{2}}^{q_{1}}\varphi_{j}\frac{\partial^{2}\varphi_{k}}{\partial q_{n}^{2}}dx=\left\{
\begin{array}{c}
-[(-1)^{j+k}\delta_{n,1}+\delta_{n,2}]\frac{4jk(j^{2}+k^{2})}{(q_{1}-q_{2})^{2}(
j^{2}-k^{2})^{2}}-[(-1)^{j+k}\delta_{n,1}+\delta_{n,2}] \frac{2jk}{(q_{1}-q_{2})^{2}(j^{2}-k^{2})},\hspace{0.5 cm} j\neq k, \\
-\frac{1}{4(q_{1}-q_{2})^{2}}-\frac{k^{2}\pi^{2}}{3(q_{1}-q_{2})^{2}},\hspace{0.5 cm} j=k.
\end{array}%
\right.
\end{equation}
To simplify the calculation, we need to seek for the relation between the second-order partial derivative function and the product of the two first-order partial derivative functions. To this end, we try to derive the integral $\int_{q_{2}}^{q_{1}}\frac{\partial\varphi_{j}}{\partial q_{n}}\frac{\partial\varphi_{k}}{\partial q_{n}}dx$ based on Eqs.~(\ref{yijiedao}) and~(\ref{matherela:1a}),
\begin{eqnarray}
\label{erjiejifjieggg}
\int_{q_{2}}^{q_{1}}\frac{\partial\varphi_{j}}{\partial q_{n}}\frac{\partial\varphi_{k}}{\partial q_{n}}dx &=&\frac{1}{4(
q_{1}-q_{2})^{2}}\delta_{j,k}-\frac{\omega_{k}}{(q_{1}-q_{2})^{3}}\int_{q_{2}}^{q_{1}}(q_{(n\:\text{mod}\:2)+1}-x)\sin[\omega_{j}(x-q_{2}) ]\cos[\omega_{k}(x-q_{2})]dx   \nonumber\\
&&-\frac{\omega_{j}}{(q_{1}-q_{2})^{3}}\int_{q_{2}}^{q_{1}}(q_{(n\:\text{mod}\:2)+1}-x)\sin[\omega_{k}(x-q_{2})]\cos[\omega_{j}(x-q_{2})]dx     \nonumber\\
&&+\frac{2\omega_{j}\omega_{k}}{(q_{1}-q_{2})^{3}}\int_{q_{2}}^{q_{1}}(q_{(n\:\text{mod}\:2)+1}-x)^{2}\cos[\omega_{j}(x-q_{2})]\cos[\omega_{k}(x-q_{2})]dx.
\end{eqnarray}
In terms of Eqs.~(\ref{scjihcjifen}) and ~(\ref{xscjihcjifen}), as well as the following integration formulas
\begin{subequations}
\label{xandxxccjifen}
    \begin{align}
\int_{q_{2}}^{q_{1}}\cos[\omega_{k}(x-q_{2})]\cos[\omega_{j}(x-q_{2})]dx&=\frac{1}{2}(q_{1}-q_{2})\delta_{j,k},  \label{ccjifen:1a} \\
\int_{q_{2}}^{q_{1}}x\cos[\omega_{k}(x-q_{2})]\cos[\omega_{j}(x-q_{2})]dx&=\left\{
\begin{array}{c}
\frac{k^{2}+j^{2}}{(k+j)^{2}(k-j)^{2}\pi^{2}}(q_{1}-q_{2})^{2}[(-1)^{k+j}-1],\hspace{0.5 cm} j\neq k, \\
\frac{q_{1}^{2}-q_{2}^{2}}{4},\hspace{0.5 cm} j=k,
\end{array}%
\right.       \label{xccjifen:1b} \\
\int_{q_{2}}^{q_{1}}x^{2}\cos [\omega_{k}(x-q_{2})]\cos[\omega_{j}(x-q_{2})]dx&=\left\{
\begin{array}{c}
\frac{2(k^{2}+j^{2})}{(k+j)^{2}(k-j)^{2}\pi^{2}}(q_{1}-q_{2})^{2}[(-1)^{k+j}q_{1}-q_{2}],\hspace{0.5 cm}  j\neq k, \\
\frac{(q_{1}-q_{2})^{3}}{4\pi ^{2}k^{2}}+\frac{q_{1}^{3}-q_{2}^{3}}{6},\hspace{0.5 cm} j=k,
\end{array}%
\right.                                                                \label{xxccjifen:1c}
\end{align}
\end{subequations}
Eq.~(\ref{erjiejifjieggg}) can be rewritten as
\begin{equation}
\label{yijieyijiejieg}
\int_{q_{2}}^{q_{1}}\frac{\partial\varphi_{j}}{\partial q_{n}}\frac{\partial\varphi_{k}}{\partial q_{n}}dx=\left\{
\begin{array}{c}
[(-1)^{k+j}\delta_{n,1}+\delta_{n,2}]\frac{4jk(j^{2}+k^{2})}{(q_{1}-q_{2})^{2}(j^{2}-k^{2})^{2}},\hspace{0.5 cm}  j\neq k, \\
\frac{1}{4(q_{1}-q_{2})^{2}}+\frac{\pi^{2}k^{2}}{3(q_{1}-q_{2})^{2}},\hspace{0.5 cm}    j=k.
\end{array}
\right.
\end{equation}
By combining Eqs.~(\ref{dimensionlesscoeff}),~(\ref{matherela:1c}), and~(\ref{erjiejifjieg}) with~(\ref{yijieyijiejieg}), Eq.~(\ref{matherela:1d}) can be obtained.

To show the relation in Eq.~(\ref{matherela:1e}), we need to derive the mixed second-order partial derivative of the function $\varphi_{k}$ with respect to $q_{1}$ and $q_{2}$,
\begin{eqnarray}
\frac{\partial^{2}\varphi_{k}}{\partial q_{1}\partial q_{2}} &=&\frac{4\omega_{k}^{2}q_{1}q_{2}-3}{4(q_{1}-q_{2})^{2}}\varphi_{k}+\frac{3\omega
_{k}(q_{1}+q_{2})}{\sqrt{2}(q_{1}-q_{2})^{\frac{5}{2}}}\cos[\omega
_{k}(x-q_{2})]-\frac{6\omega_{k}}{\sqrt{2}(q_{1}-q_{2})^{\frac{5}{2}}}x\cos[\omega_{k}(x-q_{2})]      \nonumber\\
&&-\sqrt{2}\frac{\omega_{k}^{2}(q_{1}+q_{2})}{(q_{1}-q_{2})^{\frac{5}{2}}}%
x\sin[\omega_{k}(x-q_{2})]+\sqrt{2}\frac{\omega_{k}^{2}}{(q_{1}-q_{2})^{\frac{5}{2}}}x^{2}\sin[\omega_{k}(x-q_{2})].  \label{erjiehunhpd}
\end{eqnarray}
Multiply both sides of Eq.~(\ref{erjiehunhpd}) by $\varphi_{j}$ and perform the integral with respect to $x$, we can obtain
\begin{eqnarray}
\int_{q_{2}}^{q_{1}}\varphi_{j}\frac{\partial^{2}\varphi_{k}}{\partial q_{1}\partial q_{2}}dx &=&\frac{4\omega_{k}^{2}q_{1}q_{2}-3}{4(
q_{1}-q_{2})^{2}}\delta_{j,k}+\frac{3\omega_{k}(q_{1}+q_{2})}{(q_{1}-q_{2})^{3}}\int_{q_{2}}^{q_{1}}\sin %
[\omega_{j}(x-q_{2})]\cos[\omega_{k}(x-q_{2})]dx                                      \nonumber\\
&&-\frac{6\omega_{k}}{(q_{1}-q_{2})^{3}}\int_{q_{2}}^{q_{1}}x\sin[\omega_{j}(x-q_{2})]\cos[\omega_{k}(x-q_{2})] dx      \nonumber\\
&&-2\frac{\omega_{k}^{2}(q_{1}+q_{2})}{(q_{1}-q_{2})^{3}}\int_{q_{2}}^{q_{1}}x\sin[\omega_{j}(x-q_{2})]\sin[\omega_{k}(x-q_{2})]dx    \nonumber\\
&&+\frac{2\omega_{k}^{2}}{(q_{1}-q_{2})^{3}}\int_{q_{2}}^{q_{1}}x^{2}\sin[\omega_{j}(x-q_{2})]\sin[\omega_{k}(x-q_{2})]dx, \label{erjiehunhpdgc}
\end{eqnarray}
where we used Eq.~(\ref{matherela:1a}). In terms of Eqs.~(\ref{scjihcjifen}),~(\ref{xscjihcjifen}), and~(\ref{xandxxzzjifen}), Eq.~(\ref{erjiehunhpdgc}) can be simplified to
\begin{equation}
\int_{q_{2}}^{q_{1}}\varphi_{j}\frac{\partial^{2}\varphi_{k}}{\partial q_{1}\partial q_{2}}dx=\left\{
\begin{array}{c}
\frac{jk(3j^{2}+k^{2})}{(q_{1}-q_{2})^{2}(j^{2}-k^{2})^{2}}[(-1)^{j+k}+1],\hspace{0.5 cm} j\neq k, \\
\frac{1}{4(q_{1}-q_{2})^{2}}-\frac{j^{2}\pi^{2}}{6(q_{1}-q_{2})^{2}},\hspace{0.5 cm} j=k.  \label{24ezhenm}
\end{array}
\right.
\end{equation}
According to Eqs.~(\ref{matherela:1a}) and~(\ref{yijiedao}), we have
\begin{eqnarray}
\int_{q_{2}}^{q_{1}}\frac{\partial\varphi_{j}}{\partial q_{1}}\frac{\partial\varphi_{k}}{\partial q_{2}}dx &=&-\frac{1}{4}\frac{1}{(
q_{1}-q_{2})^{2}}\delta_{j,k}+\int_{q_{2}}^{q_{1}}\frac{\omega_{k}(q_{1}-x)}{(q_{1}-q_{2})^{3}}\sin[\omega_{j}(
x-q_{2})]\cos[\omega_{k}(x-q_{2})] dx                               \nonumber\\
&&+\int_{q_{2}}^{q_{1}}\frac{\omega_{j}(q_{2}-x)}{(q_{1}-q_{2})^{3}}\sin[\omega_{k}(x-q_{2})]\cos[\omega_{j}(x-q_{2})]dx       \nonumber\\
&&-\frac{2\omega_{j}\omega_{k}}{(q_{1}-q_{2})^{3}}\int_{q_{2}}^{q_{1}}(q_{2}-x)(q_{1}-x)\cos[\omega_{j}(x-q_{2})]
\cos[\omega_{k}(x-q_{2})]dx.  \label{24ezhenmyb}
\end{eqnarray}
In terms of Eqs.~(\ref{scjihcjifen}),~(\ref{xscjihcjifen}), and~(\ref{xandxxccjifen}), we further have
\begin{equation}
\label{24ezhenmybjieg}
\int_{q_{2}}^{q_{1}}\frac{\partial\varphi_{j}}{\partial q_{1}}\frac{\partial\varphi_{k}}{\partial q_{2}}dx=\left\{
\begin{array}{c}
\frac{2jk}{(q_{1}-q_{2})^{2}(j^{2}-k^{2})}-\frac{jk(3j^{2}+k^{2})}{(q_{1}-q_{2})^{2}(j^{2}-k^{2})^{2}}[(-1)^{k+j}+1],\hspace{0.5 cm} j\neq k, \\
-\frac{1}{4(q_{1}-q_{2})^{2}}+\frac{j^{2}\pi^{2}}{6(q_{1}-q_{2})^{2}},\hspace{0.5 cm} j=k.
\end{array}
\right.
\end{equation}
Combining Eqs.~(\ref{dimensionlesscoeff}),~(\ref{matherela:1c}), and~(\ref{24ezhenm}) with~(\ref{24ezhenmybjieg}), the relation in Eq.~(\ref{matherela:1e}) can then be proved.

According to Eq.~(\ref{24ezhenm}), we know that
\begin{equation}
\int_{q_{2}}^{q_{1}}\varphi_{j}\frac{\partial^{2}\varphi_{k}}{\partial q_{1}\partial q_{2}}dx=0,
\end{equation}
when the sum of $j$ and $k$ is an odd number. In this case, Eq.~(\ref{matherela:1e}) can be reduced to Eq.~(\ref{matherela:1f}).

\section{II. The Lagrange function of the system}  \label{tlf}

In this section, we confirm the Lagrange function of the two-moving-mirror cavity optomechanical system.

\subsection{A. Confirmation of the Lagrange function of the system}

For the two-moving-mirror cavity optomechanical system, we assume that the Lagrange function for the whole system including two moving mirrors and the cavity fields can be written as
\begin{equation}
L=L_{\text{mirror}}+L_{\text{field}},   \label{miirfL}
\end{equation}
where
\begin{equation}
L_{\text{mirror}}=\frac{1}{2}m_{1}\dot{q}_{1}^{2}+\frac{1}{2}m_{2}\dot{q}_{2}^{2}-V_{1}(q_{1})-V_{2}(q_{2})    \label{miirL}
\end{equation}
is the Lagrange function of the two moving mirrors, and $L_{\text{field}}$ is the Lagrange function of the cavity fields, which takes the form as
\begin{equation}
L_{\text{field}}=\frac{1}{2}\int_{q_{2}}^{q_{1}}\left[\frac{\partial A(x,t)}{\partial t}\right]^{2}dx-\frac{1}{2}\int_{q_{2}}^{q_{1}}\left[\frac{\partial A(x,t)}{\partial x}\right]^{2}dx.    \label{filedL}
\end{equation}
Since $A(x,t)$ is regarded as a scalar, the Lagrange function of the fields can take the form of the scalar fields. In terms of Eqs.~(\ref{vpdos}), (\ref{matherela}), (\ref{miirL}), and (\ref{filedL}), the Lagrange function in Eq.~(\ref{miirfL}) of the coupled system can be derived as
\begin{eqnarray}
L &=&\frac{1}{2}\sum_{k=1}^{\infty}(\dot{Q}_{k}^{2}-\omega_{k}^{2}Q_{k}^{2})+\sum_{n,n^{\prime}=1,2}\sum_{j,k,l=1}^{\infty}\frac{%
\dot{q}_{n}\dot{q}_{n^{\prime}}g_{kl}^{(n)}g_{jl}^{(n^{\prime})}}{2(q_{1}-q_{2})^{2}}Q_{k}Q_{j}+\sum_{n=1,2}\sum_{j,k=1}^{\infty}\frac{\dot{q}%
_{n}g_{jk}^{(n)}}{q_{1}-q_{2}}\dot{Q}_{k}Q_{j}                                        \nonumber \\
&&+\frac{1}{2}m_{1}\dot{q}_{1}^{2}+\frac{1}{2}m_{2}\dot{q}_{2}^{2}-V_{1}(q_{1})-V_{2}(q_{2}).   \label{lagrangian}
\end{eqnarray}
Here, these variables have been introduced in Eqs.~(\ref{dynaequationsa}) and~(\ref{dynaequationsb}).

Next, we will confirm the correctness of the Lagrange function $L$ given in Eq.~(\ref{lagrangian}) by checking the consistence of the Euler-Lagrangian equations. For the $k$th cavity-field mode, the Euler-Lagrangian equation is given by
\begin{equation}
\frac{d}{dt}\frac{\partial L}{\partial \dot{Q}_{k}}-\frac{\partial L}{\partial Q_{k}}=0.  \label{Euler-Lagrangiank}
\end{equation}
By substituting Eq.~(\ref{lagrangian}) into Eq.~(\ref{Euler-Lagrangiank}), we can obtain the equation of motion
\begin{eqnarray}
\ddot{Q}_{k} &=&-\omega_{k}^{2}Q_{k}+\sum_{n=1,2}\sum_{j=1}^{\infty}\frac{2\dot{q}_{n}}{q_{1}-q_{2}}g_{kj}^{(n)}\dot{Q}_{j}+\sum_{n,n^{\prime
}=1,2}\sum_{j,l=1}^{\infty}\frac{\dot{q}_{n}\dot{q}_{n^{\prime}}}{(q_{1}-q_{2})^{2}}Q_{j}g_{jl}^{(n)}g_{kl}^{(n^{\prime})}            \nonumber\\
&&-\sum_{n=1,2}\frac{\ddot{q}_{n}(q_{1}-q_{2})-(\dot{q}_{1}-\dot{q}_{2})\dot{q}_{n}}{(q_{1}-q_{2})^{2}}\sum_{j=1}^{\infty}Q_{j}g_{jk}^{(n)}. \label{cavitymodek}
\end{eqnarray}
In terms of Eq.~(\ref{matherela}) and the relations
\begin{equation}
g_{jk}^{(1)}=-(-1)^{k+j}g_{jk}^{(2)},\hspace{1.0 cm} g_{jl}^{(1)}g_{kl}^{(2)}=(-1)^{j+k}g_{kl}^{(1)}g_{jl}^{(2)},
\end{equation}
we can rewrite Eq.~(\ref{cavitymodek}) as
\begin{eqnarray}
\ddot{Q}_{k}&=&-\omega_{k}^{2}Q_{k}-\sum_{n=1,2}\sum_{j=1}^{\infty}\frac{2\dot{q}_{n}}{q_{1}-q_{2}}\dot{Q}_{j}g_{jk}^{(n)}+\frac{2\dot{q}_{1}\dot{q}%
_{2}}{(q_{1}-q_{2})^{2}}\sum_{j=1,j+k=\text{even}}^{\infty}Q_{j}\left(g_{jk}^{(2)}+\sum_{l=1}^{\infty}g_{jl}^{(2)}g_{kl}^{(1)}\right)   \nonumber\\
&&+\sum_{n=1,2}\sum_{j,l=1}^{\infty}\frac{\dot{q}_{n}^{2}}{(q_{1}-q_{2})^{2}%
}Q_{j}g_{jl}^{(n)}g_{kl}^{(n)}-\sum_{n=1,2}\sum_{j=1}^{\infty}\frac{\ddot{q}%
_{n}(q_{1}-q_{2})+(-1)^{n}\dot{q}_{n}^{2}}{(q_{1}-q_{2})^{2}}Q_{j}g_{jk}^{(n)},  \label{fiezj}
\end{eqnarray}%
where $j+k=\text{even}$ means that the sum of $j$ and $k$ is an even number, and $\sum_{j=1,j+k=\text{even}}^{\infty}$ represents the sum of all positive number $j$ satisfying $j+k=\text{even}$. According to Eq.~(\ref{matherela}), we can obtain the equation
\begin{equation}
\frac{2\dot{q}_{1}\dot{q}_{2}}{(q_{1}-q_{2})^{2}}\sum_{j=1,j+k=\text{odd}}^{\infty}Q_{j}\left[g_{jk}^{(2)}
+\sum_{l=1}^{\infty}g_{jl}^{(2)}g_{kl}^{(1)}\right] =0,   \label{userala}
\end{equation}
where $\sum_{j=1,j+k=\text{odd}}^{\infty}$ represents the sum of all positive number $j$ satisfying $j+k=\text{odd}$. With the help of Eq.~(\ref{userala}), we can obtain Eq.~(\ref{dynaequationsa}) from Eq.~(\ref{fiezj}), which means that the equation of motion for the generalized coordinate $Q_{k}$ can be recovered based on the Euler-Lagrange equation and the Lagrange function $L$.

Below, we check the equation of motion for the two moving mirrors based on the Euler-Lagrangian equation and the Lagrange function. For the right moving end mirror, the Euler-Lagrangian equation is given by
\begin{equation}
\frac{d}{dt}\frac{\partial L}{\partial\dot{q}_{1}}-\frac{\partial L}{\partial q_{1}}=0.          \label{Euler-Lagrangianrighimoving}
\end{equation}
By substituting Eq.~(\ref{lagrangian}) into Eq.~(\ref{Euler-Lagrangianrighimoving}), we can obtain the following equation of motion
\begin{eqnarray}
&&m_{1}\ddot{q}_{1}-\frac{\dot{q}_{1}-\dot{q}_{2}}{(q_{1}-q_{2})^{2}}\sum_{j,k=1}^{\infty}\dot{Q}_{k}Q_{j}g_{jk}^{(1)}
+\frac{1}{(q_{1}-q_{2})}\sum_{j,k=1}^{\infty}\ddot{Q}_{k}Q_{j}g_{jk}^{(1)}+\frac{1}{(q_{1}-q_{2})}%
\sum_{j,k=1}^{\infty}\dot{Q}_{k}\dot{Q}_{j}g_{jk}^{(1)}                                                         \nonumber\\
&&+\sum_{n=1,2}\sum_{j,k,l=1}^{\infty}\frac{\ddot{q}_{n}(q_{1}-q_{2})-2(\dot{q}_{1}
-\dot{q}_{2})\dot{q}_{n}}{(q_{1}-q_{2})^{3}}Q_{k}Q_{j}g_{jl}^{(n)}g_{kl}^{(1)}+\sum_{n=1,2}\sum_{j,k,l=1}^{\infty}\frac{%
\dot{q}_{n}}{(q_{1}-q_{2})^{2}}(\dot{Q}_{k}Q_{j}+Q_{k}\dot{Q}_{j})g_{jl}^{(n)}g_{kl}^{(1)}                       \nonumber\\
&=&\frac{1}{q_{1}-q_{2}}\sum_{k=1}^{\infty}\omega_{k}^{2}Q_{k}^{2}-\frac{dV_{1}(q_{1})}{dq_{1}}-\sum_{n=1,2}\sum_{j,k=1}^{\infty}\frac{\dot{q}_{n}}{%
(q_{1}-q_{2})^{2}}\dot{Q}_{k}Q_{j}g_{jk}^{(n)}-\sum_{n,n^{\prime}=1,2}\sum_{j,k,l=1}^{\infty}\frac{\dot{q}_{n}\dot{q}_{n^{\prime}}}{%
(q_{1}-q_{2})^{3}}Q_{k}Q_{j}g_{jl}^{(n)}g_{kl}^{(n^{\prime})}.                                                    \label{fangc1}
\end{eqnarray}
By substituting Eq.~(\ref{dynaequationsa}) into Eq.~(\ref{fangc1}), and in terms of the relations
\begin{subequations}
\label{zhnemgxuy1}
\begin{align}
\sum_{j,k=1}^{\infty}\dot{Q}_{k}\dot{Q}_{j}g_{jk}^{(1)}& =0,     \label{zhnemgxuy1:1a} \\
\sum_{j,k,l,s=1}^{\infty}Q_{l}Q_{j}g_{ls}^{(1)}g_{ks}^{(1)}g_{jk}^{(1)}& =0,       \label{zhnemgxuy1:1b}
\end{align}%
\end{subequations}
we can further obtain the equation of motion
\begin{eqnarray}
m_{1}\ddot{q}_{1} &=&\frac{1}{q_{1}-q_{2}}\sum_{k=1}^{\infty}\omega_{k}^{2}Q_{k}^{2}-\frac{dV_{1}(q_{1})}{dq_{1}}-\sum_{n=1,2}\sum_{j,k,l=1}^{%
\infty}\frac{(1+\delta_{n,1})\dot{q}_{2}\dot{q}_{n}}{(q_{1}-q_{2})^{3}}Q_{k}Q_{j}g_{jl}^{(n)}\left(g_{kl}^{(1)}+g_{kl}^{(2)}\right)   \nonumber\\
&&+\frac{1}{q_{1}-q_{2}}\sum_{j,k=1}^{\infty}\omega_{k}^{2}g_{jk}^{(1)}Q_{j}Q_{k}-\sum_{n=1,2}\frac{(1
+\delta_{n,1})\dot{q}_{2}\dot{q}_{n}}{(q_{1}-q_{2})^{3}}\sum_{j,k,l,s=1}^{%
\infty}Q_{j}Q_{l}g_{jk}^{(1)}g_{ls}^{(2)}g_{ks}^{(n)}.                   \label{fangc2}
\end{eqnarray}
For the right moving end mirror, we can obtain the following two relations
\begin{subequations}
\label{zhnemgxuy2}
\begin{align}
\sum_{j,k,l,s=1}^{\infty}Q_{j}Q_{l}g_{jk}^{(2)}g_{ls}^{(2)}g_{ks}^{(1)}& =0,    \label{zhnemgxuy2:1a} \\
\sum_{j,k,l,s=1}^{\infty}Q_{j}Q_{l}g_{ls}^{(1)}g_{jk}^{(1)}g_{ks}^{(2)}& =0,     \label{zhnemgxuy2:1b}
\end{align}%
\end{subequations}
With the help of Eq.~(\ref{zhnemgxuy2}), Eq.~(\ref{fangc2}) can be further reduced to
\begin{equation}
m_{1}\ddot{q}_{1}=\frac{1}{q_{1}-q_{2}}\sum_{k=1}^{\infty}\omega_{k}^{2}Q_{k}^{2}-\frac{dV_{1}(q_{1})}{dq_{1}}
+\frac{1}{q_{1}-q_{2}}\sum_{j,k=1}^{\infty}\omega_{k}^{2}g_{jk}^{(1)}Q_{j}Q_{k}.
\end{equation}
In terms of Eq.~(\ref{dimensionlesscoeff}), we can recover the equation of motion of the right moving mirror given by Eq.~(\ref{dynaequationsb}) for $n=1$.

With a similar calculation process, we can derive Eq.~(\ref{dynaequationsb}) for $n=2$ from the Euler-Lagrangian equation for the left moving mirror, which is given by
\begin{equation}
\frac{d}{dt}\frac{\partial L}{\partial \dot{q}_{2}}-\frac{\partial L}{\partial q_{2}}=0.
\end{equation}
We can obtain Eqs.~(\ref{dynaequationsa}) and~(\ref{dynaequationsb}) through solving the Euler-Lagrange equation based on the Lagrange function $L$. Therefore, the $L$ given in Eq.~(\ref{lagrangian}) is the Lagrange function of the present system. Based on the Lagrange function, the Hamiltonian of the system can then be derived by the Legendre transformation.

\subsection{B. Derivation of Eqs.~(\ref{zhnemgxuy1}) and~(\ref{zhnemgxuy2})}

In this section, we present the detailed proof of Eqs.~(\ref{zhnemgxuy1}) and~(\ref{zhnemgxuy2}). To prove Eq.~(\ref{zhnemgxuy1:1a}), we express the left-hand side of the equation as
\begin{eqnarray}
\sum_{j,k=1}^{\infty}\dot{Q}_{k}\dot{Q}_{j}g_{jk}^{(1)}&=&\sum_{j>k}^{\infty}\dot{Q}_{k}\dot{Q}_{j}g_{jk}^{(1)}+\sum_{j<k}^{\infty
}\dot{Q}_{k}\dot{Q}_{j}g_{jk}^{(1)}               \nonumber\\
&=&\sum_{j>k}^{\infty}\dot{Q}_{k}\dot{Q}_{j}g_{jk}^{(1)}+\sum_{k<j}^{\infty}\dot{Q}_{j}\dot{Q}_{k}g_{kj}^{(1)}      \nonumber\\
&=&\sum_{j>k}^{\infty}\dot{Q}_{k}\dot{Q}_{j}(g_{jk}^{(1)}+g_{kj}^{(1)})      \nonumber\\
&=&0,
\end{eqnarray}
where we used the relation $g_{jk}^{(1)}+g_{kj}^{(1)}=0$, which can be obtained based on the definition in Eq.~(\ref{dimensionlesscoeff}).

The proof of Eq.~(\ref{zhnemgxuy1:1b}) is given as follows,
\begin{eqnarray}
&&\sum_{j,k,l,s=1}^{\infty}Q_{l}Q_{j}g_{ls}^{(1)}g_{ks}^{(1)}g_{jk}^{(1)} \nonumber\\
&=&\sum_{k,s=1}^{\infty}\sum_{j>l}^{\infty}Q_{l}Q_{j}g_{ls}^{(1)}g_{ks}^{(1)}g_{jk}^{(1)}+\sum_{k,s=1}^{\infty}\sum_{j<l}^{\infty
}Q_{l}Q_{j}g_{ls}^{(1)}g_{ks}^{(1)}g_{jk}^{(1)}+\sum_{k,s=1}^{\infty}\sum_{j=1}^{\infty}Q_{j}^{2}g_{js}^{(1)}g_{ks}^{(1)}g_{jk}^{(1)} \nonumber\\
&=&\sum_{k,s=1}^{\infty}\sum_{j>l}^{\infty}Q_{l}Q_{j}g_{ls}^{(1)}g_{ks}^{(1)}g_{jk}^{(1)}+\sum_{k,s=1}^{\infty}\sum_{r<l}^{\infty
}Q_{l}Q_{r}g_{ls}^{(1)}g_{ks}^{(1)}g_{rk}^{(1)}+\sum_{j=1}^{\infty}\sum_{k>s}^{\infty
}Q_{j}^{2}g_{js}^{(1)}g_{ks}^{(1)}g_{jk}^{(1)}+\sum_{j=1}^{\infty}\sum_{k<s}^{\infty}Q_{j}^{2}g_{js}^{(1)}g_{ks}^{(1)}g_{jk}^{(1)} \nonumber\\
&=&\sum_{k,s=1}^{\infty}\sum_{j>l}^{\infty}Q_{l}Q_{j}g_{ls}^{(1)}g_{ks}^{(1)}g_{jk}^{(1)}+\sum_{k,s=1}^{\infty
}\sum_{l<j}^{\infty}Q_{j}Q_{l}g_{jk}^{(1)}g_{sk}^{(1)}g_{ls}^{(1)}+\sum_{j=1}^{\infty}\sum_{k>s}^{\infty
}Q_{j}^{2}g_{js}^{(1)}g_{ks}^{(1)}g_{jk}^{(1)}+\sum_{j=1}^{\infty}\sum_{s<k}^{\infty }Q_{j}^{2}g_{jk}^{(1)}g_{sk}^{(1)}g_{js}^{(1)} \nonumber\\
&=&\sum_{k,s=1}^{\infty}\sum_{j>l}^{\infty}Q_{l}Q_{j}g_{jk}^{(1)}g_{ls}^{(1)}(g_{ks}^{(1)}+g_{sk}^{(1)})
+\sum_{j=1}^{\infty}\sum_{k>s}^{\infty}Q_{j}^{2}g_{jk}^{(1)}g_{js}^{(1)}(g_{ks}^{(1)}+g_{sk}^{(1)})  \nonumber\\
&=&0.
\end{eqnarray}
Here, we also used the relation $g_{ks}^{(1)}+g_{sk}^{(1)}=0$.

To prove the relation given in Eq.~(\ref{zhnemgxuy2:1a}), we also express the left-hand side of the equation as
\begin{eqnarray}
&&\sum_{j,k,l,s=1}^{\infty}Q_{j}Q_{l}g_{jk}^{(2)}g_{ls}^{(2)}g_{ks}^{(1)} \nonumber\\
&=&\sum_{k,s=1}^{\infty}\sum_{j>l}^{\infty}Q_{j}Q_{l}g_{jk}^{(2)}g_{ls}^{(2)}g_{ks}^{(1)}+\sum_{k,s=1}^{\infty}\sum_{j<l}^{\infty
}Q_{j}Q_{l}g_{jk}^{(2)}g_{ls}^{(2)}g_{ks}^{(1)}+\sum_{k,s=1}^{\infty}\sum_{j=1}^{\infty}Q_{j}^{2}g_{jk}^{(2)}g_{js}^{(2)}g_{ks}^{(1)} \nonumber\\
&=&\sum_{k,s=1}^{\infty}\sum_{j>l}^{\infty}Q_{j}Q_{l}g_{jk}^{(2)}g_{ls}^{(2)}g_{ks}^{(1)}+\sum_{k,s=1}^{\infty}\sum_{r<l}^{\infty
}Q_{r}Q_{l}g_{rk}^{(2)}g_{ls}^{(2)}g_{ks}^{(1)}+\sum_{j=1}^{\infty}\sum_{k>s}^{\infty
}Q_{j}^{2}g_{jk}^{(2)}g_{js}^{(2)}g_{ks}^{(1)}+\sum_{j=1}^{\infty}\sum_{k<s}^{\infty}Q_{j}^{2}g_{jk}^{(2)}g_{js}^{(2)}g_{ks}^{(1)} \nonumber\\
&=&\sum_{k,s=1}^{\infty}\sum_{j>l}^{\infty}Q_{j}Q_{l}g_{jk}^{(2)}g_{ls}^{(2)}g_{ks}^{(1)}+\sum_{k,s=1}^{\infty}\sum_{l<j}^{\infty
}Q_{l}Q_{j}g_{lk}^{(2)}g_{js}^{(2)}g_{ks}^{(1)}+\sum_{j=1}^{\infty}\sum_{k>s}^{\infty }Q_{j}^{2}g_{jk}^{(2)}g_{js}^{(2)}(
g_{ks}^{(1)}+g_{sk}^{(1)})  \nonumber\\
&=&\sum_{k,s=1}^{\infty}\sum_{j>l}^{\infty}Q_{j}Q_{l}g_{jk}^{(2)}g_{ls}^{(2)}(g_{ks}^{(1)}+g_{sk}^{(1)})    \nonumber\\
&=&0.
\end{eqnarray}

Using the similar method, the derivation of Eq.~(\ref{zhnemgxuy2:1b}) can be performed as follows,
\begin{eqnarray}
&&\sum_{j,k,s,l=1}^{\infty}Q_{j}Q_{l}g_{ls}^{(1)}g_{jk}^{(1)}g_{ks}^{(2)} \nonumber\\
&=&\sum_{s,k=1}^{\infty}\sum_{j>l}^{\infty}Q_{j}Q_{l}g_{ls}^{(1)}g_{jk}^{(1)}g_{ks}^{(2)}+\sum_{s,k=1}^{\infty}\sum_{j<l}^{\infty
}Q_{j}Q_{l}g_{ls}^{(1)}g_{jk}^{(1)}g_{ks}^{(2)}+\sum_{s,k=1}^{\infty}\sum_{j=1}^{\infty }Q_{j}^{2}g_{js}^{(1)}g_{jk}^{(1)}g_{ks}^{(2)} \nonumber\\
&=&\sum_{s,k=1}^{\infty }\sum_{j>l}^{\infty}Q_{j}Q_{l}g_{ls}^{(1)}g_{jk}^{(1)}g_{ks}^{(2)}+\sum_{s,k=1}^{\infty}\sum_{r<l}^{\infty
}Q_{r}Q_{l}g_{ls}^{(1)}g_{rk}^{(1)}g_{ks}^{(2)}+\sum_{j=1}^{\infty}\sum_{s>k}^{\infty
}Q_{j}^{2}g_{js}^{(1)}g_{jk}^{(1)}g_{ks}^{(2)}+\sum_{j=1}^{\infty}\sum_{s<k}^{\infty}Q_{j}^{2}g_{js}^{(1)}g_{jk}^{(1)}g_{ks}^{(2)} \nonumber\\
&=&\sum_{s,k=1}^{\infty}\sum_{j>l}^{\infty}Q_{j}Q_{l}g_{ls}^{(1)}g_{jk}^{(1)}g_{ks}^{(2)}+\sum_{s,k=1}^{\infty}\sum_{l<j}^{\infty
}Q_{l}Q_{j}g_{js}^{(1)}g_{lk}^{(1)}g_{ks}^{(2)}+\sum_{j=1}^{\infty}\sum_{s>k}^{\infty}Q_{j}^{2}g_{js}^{(1)}g_{jk}^{(1)}(
g_{ks}^{(2)}+g_{sk}^{(2)})  \nonumber\\
&=&\sum_{s,k=1}^{\infty}\sum_{j>l}^{\infty}Q_{j}Q_{l}g_{ls}^{(1)}g_{jk}^{(1)}(g_{ks}^{(1)}+g_{sk}^{(1)})  \nonumber\\
&=&0.
\end{eqnarray}

\section{III. The quantized Hamiltonian of the system described in the center-of-mass system}  \label{qhitcs}

In this section, we present the detailed derivation of the quantized Hamiltonian of the system using the method of canonical quantization. To this end, we first obtain the classical Hamiltonian of the system. By the Legendre transformation, the classical Hamiltonian of the system can be written as
\begin{equation}
H=p_{1}\dot{q}_{1}+p_{2}\dot{q}_{2}+\sum_{k=1}^{\infty}P_{k}\dot{Q}_{k}-L.  \label{legtraHamiltonian}
\end{equation}
Here, $P_{k}$ ($Q_{k}$) and $p_{l}$ ($q_{l}$) are, respectively, the canonical momenta (coordinates) for the $k$th field mode of the cavity and the $l$th moving mirror, defined by
\begin{subequations}
\begin{align}
P_{k}& =\frac{\partial L}{\partial\dot{Q}_{k}}=\dot{Q}_{k}+\sum_{n=1,2}%
\sum_{j=1}^{\infty}\frac{g_{jk}^{(n)}\dot{q}_{n}}{q_{1}-q_{2}}Q_{j},\hspace{0.5 cm}k=1,2,3,\cdots , \\
p_{l}& =\frac{\partial L}{\partial\dot{q}_{l}}=m_{l}\dot{q}_{l}-\Gamma_{l},\hspace{0.5 cm}l=1,2,
\end{align}
\end{subequations}
where
\begin{equation}
\Gamma_{l}=\frac{1}{q_{1}-q_{2}}\sum_{j,k=1}^{\infty}g_{kj}^{(l)}Q_{j}P_{k}, \hspace{1 cm} l=1,2  \label{diffemomen}
\end{equation}
describes the difference between the kinetic momentum and the canonical momentum of the $l$th ($l=1,2$) moving end mirror. After a lengthy calculation, the explicit expression of the classical Hamiltonian given in Eq.~(\ref{legtraHamiltonian}) can be written as
\begin{equation}
H=\frac{1}{2}\sum_{k=1}^{\infty}(P_{k}^{2}+\omega_{k}^{2}Q_{k}^{2})+\sum_{l=1,2}\left[\frac{1}{2m_{l}}\left(p_{l}+\Gamma
_{l}\right)^{2}+V_{l}(q_{l})\right],  \label{clssHamiltonian}
\end{equation}
which is equivalent to the total energy of the whole system. We point out that the classical Hamiltonian given in Eq.~(\ref{clssHamiltonian}) can also be expressed as
\begin{equation}
H=H_{\text{field}}+\frac{1}{2}(m_{1}\dot{q}_{1}^{2}+m_{2}\dot{q}_{2}^{2})+V_{1}(q_{1})+V_{2}(q_{2}),
\end{equation}%
where $H_{\text{field}}$ is energy of the cavity fields, taking the form as
\begin{equation}
H_{\text{field}}=\frac{1}{2}\int_{q_{2}}^{q_{1}}dx\left[(\partial_{t}A)^{2}+(\partial_{x}A)^{2}\right].
\end{equation}

Following the canonical quantization procedure, the variables $p_{1}$, $p_{2}$, $q_{1}$, $q_{2}$, $Q_{k}$, and $P_{k}$
are replaced by the corresponding operators $\hat{p}_{1}$, $\hat{p}_{2}$, $\hat{q}_{1} $, $\hat{q}_{2}$, $\hat{Q}_{k}$, and $\hat{P}_{k}$, which obey the nonzero commutation relations
\begin{equation}
[\hat{q}_{1},\hat{p}_{1}] =[\hat{q}_{2},\hat{p}_{2}]=i\hbar, \hspace{1 cm}\ [\hat{Q}_{j},\hat{P}_{k}] =i\hbar\delta_{jk}.     \label{commutationrelations}
\end{equation}
To ensure that the Hamiltonian is Hermitian after the quantization, the operator product $Q_{j}P_{k}$ in $\Gamma_{l}$ should be replaced by $(\hat{Q}_{j}\hat{P}_{k}+\hat{P}_{k}\hat{Q}_{j})/2$. Then the quantized Hamiltonian corresponding to Eq.~(\ref{clssHamiltonian}) can be written as
\begin{equation}
\hat{H}=\frac{1}{2}\sum_{k=1}^{\infty}(\hat{P}_{k}^{2}+\omega_{k}^{2}\hat{Q}_{k}^{2})
+\sum_{l=1,2}\frac{1}{2m_{l}}(\hat{p}_{l}+\hat{\Gamma}_{l})^{2}+V_{1}(\hat{q}_{1})+V_{2}(\hat{q}_{2}),    \label{quanliHamli}
\end{equation}
where the operator $\hat{\Gamma}_{l}$ has been defined in Eq.~(\ref{diffemomen}). It can be seen from Eq.~(\ref{quanliHamli}) that the Hamiltonian of the coupled system does not violate the requirement of the existing of a lower bound in the eigen-energy spectrum.

To describe the quantum state of the cavity field in the Fock-state space, we need to define the creation and annihilation operators for each cavity mode. Since the frequency of the cavity field depends on the distance between the two moving mirrors [see Eq.~(\ref{cavityfreq})], it is necessary to define the creation and annihilation operators of each cavity-field mode in the center-of-mass system. Concretely, the center-of-mass coordinate operator and the relative coordinate operator are, respectively, defined as
\begin{equation}
\hat{q}_{c}=\frac{\mu}{m_{2}}\hat{q}_{1}+\frac{\mu}{m_{1}}\hat{q}_{2}, \hspace{1 cm} \hat{q}_{r}=\hat{q}_{1}-\hat{q}_{2},       \label{crop}
\end{equation}
where $\mu =m_{1}m_{2}/m_{t}$ is the reduced mass with the total mass $m_{t}=m_{1}+m_{2}$. Correspondingly, the center-of-mass momentum operator $\hat{p}_{c}$ and the relative momentum operator $\hat{p}_{r}$ are expressed as
\begin{equation}
\hat{p}_{c}=\hat{p}_{1}+\hat{p}_{2}, \hspace{1 cm} \hat{p}_{r}=\frac{\mu}{m_{1}}\hat{p}_{1}-\frac{\mu}{m_{2}}\hat{p}_{2},       \label{crmop}
\end{equation}%
Therefore, we have the commutation relations
\begin{equation}
[\hat{q}_{c},\hat{p}_{c}]=[\hat{q}_{r},\hat{p}_{r}]=i\hbar,\hspace{1 cm} [\hat{q}_{c},\hat{q}_{r}]=[\hat{q}_{c},\hat{p}_{r}]=[\hat{q}_{r},\hat{p}_{c}]=[\hat{p}_{c},\hat{p}_{r}]=0.
\end{equation}
Based on Eq.~(\ref{crop}), the cavity-length-dependent frequency of the $k$th cavity mode can be expressed as
\begin{equation}
\omega_{k}(\hat{q}_{r})=\frac{k\pi}{\hat{q}_{r}}.
\end{equation}
Using Eqs.~(\ref{crop}) and (\ref{crmop}), the operators $\hat{q}_{1}$, $\hat{q}_{2}$, $\hat{p}_{1}$, and $\hat{p}_{2}$ can be expressed as
\begin{subequations}
    \begin{align}
\hat{q}_{1}&=\hat{q}_{c}+\frac{\mu}{m_{1}}\hat{q}_{r}, \hspace{1.0 cm} \hat{q}_{2}=\hat{q}_{c}-\frac{\mu}{m_{2}}\hat{q}_{r},\\
\hat{p}_{1}&=\frac{\mu}{m_{2}}\hat{p}_{c}+\hat{p}_{r},\hspace{1.0 cm}  \hat{p}_{2}=\frac{\mu}{m_{1}}\hat{p}_{c}-\hat{p}_{r}.
\end{align}
\end{subequations}
Thus, the quantized Hamiltonian~(\ref{quanliHamli}) can be expressed in the center-of-mass system as
\begin{eqnarray}
\hat{H} &=&\frac{1}{2}\sum_{k=1}^{\infty}\left[\hat{P}_{k}^{2}+\omega_{k}^{2}(\hat{q}_{r})\hat{Q}_{k}^{2}\right]
+V_{1}\left(\hat{q}_{c}+\frac{\mu}{m_{1}}\hat{q}_{r}\right)+V_{2}\left(\hat{q}_{c}-\frac{\mu}{m_{2}}\hat{q}_{r}\right)   \nonumber \\
&&+\sum_{l=1,2}\frac{1}{2m_{l}}\left[\frac{m_{l}}{m}\hat{p}_{c}+(-1)^{l+1}\hat{p}_{r}
+\frac{1}{\hat{q}_{r}}\sum_{j,k=1}^{\infty}g_{kj}^{(l)}\hat{Q}_{j}\hat{P}_{k}\right]^{2}.         \label{Hmiltoniancesys}
\end{eqnarray}

In the center-of-mass system, the cavity-length-dependent annihilation and creation operators of the $k$th cavity mode are defined by
\begin{equation}
\hat{a}_{k}(\hat{q}_{r})=\frac{1}{\sqrt{2\hbar\omega_{k}(\hat{q}_{r})}}\left[\omega_{k}(\hat{q}_{r})\hat{Q}_{k}+i\hat{P}_{k}\right] ,\hspace{1 cm} \hat{a}_{k}^{\dagger}(\hat{q}%
_{r})=\frac{1}{\sqrt{2\hbar\omega_{k}(\hat{q}_{r})}}\left[\omega_{k}(\hat{q}_{r})\hat{Q}_{k}-i\hat{P}_{k}\right],       \label{creationannihilation}
\end{equation}
which indicate that we have a set of Fock states associated with each relative coordinate operator $\hat{q}_{r}$ between the two moving mirrors. Based on the assumption $q_{r}\neq 0$, Eq.~(\ref{creationannihilation}) does not include the definitions of the creation and annihilation operators at $q_{r}=0$. Considering the completeness of the whole system, we label such a set of Fock states by $|\{n_{l}\},q_{r},q_{c}\rangle$ where $\{n_{l}\} =\{n_{1},n_{2},n_{3},\cdots\}$ for $n_{l}=0$, $1$, $2$, $\cdots$, $\infty$ denotes the set of occupation numbers for different cavity modes. The state vector $|\{n_{l}\},q_{r},q_{c}\rangle$ is the simultaneous eigenvector of the number operator $\hat{a}_{k}^{\dagger}(\hat{q}_{r})\hat{a}_{k}(\hat{q}_{r})$, the relative coordinate operator $\hat{q}_{r}$, and the center-of-mass coordinate operator $\hat{q}_{c}$, i.e.,
\begin{subequations}
    \begin{align}
\hat{a}_{k}^{\dagger}(\hat{q}_{r})\hat{a}_{k}(\hat{q}_{r})|\{n_{l}\},q_{r},q_{c}\rangle & =n_{k}|\{n_{l}\},q_{r},q_{c}\rangle,  \\
\hat{q}_{r}|\{n_{l}\},q_{r},q_{c}\rangle & =q_{r}|\{n_{l}\},q_{r},q_{c}\rangle,  \\
\hat{q}_{c}|\{n_{l}\},q_{r},q_{c}\rangle & =q_{c}|\{n_{l}\},q_{r},q_{c}\rangle.
\end{align}
\end{subequations}
Considering the completeness and orthogonality of these Fock states, any pure quantum state $|\varphi\rangle$ of the whole system can be expressed as a superposition of these states
\begin{equation}
|\psi\rangle =\sum_{\{ n_{l}\}}\int_{0}^{\infty}\int_{0}^{\infty}dq_{r}dq_{c}C(\{n_{l}\},q_{r},q_{c})|\{n_{l}\},q_{r},q_{c}\rangle, \label{anystate}
\end{equation}
where $C(\{ n_{l}\},q_{r},q_{c})$ is the probability amplitude.

According to Eq.~(\ref{creationannihilation}), $\hat{Q}_{j}$ and $\hat{P}_{k}$ can be expressed as
\begin{equation}
\hat{Q}_{j}=\sqrt{\frac{\hbar}{2\omega_{j}(\hat{q}_{r})}}\left[\hat{a}_{j}^{\dagger}(\hat{q}_{r})+\hat{a}_{j}(\hat{q}_{r})\right] ,\hspace{1 cm} \hat{P}_{k}=i\sqrt{\frac{\hbar\omega
_{k}(\hat{q}_{r})}{2}}\left[\hat{a}_{k}^{\dagger}(\hat{q}_{r})-\hat{a}_{k}(\hat{q}_{r})\right]. \label{zacf}
\end{equation}%
Based on Eq.~(\ref{zacf}), the quantized Hamiltonian~(\ref{Hmiltoniancesys}) can be expressed as
\begin{eqnarray}
\hat{H} &=&\sum_{k=1}^{\infty}\hbar\omega_{k}(\hat{q}_{r})\hat{a}_{k}^{\dagger}(\hat{q}_{r})\hat{a}_{k}(\hat{q}_{r})+\frac{1}{2}%
\sum_{k=1}^{\infty}\hbar\omega_{k}(\hat{q}_{r})+V_{1}\left(\hat{q}_{c}+\frac{%
\mu}{m_{1}}\hat{q}_{r}\right)+V_{2}\left(\hat{q}_{c}-\frac{\mu}{m_{2}}\hat{q}_{r}\right)  \nonumber \\
&&+\sum_{l=1,2}\frac{1}{2m_{l}}\left[\frac{m_{l}}{m}\hat{p}_{c}+(-1)^{l+1}\hat{p}_{r}+\hat{\Gamma}_{l}\right]^{2}.  \label{Haminzhanjain}
\end{eqnarray}
The second term in Eq.~(\ref{Haminzhanjain}) is the zero-point energy of the cavity field, which can be replaced by the Casimir energy $-\hbar \pi/(24\hat{q}_{r})$~\cite{LawPRA1995,Butera2022PRD}. In this work, we consider the harmonic potential wells $V_{1}(q_{1})$ and $V_{2}(q_{2})$. Concretely, the two moving end mirrors are in the harmonic potential wells with the equilibrium positions $x=l_{0}$ and $x=0$, as well as the resonant frequencies $\omega_{\text{M,}1}$ and $\omega_{\text{M,}2}$. Namely, the two moving mirrors can be described by two mechanical oscillators. In this case, Eq.~(\ref{Haminzhanjain}) can be written as
\begin{eqnarray}
\hat{H} &=&\sum_{k=1}^{\infty}\hbar\omega_{k}(\hat{q}_{r})\hat{a}_{k}^{\dagger}(\hat{q}_{r})\hat{a}_{k}(\hat{q}_{r})
+\frac{1}{2}\sum_{k=1}^{\infty}\hbar\omega_{k}(\hat{q}_{r})+\frac{1}{2m}\hat{p}_{c}^{2}+\frac{1}{2}m\omega_{\text{mc}}^{2}\hat{q}_{c}^{2}
+\frac{1}{2\mu}\hat{p}_{r}^{2}+\frac{1}{2}\mu\omega_{r}^{2}\hat{q}_{r}^{2}      \nonumber \\
&&+\hat{p}_{r}\left(\frac{\hat{\Gamma}_{1}}{2m_{1}}-\frac{\hat{\Gamma}_{2}}{2m_{2}}\right)
+\left(\frac{\hat{\Gamma}_{1}}{2m_{1}}-\frac{\hat{\Gamma}_{2}}{2m_{2}}\right)\hat{p}_{r}+\frac{1}{2m}\left[\hat{p}_{c}(\hat{\Gamma%
}_{1}+\hat{\Gamma}_{2})+(\hat{\Gamma}_{1}+\hat{\Gamma}_{2})\hat{p}_{c}\right]                        \nonumber \\
&&+\mu\left(\omega_{\text{M,}1}^{2}-\omega_{\text{M,}2}^{2}\right)\hat{q}_{r}\hat{q}_{c}-m_{1}\omega_{\text{M,}1}^{2}l_{0}\left(\hat{q}_{c}
+\frac{\mu}{m_{1}}\hat{q}_{r}\right)+\frac{1}{2}m_{1}\omega_{\text{M,}1}^{2}l_{0}^{2}
+\frac{1}{2m_{1}}\hat{\Gamma}_{1}^{2}+\frac{1}{2m_{2}}\hat{\Gamma}_{2}^{2},     \label{Hamiltonianzhangkai}
\end{eqnarray}
where $\omega_{\text{mc}}$ and $\omega_{r}$ are, respectively, the resonant frequencies of the center-of-mass mode and the relative mode for the two harmonic oscillators, which are defined by
\begin{equation}
\omega_{\text{mc}}=\sqrt{\frac{m_{1}\omega_{\text{M,}1}^{2}+m_{2}\omega_{\text{M,}2}^{2}}{m}},\hspace{1 cm} \omega_{r}=\sqrt{\frac{m_{2}\omega_{\text{M,}1}^{2}+m_{1}\omega_{\text{M,}2}^{2}}{m}}.
\end{equation}

According to Eq.~(\ref{Hamiltonianzhangkai}), the center-of-mass mode will couple with the cavity fields even when the two moving end mirrors have the same resonant frequency ($\omega_{\text{M,}1}=\omega_{\text{M,}2}$) in the case of multiple cavity modes, which means that there is no mechanical dark mode in this system. Under the single-mode approximation, however, the center-of-mass mode is decoupled from the cavity field for two degenerate ($\omega_{\text{M,}1}=\omega_{\text{M,}2}$) moving end mirrors and becomes a mechanical dark mode due to $\hat{\Gamma}_{1}=\hat{\Gamma}_{2}=0$. Thus, the single-mode approximation is the reason for the appearance of mechanical dark modes in the present two-moving-mirror optomechanical system.

Under the single-mode approximation, the Hamiltonian~(\ref{Haminzhanjain}) can be reduced to
\begin{equation}
\hat{H}=\hbar\omega_{k}(\hat{q}_{r})\hat{a}_{k}^{\dagger}(\hat{q}_{r})\hat{a}_{k}(\hat{q}_{r})+\frac{1}{2}\hbar\omega
_{k}(\hat{q}_{r})+\frac{1}{2m}\hat{p}_{c}^{2}+\frac{1}{2\mu}\hat{p}_{r}^{2}+V_{1}\left(\hat{q}_{c}+\frac{\mu }{m_{1}}\hat{q}_{r}\right)+V_{2}\left(\hat{q}_{c}-\frac{\mu}{m_{2}}\hat{q}_{r}\right).\label{smH}
\end{equation}
It can be found from Eq.~(\ref{smH}) that the energy spectrum of the system has a lower bound under the single-mode approximation. Namely, the single-mode approximation will not cause the disappearance of the lower bound in the energy spectrum of the system.

\section{IV. The quantized Hamiltonian of the system under the linear-expansion approximation}\label{qhula}

In this section, we will derive the quantized Hamiltonian of the system under the linear expansion approximation. Considering small vibrations around the equilibrium positions of the two moving mirrors, i.e.,
\begin{subequations}
\begin{align}
\langle\hat{x}_{1}\rangle&\equiv\langle\hat{q}_{1}-l_{0}\rangle\ll l_{0},   \\
\langle\hat{x}_{2}\rangle&\equiv\langle\hat{q}_{2}\rangle\ll l_{0}.
\end{align}
\end{subequations}
Then we can perform the Taylor expansion and keep both $\omega_{k}(\hat{q}_{r})$ and $\hat{a}_{k}(\hat{q}_{r})$ upto the first order of a small amount $(\hat{x}_{1}-\hat{x}_{2})/l_{0}$,
\begin{equation}
\omega_{k}(\hat{q}_{r}) \simeq  \omega_{k0}-\frac{\omega_{k0}}{l_{0}}\hat{x}_{r}, \hspace{1 cm} \hat{a}_{k}(\hat{q}_{r}) \simeq \hat{a}_{k0}-\frac{\hat{x}_{r}}{2l_{0}}\hat{a}_{k0}^{\dagger}, \label{frquencyoperatorpotentialfunctions}
\end{equation}
where $\omega_{k0}=k\pi/l_{0}$ and $\hat{x}_{r}=\hat{x}_{1}-\hat{x}_{2}$. To be concise, we introduce the notations $\hat{a}_{k0}\equiv\hat{a}_{k}(l_{0})$ and $\hat{a}_{k0}^{\dagger}\equiv\hat{a}_{k}^{\dagger}(l_{0})$, which denote the annihilation and creation operators corresponding to the equilibrium position, respectively. Under the linear expansion approximation, the Hamiltonian in Eq.~(\ref{Hamiltonianzhangkai}) can be approximated as
\begin{eqnarray}
\hat{H} &=&\sum_{k=1}^{\infty}\hbar\omega_{k0}\left(\hat{a}_{k0}^{\dagger}\hat{a}_{k0}+\frac{1}{2}\right)
+\frac{1}{2m_{1}}\left(\frac{\mu}{m_{2}}\hat{p}_{c}+\hat{p}_{r}+\hat{\Gamma}_{10}\right)^{2}+\frac{1}{2m_{2}}\left(\frac{\mu}{m_{1}}\hat{p%
}_{c}-\hat{p}_{r}+\hat{\Gamma}_{20}\right)^{2}             \nonumber \\
&&-\sum_{k=1}^{\infty}\hbar\omega_{k0}\frac{\hat{x}_{r}}{2l_{0}}(\hat{a}_{k0}^{\dagger}+\hat{a}_{k0})^{2}+\frac{1}{2}m\omega_{c}^{2}\hat{x}%
_{c}^{2}+\frac{1}{2}\mu\omega_{r}^{2}\hat{x}_{r}^{2}+\mu(\omega_{\text{M,}1}^{2}-\omega_{\text{M,}2}^{2})\hat{x}_{c}\hat{x}_{r},
\end{eqnarray}
where $\hat{\Gamma}_{n0}\simeq \hat{\Gamma}_{n}$ is defined by
\begin{equation}
\label{parametersofmasterequation}
\hat{\Gamma}_{n0}=\frac{i\hbar}{2l_{0}}\sum_{j,k}^{\infty}g_{kj}^{(n)}\sqrt{\frac{k}{j}}(\hat{a}_{k0}^{\dagger}
-\hat{a}_{k0})(\hat{a}_{j0}^{\dagger}+\hat{a}_{j0})
\end{equation}
for $n=1,2$, and $\hat{x}_{c}=(m_{1}\hat{x}_{1}+m_{2}\hat{x}_{2})/m_{t}$. The commutator of $\hat{\Gamma}_{10}$ and $\hat{\Gamma}_{20}$ can be further obtained as
\begin{equation}
\label{commutor12}
[\hat{\Gamma}_{10},\hat{\Gamma}_{20}]=-\frac{\hbar^{2}}{l_{0}^{2}}\sum_{j+s=\text{odd}}^{\infty}g_{js}^{(2)}\sqrt{%
\frac{s}{j}}(\hat{a}_{j0}^{\dagger}\hat{a}_{s0}^{\dagger}-\hat{a}_{j0}^{\dagger}\hat{a}_{s0}
+\hat{a}_{j0}\hat{a}_{s0}^{\dagger}-\hat{a}_{j0}\hat{a}_{s0}).
\end{equation}
According to Eqs.~(\ref{dimensionlesscoeff}) and~(\ref{commutor12}), we know $[\hat{\Gamma}_{10},\hat{\Gamma}_{20}]=0$ under the single-mode approximation. In the multimode case, differently, we find that $[\hat{\Gamma}_{10},\hat{\Gamma}_{20}]$ is not equal to zero. However, we can check that the relation
\begin{equation}
[\hat{\Gamma}_{10},\hat{\Gamma}_{20}] \simeq 0
\end{equation}
can be approximately obtained under the linear expansion approximation. In addition, for the vacuum-cavity-field case under our consideration, we can obtain the relation $\langle[\hat{\Gamma}_{10},\hat{\Gamma}_{20}]\rangle =0$. It is indicated that, in the vacuum cavity field, the average value of the canonical momenta of the two moving end mirrors are equal to their mechanical momenta.

To make sure that the canonical momenta of the two moving end mirrors are equal to their mechanical momenta in this representation, we perform the transformation $\hat{H}^{\prime}=\hat{T}^{\dagger}\hat{H}\hat{T}$ with
\begin{equation}
\hat{T}=\exp\left[\frac{i}{\hbar}\left(\frac{m_{1}}{m}\hat{x}_{r}-\hat{x}_{c}\right)\hat{\Gamma}_{20}
-\frac{i}{\hbar}\left(\frac{m_{2}}{m}\hat{x}_{r}+\hat{x}_{c}\right)\hat{\Gamma}_{10}\right].
\end{equation}
The transformed Hamiltonian can be expressed as
\begin{eqnarray}
\hat{H}^{\prime} &=&\hbar\sum_{k=1}^{\infty}\omega_{k0}(\hat{a}_{k0}^{\dagger}\hat{a}_{k0}+1/2)+\frac{\hat{p}_{c}^{2}}{2m}+\frac{m\omega_{c}^{2}}{2}\hat{x}_{c}^{2}
+\frac{\hat{p}_{r}^{2}}{2\mu}+\frac{\mu\omega_{r}^{2}}{2}\hat{x}_{r}^{2}
+\hbar\hat{x}_{c}\sum_{k+j=\text{odd}}^{\infty}\hat{F}_{j,k}      \nonumber \\
&&-\sum_{k=1}^{\infty}\frac{\hbar\omega_{k0}}{2l_{0}}\hat{x}_{r}-\frac{\hbar\hat{x}_{r}}{2m}\sum_{k,j=1}^{\infty}[m_{1}+(-1)^{j+k}m_{2}]\hat{F}_{j,k} +\mu(\omega_{\text{M,}1}^{2}-\omega_{\text{M,}2}^{2})\hat{x}_{c}\hat{x}_{r},   \label{Hamiltonianlast}
\end{eqnarray}
where we introduce
\begin{equation}
\hat{F}_{j,k}\equiv\frac{1}{l_{0}}\sqrt{\omega_{j0}\omega_{k0}}(\hat{a}_{k0}^{\dagger}\hat{a}_{j0}^{\dagger}+\hat{a}_{j0}\hat{a}_{k0}
+\hat{a}_{k0}^{\dagger}\hat{a}_{j0}+\hat{a}_{j0}^{\dagger}\hat{a}_{k0}). \label{fdinf}
\end{equation}
Based on Eq.~(\ref{fdinf}), we can prove the relation $\hat{F}_{j,k}=\hat{F}_{k,j}$. According to Eq.~(\ref{Hamiltonianlast}), we know that the center-of-mass mode is decoupled from the cavity field and become the dark mode for the two degenerate moving end mirrors under the single-mode approximation.

Under the linear expansion approximation, it is more convenient to study the phonon transfer in the original coordinate system. In terms of the relations
\begin{equation}
\hat{x}_{r}=\hat{x}_{1}-\hat{x}_{2}, \hspace{1 cm} \hat{x}_{c}=\frac{1}{m_{t}}(m_{1}\hat{x}_{1}+m_{2}\hat{x}_{2}),
\end{equation}
and
\begin{equation}
\hat{x}_{l}=x_{l,\text{zpf}}(\hat{b}_{l}^{\dagger}+\hat{b}_{l}), \hspace{1 cm} \hat{p}_{l}=im_{l}\omega_{\text{M,}l}x_{l,\text{zpf}}(\hat{b}_{l}^{\dagger}-\hat{b}_{l}),  \label{phcadefined}
\end{equation}
the Hamiltonian of the system can be written as
\begin{eqnarray}
\hat{H} &=&\hbar\sum_{n=1}^{\infty}\omega_{n0}\hat{a}_{n0}^{\dagger}\hat{a}_{n0}+\hbar\omega_{\text{M,}1}\hat{b}_{1}^{\dagger}\hat{b}_{1}
+\hbar\omega_{\text{M,}2}\hat{b}_{2}^{\dagger}\hat{b}_{2}+\frac{\hbar}{2}\sum_{k,n=1}^{\infty}[x_{\text{2,zpf}}(\hat{b}_{2}^{\dagger}+%
\hat{b}_{2})-(-1)^{n+k}x_{\text{1,zpf}}(\hat{b}_{1}^{\dagger}+\hat{b}_{1})]\hat{F}_{n,k}            \nonumber \\
&&+\sum_{n=1}^{\infty}\frac{\hbar\omega_{n0}}{2}+\sum_{n=1}^{\infty}\frac{\hbar\omega_{n0}}{2l_{0}}[x_{\text{2,zpf}}(\hat{b}_{2}^{\dagger}+%
\hat{b}_{2})-x_{\text{1,zpf}}(\hat{b}_{1}^{\dagger}+\hat{b}_{1})],          \label{Hamiltonianlastinorg}
\end{eqnarray}
where $x_{l,\text{zpf}}=\sqrt{\hbar/(2m_{l}\omega_{\text{M,}l})}$ is the zero-point-fluctuation amplitude of the $l$th ($l=1,2$) moving end mirror. In Eq.~(\ref{Hamiltonianlastinorg}), the first, second, and third terms in the first line represent the free Hamiltonians of the multimode cavity fields, the right moving mirror, and the left moving mirror, respectively. Moreover, the fourth term in the first line represents the interactions between the multimode cavity fields and the two moving end mirrors. The first term of the second line represents the zero-point energy of the multimode cavity fields and it is a constant. The second term of the second line describes the motion of the two moving end mirrors caused by the zero-point energy of the cavity fields, which can be balanced by properly introducing external forces. Thus, Eq.~(\ref{Hamiltonianlastinorg}) can be written as
\begin{equation}
\hat{H} =\hbar\sum_{n=1}^{\infty}\omega_{n0}\hat{a}_{n0}^{\dagger}\hat{a}_{n0}+\hbar\omega_{\text{M,}1}\hat{b}_{1}^{\dagger}\hat{b}_{1}+\hbar\omega
_{\text{M,}2}\hat{b}_{2}^{\dagger}\hat{b}_{2}+
\frac{\hbar}{2}\sum_{k,n=1}^{\infty}[x_{\text{2,zpf}}(\hat{b}_{2}^{\dagger}+\hat{b}_{2})-(-1)^{n+k}x_{%
\text{1,zpf}}(\hat{b}_{1}^{\dagger}+\hat{b}_{1})]\hat{F}_{n,k}.   \label{Hamiltonianlastinorg1}
\end{equation}

Under the single-mode approximation, we can only consider the $k$th cavity mode. Then Eq.~(\ref{Hamiltonianlastinorg}) is reduced to
\begin{eqnarray}
\hat{H} &=&\hbar\omega_{k0}\hat{a}_{k0}^{\dagger}\hat{a}_{k0}+\hbar\omega_{\text{M,}1}\hat{b}_{1}^{\dagger}\hat{b}_{1}+\hbar\omega_{\text{M,%
}2}\hat{b}_{2}^{\dagger}\hat{b}_{2}+\hbar g_{k,2}\hat{a}_{k0}^{\dagger}\hat{a}_{k0}(\hat{b}_{2}^{\dagger}+\hat{b}_{2})-\hbar g_{k,1}\hat{a}%
_{k0}^{\dagger}\hat{a}_{k0}(\hat{b}_{1}^{\dagger}+\hat{b}_{1})                         \nonumber \\
&&+\frac{\hbar}{2}g_{k,2}(\hat{a}_{k0}^{\dagger}\hat{a}_{k0}^{\dagger}+\hat{a}_{k0}\hat{a}_{k0})(\hat{b}_{2}^{\dagger}+%
\hat{b}_{2})-\frac{\hbar}{2}g_{k,1}(\hat{a}_{k0}^{\dagger}\hat{a}_{k0}^{\dagger}+\hat{a}_{k0}\hat{a}_{k0})(\hat{b}_{1}^{\dagger}+\hat{b}_{1})     \nonumber \\
&&+\frac{\hbar}{2}[g_{k,2}(\hat{b}_{2}^{\dagger}+\hat{b}_{2})-g_{k,1}(\hat{b}_{1}^{\dagger}+\hat{b}_{1})],    \label{singleHamiltonianlastinorg}
\end{eqnarray}
where we discard the constant term $\hbar\omega_{k0}/2$, and introduce the parameters
\begin{equation}
g_{k,1}=\frac{\omega_{k0}}{l_{0}}x_{\text{1,zpf}}, \hspace{1 cm}  g_{k,2}=\frac{\omega_{k0}}{l_{0}}x_{\text{2,zpf}}.   \label{couplingstrong}
\end{equation}
We can introduce external forces or define new photon and phonon operators to eliminate the third line of Eq.~(\ref{singleHamiltonianlastinorg}). For typical cavity optomechanical systems, the cavity frequency is much larger than the mechanical frequency, then the second line of Eq.~(\ref{singleHamiltonianlastinorg}) determining the dynamical Casimir effect can be safely neglected. As a result, Eq.~(\ref{singleHamiltonianlastinorg}) can be reduced to
\begin{equation}
\hat{H}=\hbar\omega_{k0}\hat{a}_{k0}^{\dagger}\hat{a}_{k0}+\hbar\omega_{\text{M,}1}\hat{b}_{1}^{\dagger}\hat{b}_{1}+\hbar\omega_{\text{M,}2}\hat{b%
}_{2}^{\dagger}\hat{b}_{2}+\hbar g_{k,2}\hat{a}_{k0}^{\dagger}\hat{a}_{k0}(\hat{b}_{2}^{\dagger}+\hat{b}_{2})-\hbar g_{k,1}\hat{a}_{k0}^{\dagger}\hat{%
a}_{k0}(\hat{b}_{1}^{\dagger}+\hat{b}_{1}),   \label{singleHamra}
\end{equation}
which can be extended from the Hamiltonian of the single-moving-mirror cavity optomechanical system. Therefore, it is necessary to eliminate the misunderstanding that the Hamiltonian~(\ref{singleHamra}) has no a lower bound in the eigen-energy spectrum.

The Hamiltonian~(\ref{singleHamra}) can be diagonalized, and the eigen-equation can be expressed as
\begin{equation}
\label{eigenequation}
\hat{H}|n\rangle_{c}|\tilde{j}(n)\rangle_{b_{1}}|\tilde{s}(n)\rangle_{b_{2}}
=E_{n,j,s}|n\rangle_{c}|\tilde{j}(n)\rangle_{b_{1}}|\tilde{s}(n)\rangle_{b_{2}}.
\end{equation}%
Here, the eigenvalue $E_{n,j,s}$ is given by
\begin{equation}
E_{n,j,s}=\hbar\left(j\omega_{\text{M,}1}+s\omega_{\text{M,}2}+n\omega_{k0}-n^{2}\frac{\omega_{\text{M,}1}g_{k,2}^{2}+\omega_{\text{M,}%
2}g_{k,1}^{2}}{\omega_{\text{M,}1}\omega_{\text{M,}2}}\right),   \label{eigenvalue}
\end{equation}
and $|\tilde{j}(n)\rangle_{b_{1}}\equiv\hat{D}_{b_{1}}(n\alpha)|j\rangle_{b_{1}}$ ($|\tilde{s}(n)\rangle_{b_{2}}\equiv\hat{D}_{b_{2}}(n\beta)|s\rangle_{b_{1}}$) is an $n$-photon displaced number state of the right (left) moving end mirror, where
$\hat{D}_{b_{1}}(n\alpha)=\exp[n\alpha(\hat{b}_{1}^{\dagger}-\hat{b}_{1})]$ and $\hat{D}_{b_{2}}(n\beta)=\exp[n\beta(\hat{b}_{2}^{\dagger}-\hat{b}_{2})]$ are the displacement operators with $\alpha=g_{k,1}/\omega_{\text{M,}1}$ and $\beta=-g_{k,2}/\omega_{\text{M,}2}$. According to Eq.~(\ref{eigenvalue}), we find that there exists a critical photon number
\begin{equation}
n_{c}=\frac{\omega_{\text{M,}1}\omega_{\text{M,}2}\omega_{k0}}{2(\omega_{\text{M,}1}g_{k,2}^{2}+\omega_{\text{M,}2}g_{k,1}^{2})}    \label{criticalphoton}
\end{equation}
for given phonon numbers $j$ and $s$. In this case, the energy $E_{n,j,s}$ increases with the increase of the photon number $n$ under the condition $n<n_{c}$, and it decreases with the increase of $n$ when $n>n_{c}$. Therefore, the Hamiltonian~(\ref{singleHamra}) seems wrong because the energy spectrum has no a lower bound.

In addition, under a given number of photon $n$, we know from Eqs.~(\ref{eigenequation}) and~(\ref{eigenvalue}) that the eigenstate corresponding to the lowest energy is $|n\rangle_{c}|\tilde{0}(n)\rangle_{b_{1}}|\tilde{0}(n)\rangle_{b_{2}}$. When the system is in the eigenstate $|n\rangle_{c}|\tilde{0}(n)\rangle_{b_{1}}|\tilde{0}(n)\rangle_{b_{2}}$, we can calculate the average value and standard deviation of the operator $\hat{x}_{r}$ as follows,
\begin{subequations}
\label{avergenvariance}
    \begin{align}
\langle\hat{x}_{r}\rangle &=n\left(\frac{\hbar}{m_{1}\omega_{\text{M,}1}^{2}}+\frac{\hbar}{m_{2}\omega_{\text{M,}2}^{2}}\right)\frac{\omega
_{k0}}{l_{0}},     \\
\Delta x_{r}&=\sqrt{\langle\hat{x}_{r}^{2}\rangle-\langle\hat{x}_{r}\rangle^{2}}=\sqrt{x_{\text{1,zpf}}^{2}+x_{\text{2,zpf}}^{2}}.
\end{align}
\end{subequations}
According to Eq.~(\ref{avergenvariance}), the average value $\langle\hat{x}_{r}\rangle$ increases with the photon number $n$. When the photon number $n$ reaches the critical photon number $n_{c}$, the average value $\langle\hat{x}_{r}\rangle$ is given by
\begin{equation}
\langle\hat{x}_{r}\rangle =l_{0}.    \label{avergen}
\end{equation}
In this case, the average value $\langle\hat{x}_{r}\rangle$ is equal to the length $l_{0}$ of the bare cavity. Therefore, the linear expansion condition is seriously destroyed when the involved photon number is equal to or greater than the critical photon number $n_{c}$. At the same time, the bosonic commutation relation between the creation and annihilation operators [see Eq.~(\ref{frquencyoperatorpotentialfunctions})] of the $k$th cavity mode is seriously destroyed. Therefore, when the condition under which the Hamiltonian~(\ref{singleHamra}) keeps valid is satisfied, the Hamiltonian~(\ref{singleHamra}) does not violate the requirement that the energy spectrum of a quantum system should have a lower bound.

\section{V. The Effective Hamiltonian describing the dynamics of the two moving mirrors \label{HapproxSM}}

In this section, we derive the effective Hamiltonian, which governs the evolution of the two moving mirrors. To be concise, we introduce the parameters
\begin{subequations}
\label{Ckn12exp}
\begin{align}
C_{k,n}^{(1)}&=C_{n,k}^{(1)}\equiv -(-1)^{n+k}\left(\frac{\hbar}{2}\right)^{3/2}\frac{1}{l_{0}\sqrt{m_{1}}}\sqrt{\frac{\omega_{n0}\omega_{k0}}{%
\omega_{\text{M},1}}},   \\
C_{k,n}^{(2)}&=C_{n,k}^{(2)}\equiv \left(\frac{\hbar}{2}\right)^{3/2}\frac{1}{l_{0}\sqrt{m_{2}}}\sqrt{\frac{%
\omega_{n0}\omega_{k0}}{\omega_{\text{M},2}}},
\end{align}
\end{subequations}
then the Hamiltonian in Eq.~(\ref{Hamiltonianlastinorg1}) of the system can be expressed as
\begin{equation}
\hat{H}=\hbar\sum_{n=1}^{\infty}\omega_{n0}\hat{a}_{n0}^{\dagger}\hat{a}_{n0}+\sum_{\mu =1,2}\hbar\omega_{\text{M,}\mu}%
\hat{b}_{\mu}^{\dagger}\hat{b}_{\mu }+\sum_{\mu =1,2}\sum_{k,n=1}^{\infty}C_{k,n}^{(\mu)}(\hat{b}_{\mu}^{\dagger}+\hat{b}_{\mu
})(\hat{a}_{k0}^{\dagger}\hat{a}_{n0}^{\dagger}+\hat{a}_{n0}\hat{a}%
_{k0}+\hat{a}_{k0}^{\dagger}\hat{a}_{n0}+\hat{a}_{n0}^{\dagger}\hat{a}_{k0}).    \label{Hamiltoniansimply}
\end{equation}
This Hamiltonian describes the interactions among the two moving mirrors and these cavity fields.

To expound the microscopic physical processes involving the photons and phonons in this system, we first analyze the physical processes associated with the interactions in the Hamiltonian $\hat{H}$ given in Eq.~(\ref{Hamiltoniansimply}). We can see from Eq.~(\ref{Hamiltoniansimply}) that there exist four kinds of physical processes, which are described by the interaction terms $\hat{b}_{\mu}^{\dagger}\hat{a}_{k0}^{\dagger}\hat{a}_{n0}^{\dagger}$, $\hat{b}_{\mu}^{\dagger}\hat{a}_{n0}\hat{a}_{k0}$, $\hat{b}_{\mu}^{\dagger
}\hat{a}_{k0}^{\dagger}\hat{a}_{n0}$, $\hat{b}_{\mu}^{\dagger}\hat{a}_{n0}^{\dagger}\hat{a}_{k0}$, and their Hermitian conjugates. Especially, when $n=k$, we have these terms $\hat{b}_{\mu}^{\dagger}\hat{a}_{k0}^{\dagger}\hat{a}_{k0}^{\dagger}$, $\hat{b}_{\mu}^{\dagger}\hat{a}_{k0}\hat{a}_{k0}$, $\hat{b}_{\mu}^{\dagger}\hat{a}_{k0}^{\dagger}\hat{a}_{k0}$, and their Hermitian conjugates. The detunings correspond to the former terms are given by $(\omega_{M}+\omega_{k0}+\omega_{n0})$, $(\omega_{M}-\omega_{k0}-\omega_{n0})$, $(\omega_{M}+\omega_{k0}-\omega_{n0})$, and $(\omega_{M}-\omega_{k0}+\omega_{n0})$. When $k\neq n$, the detunings are $\Delta\omega_{c}\pm s\omega_{M}$ for nonzero integers, which are much larger than the coupling strengths. We should point out that, when $n=k$, there are two special cases, i.e., the terms $\hat{a}_{k0}^{\dagger}\hat{a}_{k0}(\hat{b}_{\mu}^{\dagger}+\hat{b}_{\mu})$, with the detunings $\pm\omega_{M}$. These two detunings could
be comparable even larger than the corresponding coupling strengths. However, for our present considered vacuum cavity, these two terms will disappear because the photon number is zero, i.e., $\langle\hat{a}_{k0}^{\dagger}\hat{a}_{k0}\rangle=0$, and hence the radiation pressure couplings disappear.

The Hamiltonian~(\ref{Hamiltoniansimply}) can be written as $\hat{H}=\hat{H}_{0}+\hat{H}_{I}$, where
\begin{equation}
\hat{H}_{0}=\sum_{n=1}^{\infty}\hbar\omega_{n0}\hat{a}_{n0}^{\dagger}\hat{a}_{n0}
+\sum_{\mu=1,2}\hbar\omega_{\text{M,}\mu}\hat{b}_{\mu}^{\dagger}\hat{b}_{\mu}
\end{equation}%
is the free Hamiltonian of the multimode cavity fields and the two mechanical mode, and
\begin{equation}
\hat{H}_{I}=\sum_{\mu =1,2}\sum_{k,n=1}^{\infty}C_{k,n}^{(\mu)}(\hat{b}_{\mu}^{\dagger}+\hat{b}_{\mu})(\hat{a}_{k0}^{\dagger}%
\hat{a}_{n0}^{\dagger}+\hat{a}_{n0}\hat{a}_{k0}+\hat{a}_{k0}^{\dagger}\hat{a}_{n0}+\hat{a}_{n0}^{\dagger}\hat{a}_{k0})
\end{equation}%
describes the interactions between the two moving mirrors and these cavity fields.

In the interaction picture with respect to $\hat{H}_{0}=\hbar\sum_{n=1}^{\infty}\omega_{n0}\hat{a}_{n0}^{\dagger}\hat{a}_{n0}
+\hbar\sum_{\mu=1,2}\omega_{\text{M,}\mu}\hat{b}_{\mu}^{\dagger}\hat{b}_{\mu}$, the Hamiltonian~(\ref{Hamiltoniansimply}) can be expressed as
\begin{eqnarray}
\label{interHamiltonian}
\hat{H}_{I}^{(I)}(t) &=&\sum_{\mu=1,2}\sum_{k,n=1}^{\infty}C_{k,n}^{(\mu)}(\hat{b}_{\mu}^{\dagger}e^{i\omega_{\text{M,}\mu}t}+\hat{b}_{\mu}e^{-i\omega_{\text{M,}\mu}t})
\left(\hat{a}_{k0}^{\dagger}\hat{a}_{n0}^{\dagger}e^{i(\omega_{n0}+\omega_{k0})t}+\hat{a}_{n0}\hat{a}_{k0}e^{-i(\omega_{n0}+\omega_{k0})t}\right. \nonumber \\
&&\left.+\hat{a}_{k0}^{\dagger}\hat{a}_{n0}e^{-i(\omega_{n0}-\omega_{k0})t}+\hat{a}_{n0}^{\dagger}\hat{a}_{k0}e^{i(\omega_{n0}-\omega_{k0})t}\right).
\end{eqnarray}
It can be seen from Eq.~(\ref{interHamiltonian}) that the oscillating frequencies $(\pm\omega_{\text{M,}\mu}\pm\omega_{n0}\pm\omega_{k0})$ of these terms are distinct, because the free spectral range $\Delta \omega _{c}$ is much larger than the resonance frequency $\omega_{\text{M,}\mu}$ of the moving mirrors. Then the first-order physical processes are largely detuned. Nevertheless, the physical system can be approximately described by the second-order effective Hamiltonian when the second-order physical processes are resonant. Below, we derive the second-order effective Hamiltonian of the system using the James effective Hamiltonian method.

In our following considerations, we only study the degenerate-mirror case, namely, the two movable end mirrors have the same frequency, i.e., $\omega_{\text{M,}1}=\omega_{\text{M,}2}=\omega_{M}$. Using the relationship $C_{k,n}^{(\mu)}=C_{n,k}^{(\mu )}$, the Hamiltonian~(\ref{interHamiltonian}) can be expressed as
\begin{eqnarray}
\hat{H}_{I}^{(I)}(t) &=&\sum_{k,n=1}^{\infty}\sum_{\mu =1,2}C_{k,n}^{(\mu
)}\left(\hat{b}_{\mu}^{\dagger}\hat{a}_{k0}^{\dagger}\hat{a}_{n0}^{\dagger}e^{i(\omega_{M}+\omega_{n0}+\omega_{k0})t}+\hat{b}_{\mu}%
\hat{a}_{n0}\hat{a}_{k0}e^{-i(\omega_{M}+\omega_{n0}+\omega_{k0})t}\right.   \nonumber\\
&&\left.+\hat{b}_{\mu}^{\dagger}\hat{a}_{n0}\hat{a}_{k0}e^{i(\omega_{M}-\omega_{n0}-\omega_{k0})t}+\hat{b}_{\mu}\hat{a}_{k0}^{\dagger}\hat{a%
}_{n0}^{\dagger}e^{-i(\omega_{M}-\omega_{n0}-\omega_{k0})t}\right.    \nonumber\\
&&\left.+2\hat{b}_{\mu}^{\dagger}\hat{a}_{k0}^{\dagger}\hat{a}_{n0}e^{i(\omega_{M}-\omega_{n0}+\omega_{k0})t}+2\hat{b}_{\mu}\hat{a}%
_{n0}^{\dagger}\hat{a}_{k0}e^{-i(\omega_{M}-\omega_{n0}+\omega_{k0})t}\right).   \label{interHinpic}
\end{eqnarray}%
To adopt the James effective Hamiltonian method, we denote the interaction terms in the Hamiltonian $\hat{H}_{I}^{(I)}(t)$ as
\begin{eqnarray}
\label{omeganunk}
\hat{h}_{\mu,n,k}^{(1)}=C_{k,n}^{(\mu)}\hat{b}_{\mu}\hat{a}_{n0}\hat{a}_{k0},\hspace{1 cm}
\hat{h}_{\mu ,n,k}^{(2)}=C_{k,n}^{(\mu)}\hat{b}_{\mu}\hat{a}_{k0}^{\dagger}\hat{a}_{n0}^{\dagger},\hspace{1 cm}
\hat{h}_{\mu,n,k}^{(3)}=2 C_{k,n}^{(\mu)}\hat{b}_{\mu}\hat{a}_{n0}^{\dagger}\hat{a}_{k0},
\end{eqnarray}
with the corresponding oscillating frequencies
\begin{eqnarray}
\label{omeganunkfuls}
\omega_{\mu,n,k}^{(1)}=\omega_{M}+\omega_{n0}+\omega_{k0},\hspace{1 cm}
\omega_{\mu,n,k}^{(2)}=\omega_{M}-\omega_{n0}-\omega_{k0},\hspace{1 cm}
\omega_{\mu,n,k}^{(3)}=\omega_{M}-\omega_{n0}+\omega_{k0}.
\end{eqnarray}
Therefore, the Hamiltonian $\hat{H}_{I}^{(I)}(t)$ can be expressed as
\begin{equation}
\hat{H}_{I}^{(I)}(t)=\sum_{k,n=1}^{\infty}\sum_{\mu =1,2}\sum_{\alpha=1,2,3}\left(\hat{h}_{\mu ,n,k}^{(\alpha)\dagger }e^{i\omega_{\mu
,n,k}^{(\alpha)}t}+\hat{h}_{\mu ,n,k}^{(\alpha)}e^{-i\omega_{\mu,n,k}^{(\alpha)}t}\right).
\end{equation}
Based on the James effective Hamiltonian method~\cite{Gamel2010PRA}, the second-order Hamiltonian
can be written as
\begin{equation}
\label{Heffectiveformulain}
\hat{H}^{(2)}_{\text{eff}}(t)\approx\sum_{k,n=1}^{\infty}\sum_{r,s=1}^{\infty}\sum_{\mu,\nu =1,2}\sum_{\alpha,\beta=1,2,3}\frac{1
}{2\hbar}\left(\frac{1}{\omega_{\mu,n,k}^{(\alpha)}}+\frac{1}{\omega_{\nu,r,s}^{(\beta)}}\right)
\left[\hat{h}_{\nu,r,s}^{(\beta)\dagger},\hat{h}_{\mu,n,k}^{(\alpha)}\right]e^{i(\omega_{\nu,r,s}^{(\beta)}-\omega_{\mu,n,k}^{(\alpha)})t}.
\end{equation}
Based on the oscillating frequencies $\omega_{\mu ,n,k}^{(\alpha)}$ for $\alpha=1,2,3$ given in Eq.~(\ref{omeganunkfuls}), we can obtain the following oscillating frequencies
\begin{subequations}
\label{omegars1unk1}
\begin{align}
\omega_{\nu,r,s}^{(1)}-\omega_{\mu,n,k}^{(1)}=&\omega_{r0}+\omega_{s0}-\omega_{n0}-\omega_{k0},\\
\omega_{\nu,r,s}^{(2)}-\omega_{\mu,n,k}^{(1)}=&-\omega_{r0}-\omega_{s0}-\omega_{n0}-\omega_{k0},\\
\omega_{\nu,r,s}^{(3)}-\omega_{\mu,n,k}^{(1)}=&\omega_{s0}-\omega_{r0}-\omega_{n0}-\omega_{k0},\\
\omega_{\nu,r,s}^{(1)}-\omega_{\mu,n,k}^{(2)}=&\omega_{r0}+\omega_{s0}+\omega_{n0}+\omega_{k0},\\
\omega_{\nu,r,s}^{(2)}-\omega_{\mu,n,k}^{(2)}=&\omega_{n0}+\omega_{k0}-\omega_{r0}-\omega_{s0},\\
\omega_{\nu,r,s}^{(3)}-\omega_{\mu,n,k}^{(2)}=&\omega_{s0}+\omega_{n0}+\omega_{k0}-\omega_{r0},\\
\omega_{\nu,r,s}^{(1)}-\omega_{\mu,n,k}^{(3)}=&\omega_{r0}+\omega_{s0}+\omega_{n0}-\omega_{k0},\\
\omega_{\nu,r,s}^{(2)}-\omega_{\mu,n,k}^{(3)}=&\omega_{n0}-\omega_{r0}-\omega_{s0}-\omega_{k0},\\
\omega_{\nu,r,s}^{(3)}-\omega_{\mu,n,k}^{(3)}=&\omega_{s0}+\omega_{n0}-\omega_{r0}-\omega_{k0}.
\end{align}
\end{subequations}

It can be seen from the relation $\omega_{n0}=n\Delta\omega_{c}$ with $\Delta\omega_{c}=\pi c/l_{0}$ that, these oscillating frequencies could be zero with proper values of $r$, $s$, $n$, $k$, and these terms corresponding to the resonant second-order physical processes. In order to use the International System of Units in specific numerical calculations, henceforth, we  restore the value of the speed of light $c$. In the derivation of the second-order effective Hamiltonian, we only keep the resonant coupling terms. By inspecting Eqs.~(\ref{Heffectiveformulain}) and~(\ref{omegars1unk1}), we find that the oscillating frequencies of these terms take the form as
$(\pm\omega_{s0}\pm\omega_{r0}\pm\omega_{n0}\pm\omega_{k0})$. On one hand, since the free spectrum range of the cavity field is
much larger than the resonance frequency of the moving mirrors [see Fig.~\ref{mode}(b)], then only the terms associated with the relation $\pm\omega_{s0}\pm\omega_{r0}\pm\omega_{n0}\pm\omega_{k0}=0$ could be kept. Based on these analyses, we know that, for the phononic mode part, only the terms containing $\hat{b}_{\nu}\hat{b}_{\mu}^{\dagger}$ and $\hat{b}_{\nu}^{\dagger}\hat{b}_{\mu}$ can be kept. For the photonic mode part, the
kept terms should contain one annihilation operator and three creation operators, two annihilation operators and two creation operators, three annihilation operators and one creation operator. For the left two cases, namely the terms containing four annihilation operators or four creation operators, the physical process cannot be resonant. In particular, for the kept terms, the summation of the oscillating frequencies should be zero, such that the interaction terms are resonant. This resonance effect can be described by introducing the Kronecker delta function. As a result, the second-order effective Hamiltonian can be reduced to the following form
\begin{eqnarray}
\label{HNeectiveHamil01}
\hat{H}^{(2)}_{\text{eff}}(t)&\simeq&\sum_{k,n=1}^{\infty}\sum_{r,s=1}^{\infty}\sum_{\mu,\nu=1,2}\frac{1}{2\hbar}\left(\frac{1}{\omega_{\mu,n,k}^{(1)}}+\frac{1}{\omega_{\nu,r,s}^{(1)}}\right)
\left[\hat{h}_{\nu,r,s}^{(1)\dagger},\hat{h}_{\mu ,n,k}^{(1)}\right]\delta_{r+s,n+k}\nonumber\\
&&+\sum_{k,n=1}^{\infty}\sum_{r,s=1}^{\infty}\sum_{\mu,\nu=1,2}\frac{1}{2\hbar}\left(\frac{1}{\omega_{\mu,n,k}^{(1)}}+\frac{1}{\omega_{\nu,r,s}^{(3)}}\right)
\left[\hat{h}_{\nu,r,s}^{(3)\dagger},\hat{h}_{\mu ,n,k}^{(1)}\right]\delta_{s,r+n+k}\nonumber\\
&&+\sum_{k,n=1}^{\infty}\sum_{r,s=1}^{\infty}\sum_{\mu,\nu=1,2}\frac{1}{2\hbar}\left(\frac{1}{\omega_{\mu,n,k}^{(2)}}+\frac{1}{\omega_{\nu,r,s}^{(2)}}\right)
\left[\hat{h}_{\nu,r,s}^{(2)\dagger},\hat{h}_{\mu,n,k}^{(2)}\right]\delta_{n+k,r+s}\nonumber\\
&&+\sum_{k,n=1}^{\infty}\sum_{r,s=1}^{\infty}\sum_{\mu,\nu=1,2}\frac{1}{2\hbar}\left(\frac{1}{\omega_{\mu,n,k}^{(2)}}+\frac{1}{\omega_{\nu,r,s}^{(3)}}\right)
\left[\hat{h}_{\nu,r,s}^{(3)\dagger},\hat{h}_{\mu,n,k}^{(2)}\right]\delta_{s+n+k,r}\nonumber\\
&&+\sum_{k,n=1}^{\infty}\sum_{r,s=1}^{\infty}\sum_{\mu,\nu=1,2}\frac{1}{2\hbar}\left(\frac{1}{\omega_{\mu,n,k}^{(3)}}+\frac{1}{\omega_{\nu,r,s}^{(1)}}\right)
\left[\hat{h}_{\nu,r,s}^{(1)\dagger},\hat{h}_{\mu ,n,k}^{(3)}\right]\delta_{r+s+n,k}\nonumber\\
&&+\sum_{k,n=1}^{\infty}\sum_{r,s=1}^{\infty}\sum_{\mu,\nu=1,2}\frac{1}{2\hbar}\left(\frac{1}{\omega_{\mu,n,k}^{(3)}}+\frac{1}{\omega_{\nu,r,s}^{(2)}}\right)
\left[\hat{h}_{\nu,r,s}^{(2)\dagger},\hat{h}_{\mu,n,k}^{(3)}\right]\delta_{r+s+k,n}\nonumber\\
&&+\sum_{k,n=1}^{\infty}\sum_{r,s=1}^{\infty}\sum_{\mu,\nu=1,2}\frac{1}{2\hbar}\left(\frac{1}{\omega_{\mu,n,k}^{(3)}}+\frac{1}{\omega_{\nu,r,s}^{(3)}}\right)
\left[\hat{h}_{\nu,r,s}^{(3)\dagger},\hat{h}_{\mu,n,k}^{(3)}\right]\delta_{s+n,r+k},
\end{eqnarray}
where we introduce the Kronecker delta function to pick up the resonant terms. Further, in terms of the oscillating frequencies $\omega_{\mu,n,k}^{(\alpha)}$ for $\alpha=1,2,3$, the coefficients can be obtained as
\begin{subequations}
\label{omegaresonance}
\begin{align}
\frac{1}{2}\left(\frac{1}{\omega_{\mu,n,k}^{(1)}}+\frac{1}{\omega_{\nu,r,s}^{(1)}}\right)\delta_{r+s,n+k}=&\frac{1}{\omega_{M}+\omega_{n0}+\omega_{k0}}\delta_{r+s,n+k},\\
\frac{1}{2}\left(\frac{1}{\omega_{\mu,n,k}^{(1)}}+\frac{1}{\omega_{\nu,r,s}^{(3)}}\right)\delta_{s,r+n+k}=&\frac{1}{\omega_{M}+\omega_{n0}+\omega_{k0}}\delta_{s,r+n+k},\\
\frac{1}{2}\left(\frac{1}{\omega_{\mu,n,k}^{(2)}}+\frac{1}{\omega_{\nu,r,s}^{(2)}}\right)\delta_{n+k,r+s}=&\frac{1}{\omega_{M}-\omega_{n0}-\omega_{k0}}\delta_{n+k,r+s},\\
\frac{1}{2}\left(\frac{1}{\omega_{\mu,n,k}^{(2)}}+\frac{1}{\omega_{\nu,r,s}^{(3)}}\right)\delta_{s+n+k,r}=&\frac{1}{\omega_{M}-\omega_{n0}-\omega_{k0}}\delta_{s+n+k,r},\\
\frac{1}{2}\left(\frac{1}{\omega_{\mu,n,k}^{(3)}}+\frac{1}{\omega_{\nu,r,s}^{(1)}}\right)\delta_{r+s+n,k}=&\frac{1}{\omega_{M}-\omega_{n0}+\omega_{k0}}\delta_{r+s+n,k},\\
\frac{1}{2}\left(\frac{1}{\omega_{\mu,n,k}^{(3)}}+\frac{1}{\omega_{\nu,r,s}^{(2)}}\right)\delta_{r+s+k,n}=&\frac{1}{\omega_{M}-\omega_{n0}+\omega_{k0}}\delta_{r+s+k,n},\\
\frac{1}{2}\left(\frac{1}{\omega_{\mu,n,k}^{(3)}}+\frac{1}{\omega_{\nu,r,s}^{(3)}}\right)\delta_{s+n,r+k}=&\frac{1}{\omega_{M}-\omega_{n0}+\omega_{k0}}\delta_{s+n,r+k}.
\end{align}
\end{subequations}
Based on Eqs.~(\ref{omeganunk}), (\ref{HNeectiveHamil01}), and (\ref{omegaresonance}), the effective Hamiltonian can be
written as
\begin{eqnarray}
\label{HNeectiveHamil2}
\hat{H}^{(2)}_{\text{eff}}&=&\sum_{k,n=1}^{\infty}\sum_{r,s=1}^{\infty}\sum_{\mu,\nu=1,2}
\frac{1}{\hbar(\omega_{M}+\omega_{n0}+\omega_{k0})}C_{s,r}^{(\nu)}C_{k,n}^{(\mu)}\left(\hat{b}_{\nu}^{\dagger}\hat{b}_{\mu}
\hat{a}_{s0}^{\dagger}\hat{a}_{r0}^{\dagger}\hat{a}_{n0}\hat{a}_{k0}
-\textcolor{red}{\hat{b}_{\mu}\hat{b}_{\nu}^{\dagger}\hat{a}_{n0}\hat{a}_{k0}\hat{a}_{s0}^{\dagger}\hat{a}_{r0}^{\dagger}}\right)\delta_{r+s,n+k} \notag\\
&&+\sum_{k,n=1}^{\infty}\sum_{r,s=1}^{\infty}\sum_{\mu,\nu=1,2}\frac{2}{\hbar(\omega_{M}+\omega_{n0}+\omega_{k0})}C_{s,r}^{(\nu)}C_{k,n}^{(\mu)}\left(\hat{b}_{\nu}^{\dagger}
\hat{b}_{\mu}\hat{a}_{s0}^{\dagger}\hat{a}_{r0}\hat{a}_{n0}\hat{a}_{k0}
-\hat{b}_{\mu}\hat{b}_{\nu}^{\dagger}\hat{a}_{n0}\hat{a}_{k0}\hat{a}_{s0}^{\dagger}\hat{a}_{r0}\right)\delta_{s,r+n+k} \notag\\
&&+\sum_{k,n=1}^{\infty }\sum_{r,s=1}^{\infty}\sum_{\mu,\nu=1,2}\frac{1}{\hbar(\omega_{M}-\omega_{n0}-\omega_{k0})}C_{s,r}^{(\nu)}C_{k,n}^{(\mu)}\left(\textcolor{red}{\hat{b}_{\nu }^{\dagger}\hat{b}_{\mu }\hat{a}_{r0}\hat{a}_{s0}\hat{a}_{k0}^{\dagger}\hat{a}_{n0}^{\dagger}}
-\hat{b}_{\mu}\hat{b}_{\nu}^{\dagger}\hat{a}_{k0}^{\dagger}\hat{a}_{n0}^{\dagger}\hat{a}_{r0}\hat{a}_{s0}\right)\delta_{n+k,r+s} \notag\\
&&+\sum_{k,n=1}^{\infty}\sum_{r,s=1}^{\infty}\sum_{\mu,\nu=1,2}\frac{2}{\hbar(\omega_{M}-\omega_{n0}-\omega_{k0})}C_{s,r}^{(\nu)}C_{k,n}^{(\mu)}\left(\hat{b}_{\nu}^{\dagger}\hat{b}_{\mu}\hat{a}_{s0}^{\dagger}
\hat{a}_{r0}\hat{a}_{k0}^{\dagger}\hat{a}_{n0}^{\dagger}
-\hat{b}_{\mu }\hat{b}_{\nu}^{\dagger}\hat{a}_{k0}^{\dagger}\hat{a}_{n0}^{\dagger}\hat{a}_{s0}^{\dagger }\hat{a}_{r0}\right)\delta_{s+n+k,r} \notag\\
&&+\sum_{k,n=1}^{\infty}\sum_{r,s=1}^{\infty}\sum_{\mu,\nu=1,2}\frac{2}{\hbar(\omega_{M}-\omega_{n0}+\omega_{k0})}C_{s,r}^{(\nu)}C_{k,n}^{(\mu)}\left(\hat{b}_{\nu}^{\dagger}
\hat{b}_{\mu}\hat{a}_{s0}^{\dagger}\hat{a}_{r0}^{\dagger}\hat{a}_{n0}^{\dagger}\hat{a}_{k0}
-\hat{b}_{\mu}\hat{b}_{\nu}^{\dagger}\hat{a}_{n0}^{\dagger}\hat{a}_{k0}\hat{a}_{s0}^{\dagger}\hat{a}_{r0}^{\dagger}\right)\delta_{r+s+n,k}\notag \\
&&+\sum_{k,n=1}^{\infty}\sum_{r,s=1}^{\infty}\sum_{\mu,\nu =1,2}\frac{2}{\hbar(\omega_{M}-\omega_{n0}+\omega_{k0})}C_{s,r}^{(\nu)}C_{k,n}^{(\mu)}\left(\hat{b}_{\nu}^{\dagger}
\hat{b}_{\mu }\hat{a}_{r0}\hat{a}_{s0}\hat{a}_{n0}^{\dagger}\hat{a}_{k0}
-\hat{b}_{\mu}\hat{b}_{\nu }^{\dagger }\hat{a}_{n0}^{\dagger}\hat{a}_{k0}\hat{a}_{r0}\hat{a}_{s0}\right)\delta_{r+s+k,n} \notag\\
&&+\sum_{k,n=1}^{\infty}\sum_{r,s=1}^{\infty}\sum_{\mu,\nu =1,2}\frac{4}{\hbar(\omega_{M}-\omega_{n0}+\omega_{k0})}C_{s,r}^{(\nu)}C_{k,n}^{(\mu)}\left(\hat{b}_{\nu}^{\dagger}
\hat{b}_{\mu}\hat{a}_{s0}^{\dagger}\hat{a}_{r0}\hat{a}_{n0}^{\dagger}\hat{a}_{k0}
-\hat{b}_{\mu}\hat{b}_{\nu}^{\dagger}\hat{a}_{n0}^{\dagger}\hat{a}_{k0}\hat{a}_{s0}^{\dagger}\hat{a}_{r0}\right)\delta_{s+n,r+k}.
\end{eqnarray}

Though these first-order interactions in Eq.~(\ref{interHinpic}) are off-resonant, the second-order physical processes described by Eq.~(\ref{HNeectiveHamil2}) could be resonant. The second-order Hamiltonian~(\ref{HNeectiveHamil2}) describes the physical processes involving four photons and two phonons. Here, we see that there exist many resonant physical processes involving four photons and two phonons. In particular, when the two mechanical modes are resonant, the four photon processes must also be resonant. These four-photon processes contain two cases of interaction. The first case corresponds the creation of three photons and annihilation of one photon. In this case, the frequency of the single photon should be equal to the total frequency sum of these three photons. The second case corresponds the creation of two photons and annihilation of two photons. Here, the total frequency of the two created photons is equal to that of the two annihilated photons. The resonance of these four-photon processes can be confirmed from the delta functions in Eq.~(\ref{HNeectiveHamil2}).

In this work, we aim to study the phonon heat transfer across the vacuum between the two moving end mirrors. Therefore, we assume that the cavity fields are in a vacuum state. Then only the physical process corresponding to the above mentioned second case can exist. Namely, only the process corresponding to the case in which the two created photons are completely annihilated can exist. In this special case, the two-phonon process is also resonant. When these two phonons come from different mechanical resonator, then this physical process contributes to the resonant phonon exchange physical process, which provides the physical process for the phonon heat transfer in this system.

When the system is in the initial state $\rho_{\text{sys}}(0)=|\emptyset\rangle_{c\:c}\langle\emptyset|\otimes\rho(0)$, where $|\emptyset\rangle_{c}$ denotes the vacuum state of the cavity field and $\rho(0)$ is an any initial state of the two moving mirrors, then the density matrix of the system at time $t$ can be expressed as
\begin{eqnarray}
\rho_{\text{sys}}(t)&=&\hat{U}_{I}(t)\rho_{\text{sys}}(0)\hat{U}_{I}^{\dagger}(t) \nonumber \\
&=&\hat{U}_{I}(t)\rho(0)\otimes|\emptyset\rangle_{c}\;_{c}\langle \emptyset\vert \hat{U}_{I}^{\dagger}(t) \nonumber \\
&=&\sum_{j,k,n,m=0}^{\infty}\rho_{jknm}(0)\hat{U}_{I}(t)
\vert j\rangle_{b_{1}}\vert k\rangle_{b_{2}}|\emptyset\rangle_{c}\;_{b_{1}}\!\langle n\vert_{b_{2}}\!\langle m\vert_{c}\!\langle \emptyset\vert\hat{U}_{I}^{\dagger}(t),
\end{eqnarray}
where $\hat{U}_{I}(t)$ is the unitary evolution operator associated with the Hamiltonian $\hat{H}^{(2)}_{\text{eff}}$. In addition, we expand the density matrix of the two mirrors in the number-state representation, with $\rho_{jknm}(0)=\;_{b_{1}}\!\langle j|_{b_{2}}\!\langle k |\rho(0)|n \rangle_{b_{1}}|m\rangle_{b_{2}}$ being the density matrix elements. The operation of the unitary evolution operator on the basis state can be expressed as
\begin{eqnarray}
\hat{U}_{I}(t)\vert j\rangle_{b_{1}}\vert
k\rangle_{b_{2}}|\emptyset \rangle_{c} &=&e^{\frac{-it}{\hbar}\hat{H}_{\text{eff}}^{(2)}}\vert j\rangle_{b_{1}}\vert k\rangle_{b_{2}}|\emptyset\rangle_{c}\nonumber \\
&=&\left(1+\frac{t}{i\hbar}\hat{H}_{\text{eff}}^{(2)}+\frac{1}{2!}\left(\frac{t}{i\hbar}\right)^{2}[\hat{H}_{\text{eff}}^{(2)}]^{2}+\cdots +\frac{1}{n!}\left(\frac{t}{i\hbar}\right)
^{n}[\hat{H}_{\text{eff}}^{(2)}]^{n}+\cdots\right)|\emptyset\rangle_{c}\vert j\rangle_{b_{1}}\vert
k\rangle_{b_{2}}.\label{UITonbasis}
\end{eqnarray}%
To derive the expression of the above equation, below we calculate the action of the Hamiltonian upon the vacuum state of the cavity field modes,
\begin{eqnarray}
\hat{H}_{\text{eff}}^{(2)}|\emptyset \rangle_{c} &=&\frac{1}{i\hbar}\sum_{r,s,k,n=1}^{\infty}\sum_{\nu,\mu =1,2}C_{r,s}^{(\nu)}C_{k,n}^{(\mu)}\hat{b}_{\nu}^{\dagger}\hat{b}_{\mu}
\hat{a}_{s0}\hat{a}_{r0}\hat{a}_{k0}^{\dagger}\hat{a}_{n0}^{\dagger}\frac{1}{-i(\omega_{M}-\omega_{n0}-\omega_{k0})}\delta_{s+r,n+k}|\emptyset\rangle_{c}\nonumber\\
&&+\frac{1}{i\hbar}\sum_{r,s,k,n=1}^{\infty}\sum_{\nu,\mu=1,2}C_{r,s}^{(\nu)}C_{k,n}^{(\mu)}\hat{b}_{\nu}
\hat{b}_{\mu}^{\dagger}\hat{a}_{s0}\hat{a}_{r0}\hat{a}_{k0}^{\dagger}\hat{a}_{n0}^{\dagger}\frac{1}{i(\omega_{M}+\omega_{n0}+\omega_{k0})}\delta_{s+r,n+k}|\emptyset\rangle_{c}.
\end{eqnarray}
Namely, only the two terms marked by the red fonts in Eq.~(\ref{HNeectiveHamil2}) can be kept, other terms will disappear. Acting of the cavity field operators on the vacuum state, we obtain
\begin{equation}
\hat{H}_{\text{eff}}^{(2)}|\emptyset\rangle_{c}=I_{c}|\emptyset\rangle_{c}\otimes\hat{H}^{\text{inter}},\label{HeffHinterdef}
\end{equation}
where $I_{c}$ is the identity operator in the Hilbert space of the cavity modes, and
\begin{eqnarray}
\hat{H}^{\text{inter}}&=&(\lambda_{1,1}+\eta_{1,1})\hat{b}_{1}^{\dagger}\hat{b}_{1}+(\lambda_{2,2}+\eta_{2,2})\hat{b}
_{2}^{\dagger}\hat{b}_{2}+(\lambda_{1,2}+\eta_{1,2})(\hat{b}_{2}\hat{b}_{1}^{\dagger}+\hat{b}_{1}\hat{b}_{2}^{\dagger})+\eta_{1,1}+\eta_{2,2}\nonumber\\
&\equiv &\hat{H}_{\text{eff}}+\eta_{1,1}+\eta_{2,2},
\end{eqnarray}
with
\begin{equation}
\hat{H}_{\text{eff}}=(\lambda_{1,1}+\eta_{1,1})\hat{b}_{1}^{\dagger}\hat{b}_{1}+(\lambda_{2,2}+\eta_{2,2})\hat{b}_{2}^{\dagger}
\hat{b}_{2}+(\lambda_{1,2}+\eta_{1,2})(\hat{b}_{2}\hat{b}_{1}^{\dagger}+\hat{b}_{1}\hat{b}_{2}^{\dagger}).\label{Hmdef}
\end{equation}
The parameters in Eq.~(\ref{Hmdef}) are defined by
\begin{subequations}
\label{lambdaetajjjjprime}
\begin{align}
\lambda_{j,j}=&\frac{2}{\hbar}\sum_{k,n=1}^{\infty}C_{k,n}^{(j)}C_{k,n}^{(j)}\frac{1}{(\omega_{M}-\omega_{n0}-\omega_{k0})},\hspace{0.5 cm}j=1,2,\\
\eta_{j,j}=&\frac{2}{-\hbar }\sum_{k,n=1}^{\infty }C_{k,n}^{(j)}C_{k,n}^{(j)}\frac{1}{(\omega_{M}+\omega_{n0}+\omega_{k0})},\hspace{0.5 cm}j=1,2,\\
\lambda_{1,2}=&\frac{2}{\hbar}\sum_{k,n=1}^{\infty }C_{k,n}^{(2)}C_{k,n}^{(1)}\frac{1}{(\omega_{M}-\omega_{n0}-\omega_{k0})}=\lambda_{2,1},\\
\eta_{1,2}=&\frac{2}{-\hbar}\sum_{k,n=1}^{\infty}C_{k,n}^{(2)}C_{k,n}^{(1)}\frac{1}{(\omega_{M}+\omega_{n0}+\omega_{k0})}=\eta_{2,1}.
\end{align}
\end{subequations}
In terms of Eq.~(\ref{HeffHinterdef}), we can obtain
\begin{equation}
\hat{H}_{\text{eff}}^{(2)}|\emptyset\rangle_{c}\vert
j\rangle_{b_{1}}\vert k\rangle_{b_{2}}=|\emptyset\rangle_{c}\hat{H}^{\text{inter}}\vert j\rangle_{b_{1}}
\vert k\rangle_{b_{2}}.
\end{equation}
Further, we have
\begin{eqnarray}
[\hat{H}_{\text{eff}}^{(2)}]^{2}|\emptyset\rangle_{c}\vert j\rangle_{b_{1}}\vert k\rangle_{b_{2}}
&=&\hat{H}_{\text{eff}}^{(2)}\hat{H}_{\text{eff}}^{(2)}|\emptyset\rangle_{c}\vert j\rangle_{b_{1}}
\vert k\rangle_{b_{2}}=\hat{H}_{\text{eff}}^{(2)}|\emptyset\rangle_{c}\hat{H}_{m}^{\text{inter}}\vert j\rangle_{b_{1}}\vert k\rangle_{b_{2}}\nonumber\\
&=&|\emptyset\rangle_{c}(\hat{H}^{\text{inter}})^{2}\vert j\rangle_{b_{1}}\vert k\rangle_{b_{2}},
\end{eqnarray}
and
\begin{equation}
[\hat{H}_{\text{eff}}^{(2)}]^{n}|\emptyset\rangle_{c}\vert j\rangle_{b_{1}}\vert k\rangle_{b_{2}}
=|\emptyset\rangle_{c}(\hat{H}^{\text{inter}})^{n}\vert j\rangle_{b_{1}}\vert k\rangle_{b_{2}}.
\end{equation}
Then, Eq.~(\ref{UITonbasis}) can be further expressed as
\begin{eqnarray}
\hat{U}_{I}(t)\vert j \rangle_{b_{1}}\vert
k\rangle_{b_{2}}|\emptyset\rangle_{c}&=&e^{\frac{-it}{\hbar}\hat{H}_{\text{eff}}^{(2)}}\vert j\rangle
_{b_{1}}\vert k\rangle_{b_{2}}|\emptyset\rangle_{c}\nonumber\\
&=&|\emptyset\rangle_{c}e^{\frac{t}{i\hbar}\hat{H}^{\text{inter}}}\vert j\rangle_{b_{1}}\vert k\rangle_{b_{2}}.
\end{eqnarray}
Therefore, we have
\begin{eqnarray}
\rho_{sys}^{(I)}(t) &=&\hat{U}_{I}(t)\rho(0)\otimes|\emptyset\rangle_{c}\,_{c}\langle\emptyset|\hat{U}_{I}^{\dagger}(t)  \nonumber\\
&=&\sum_{j,k,n,m=0}^{\infty}\rho_{jknm}|\emptyset\rangle_{c}e^{-i\frac{t}{\hbar}\hat{H}^{\text{inter}}}|j\rangle_{b_{1}}|k\rangle
_{b_{2}}\,_{b_{1}}\langle n|\,_{b_{2}}\langle m|\,_{c}\langle\emptyset|e^{i\frac{t}{\hbar}\hat{H}^{\text{inter}}}  \nonumber\\
&=&e^{-i\frac{t}{\hbar}\hat{H}^{\text{inter}}}\rho(0)e^{i\frac{t}{\hbar}\hat{H}^{\text{inter}}}\otimes|\emptyset\rangle_{c}\,_{c}\langle
\emptyset|  \nonumber\\
&=&e^{-i\frac{t}{\hbar}\hat{H}_{\text{eff}}}\rho(0)e^{i\frac{t}{\hbar}\hat{H}_{\text{eff}}}\otimes|\emptyset\rangle_{c}\,_{c}\langle\emptyset|.
\end{eqnarray}
Based on Eq.~(\ref{lambdaetajjjjprime}), the Hamiltonian (\ref{Hmdef}) can be expressed as
\begin{equation}
\label{effHamwi}
\hat{H}_{\text{eff}}=\hbar\Delta_{b,1}\hat{b}_{1}^{\dagger}\hat{b}_{1}+\hbar\Delta_{b,2}\hat{b}_{2}^{\dagger}\hat{b}_{2}+\hbar \xi(\hat{b}_{2}\hat{b}_{1}^{\dagger}+\hat{b}_{1}\hat{b}_{2}^{\dagger}),
\end{equation}
where we introduce
\begin{subequations}
\label{Deltab1b2lambdadelta}
\begin{align}
\Delta_{b,1}\equiv& \frac{(\lambda_{1,1}+\eta_{1,1})}{\hbar}=
\frac{\hbar}{2}\frac{1}{l_{0}^{2}m_{1}}\sum_{k,n=1}^{\infty}\frac{\omega_{n0}\omega_{k0}}{\omega_{M}}\frac{\omega_{n0}+\omega_{k0}}{\omega_{M}^{2}-(\omega_{n0}+\omega_{k0})^{2}},\\
\Delta_{b,2}\equiv &\frac{\lambda_{2,2}+\eta_{2,2}}{\hbar}=\frac{\hbar}{2}\frac{1}{l_{0}^{2}m_{2}}\sum_{k,n=1}^{\infty}\frac{\omega_{n0}\omega_{k0}
}{\omega_{M}}\frac{\omega_{n0}+\omega_{k0}}{\omega_{M}^{2}-(\omega_{n0}+\omega_{k0})^{2}},\\
\xi\equiv &\frac{\lambda_{1,2}+\eta_{1,2}}{\hbar}=-\frac{\hbar}{2}\frac{1}{l_{0}^{2}\sqrt{m_{1}m_{2}}}\sum_{k,n=1}^{\infty}(-1)
^{n+k}\frac{\omega_{n0}\omega_{k0}}{\omega_{M}}\frac{\omega_{n0}+\omega_{k0}}{\omega_{M}^{2}-(\omega_{n0}+\omega_{k0})^{2}}.
\end{align}
\end{subequations}
The effective Hamiltonian $\hat{H}_{\text{eff}}$ governs the evolution of the two moving end mirrors. In Eq.~(\ref{effHamwi}),
$\Delta_{b,n}$ denotes the frequency shift of the $n$th ($n=1,2$) moving end mirror, and $\xi$ is the effective coupling strength between the two mirrors. Using the zero-point-fluctuation amplitude $x_{n,\text{zpf}}=\sqrt{\frac{\hbar}{2m_{n}\omega_{M}}}$ of the $n$th ($n=1,2$) moving end mirror, Eq.~(\ref{Deltab1b2lambdadelta}) can be written as
\begin{subequations}
\label{frshifandcoup1}
\begin{align}
\Delta_{b,n}& =\frac{x_{n,\text{zpf}}^{2}}{l_{0}^{2}}\sum_{k,j=1}^{\infty}%
\frac{\omega_{j0}\omega_{k0}(\omega_{j0}+\omega_{k0})}{\omega_{M}^{2}-(\omega_{j0}+\omega_{k0})^{2}}, \hspace{1 cm} n=1,2, \\
\xi & =-\frac{x_{1,\text{zpf}}x_{2,\text{zpf}}}{l_{0}^{2}}\sum_{k,j=1}^{\infty}\frac{(-1)^{j+k}\omega_{j0}\omega_{k0}(\omega
_{j0}+\omega_{k0})}{\omega_{M}^{2}-(\omega_{j0}+\omega_{k0})^{2}}.
\end{align}
\end{subequations}
Under the condition $\omega_{M}\ll\omega_{k0}, \omega_{j0}$, we can neglect the term $\omega_{M}^{2}$ in the denominator. In terms of the cavity frequency $\omega_{k0}=k\pi c/l_{0}$, Eq.~(\ref{frshifandcoup1}) can be reduced to
\begin{subequations}
\begin{align}
\Delta_{b,n}&\simeq -\pi c\frac{x_{n,\text{zpf}}^{2}}{l_{0}^{3}}\sum_{k,j=1}^{\infty}\frac{jk}{j+k}, \hspace{1 cm} n=1,2, \\
\xi&\simeq\pi c\frac{x_{1,\text{zpf}}x_{2,\text{zpf}}}{l_{0}^{3}}\sum_{k,j=1}^{\infty}(-1)^{j+k}\frac{jk}{j+k}. \label{couplsva}
\end{align}
\end{subequations}
Obviously, both the frequency shift and the coupling strength do not converge. Therefore, a truncation frequency needs to be introduced in the numerical calculations. The dependence of both the frequency shift and the coupling strength on the system parameters will be analyzed in Sec.~\ref{atep}.

\section{VI. The mode temperatures of the two moving end mirrors}  \label{ttem}

To include the dissipation of the system, we assume that the two moving mirrors are connected to two separating heat baths. Then the evolution of the two moving end mirrors is governed by the quantum master equation
\begin{equation}
\frac{\partial\hat{\rho}}{\partial t}=-\frac{i}{\hbar}[\hat{H}_{\text{eff}},\hat{\rho}]-\sum_{n=1,2}[\gamma_{n}(\bar{n}_{\text{th},n}+1)\mathcal{%
\hat{D}}(\hat{b}_{n})\hat{\rho}+\gamma_{n}\bar{n}_{\text{th},n}\mathcal{\hat{D}}(\hat{b}_{n}^{\dagger})\hat{\rho}], \label{pmasterequation}
\end{equation}
where $\hat{H}_{\text{eff}}$ is the effective Hamiltonian given in Eq.~(\ref{effHamwi}), $\mathcal{\hat{D}}(\hat{o})\hat{\rho}=(\hat{o}^{\dagger}\hat{o}\hat{\rho}+\hat{\rho}\hat{o}^{\dagger}\hat{o}) /2-\hat{o}\hat{\rho}\hat{o}^{\dagger}$ is the Lindbland superoperators, $\gamma_{n}$ is the decay rate of the $n$th ($n=1,2$) moving end mirror, and $\bar{n}_{\text{th},n}=1/(e^{\hbar\omega_{M}/k_{B}T_{n}}-1)$ is the mean thermal phonon occupation corresponding to the environmental temperature $T_{n}$. According to Eq.~(\ref{pmasterequation}), the equations of motion for these density matrix elements are given by
\begin{eqnarray}
\frac{\partial\rho_{jknm}}{\partial t}&=&\left\{-i[\Delta_{b,1}(j-n)+\Delta
_{b,2}(k-m)]-\frac{\gamma_{1}}{2}[2\bar{n}_{\text{th},1}(j+n+1)+j+n]-\frac{\gamma_{2}}{2}[2\bar{n}_{\text{th},2}(k+m+1)+k+m]\right\}\rho_{jknm} \nonumber \\
&&-i\xi \sqrt{j(k+1)}\rho_{(j-1)(k+1)nm} -i\xi\sqrt{(j+1)k}\rho_{(j+1)(k-1)nm} \nonumber \\
&&+i\xi\sqrt{(n+1)m}\rho_{jk(n+1)(m-1)}+i\xi\sqrt{n(m+1)}\rho_{jk(n-1)(m+1)}\nonumber \\
&&+\gamma_{1}\bar{n}_{\text{th},1}\sqrt{jn}\rho_{(j-1)k(n-1)m}+\gamma_{2}\bar{n}_{\text{th},2}\sqrt{km}\rho_{j(k-1)n(m-1)} \nonumber \\
&&+\gamma_{1}(\bar{n}_{\text{th},1}+1)\sqrt{(j+1)(n+1)}\rho_{(j+1)k(n+1)m}+\gamma_{2}(\bar{n}_{\text{th},2}+1)\sqrt{(k+1)(m+1)}\rho_{j(k+1)n(m+1)},  \label{meeq}
\end{eqnarray}
where $\rho_{jknm}=\,_{b_{1}}\langle j|\,_{b_{2}}\langle k|\rho|n\rangle_{b_{1}}|m\rangle_{b_{2}}$ is the density matrix element in the Fock state representation. By numerically solving the equations of motion~(\ref{meeq}) under the initial condition, the transient state density matrix $\rho(t)$ of the two moving mirrors can be obtained. In our following discussions, we focus on the steady state of the system.

Below, we will confirm that the steady states of the two moving mirrors are the thermal states. Based on the quantum master equation, we can obtain the steady state density matrix $\rho_{ss}$ of the system. By performing the partial trace, we obtain the reduced density matrices of the each moving mirror as
\begin{subequations}
\label{reddenm}
\begin{align}
\rho_{b_{1}}&=\text{Tr}_{b_{2}}(\rho_{ss})=\sum_{k=0}^{n_{b_{2}}-1}\,_{b_{2}}\langle k|\rho_{ss}|k\rangle_{b_{2}},     \\
\rho_{b_{2}}&=\text{Tr}_{b_{1}}(\rho_{ss})=\sum_{k=0}^{n_{b_{1}}-1}\,_{b_{1}}\langle k|\rho_{ss}|k\rangle_{b_{1}},
\end{align}
\end{subequations}
where $n_{b_{1}}$ and $n_{b_{2}}$ represent the truncation dimensions of the Hilbert space for the right and left moving mirrors, respectively. These two reduced density matrices are diagonal matrix in the Fock state representation, and the phonon-number distributions of the two moving mirrors are given by
\begin{subequations}
\label{prodis}
\begin{align}
p_{n_{1}}&=\,_{b_{1}}\langle n|\,\rho_{b_{1}}|n\rangle_{b_{1}}=\sum_{k=0}^{n_{b_{2}}-1}\rho^{(ss)}_{nknk}, \\
p_{n_{2}}&=\,_{b_{2}}\langle n|\,\rho_{b_{2}}|n\rangle_{b_{2}}=\sum_{k=0}^{n_{b_{1}}-1}\rho^{(ss)}_{knkn}.
\end{align}
\end{subequations}
\begin{figure}[tbp]
\center
\includegraphics[bb=0 0 755 550, width=0.6 \textwidth]{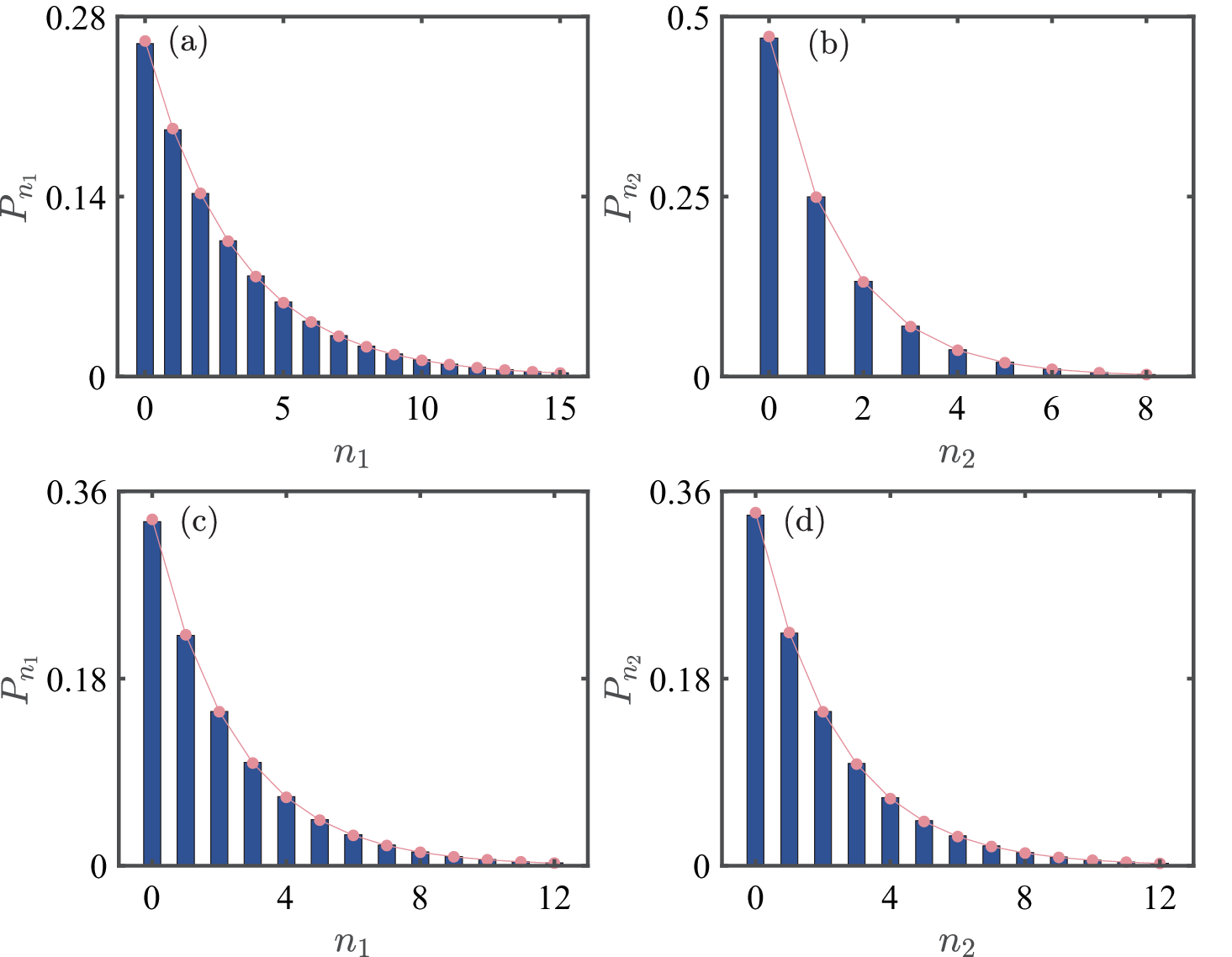}
\caption{(Color online) The phonon number distributions (bars) of the two moving mirrors in the (a, b) weak-coupling ($\xi/\omega_{M}=0.2$) and (c,d) strong-coupling ($\xi/\omega_{M}=3$) regimes. The exact phonon number distributions $\langle n_{1}\rangle^{n}/(\langle n_{1}\rangle+1)^{n+1}$ (the jointed dots) under the average phonon numbers of the two mirrors are presented for comparison. Other parameters used are $\bar{n}_{\text{th},1}=3$ and $\bar{n}_{\text{th},2}=1$.}\label{Sigmasss}
\end{figure}
By solving the quantum master equation, the phonon number distributions of the two moving mirrors can be obtained.
In Fig.~\ref{Sigmasss}, we show the phonon number distributions of the two moving mirrors in the steady states when the coupling strength between the two mirrors works in both the weak- ($\xi/\omega_{M}=0.2$) and strong-coupling ($\xi/\omega_{M}=3$) regimes. We also present the exact phonon number distributions $\langle n_{1}\rangle^{n}/(\langle n_{1}\rangle+1)^{n+1}$ for the thermal state for comparison. Here we can see that the numerical results are consistent with the exact results in both the weak- and strong-coupling regimes. In addition, we see that the phonon number distributions exhibit clear feature for the thermal states. In the strong-coupling regime, the final average phonon numbers in the two moving mirrors are approximately identical. This is because the strong coupling will enhance the heat transfer between the two mirrors, and then the steady-state temperatures of the two moving mirrors will be the same.

To further describe the excitation transfer between the two moving mirrors, we also investigate the second-order moments of the two-moving-mirror system. Based on Eq.~(\ref{pmasterequation}), the equations of motion for the second-order moments can be obtained as
\begin{subequations}
\label{avergequ1}
    \begin{align}
\frac{d}{dt}\langle \hat{b}_{1}^{\dagger}\hat{b}_{1}\rangle &=-i\xi\langle \hat{b}_{1}^{\dagger}\hat{b}_{2}\rangle +i\xi
\langle \hat{b}_{1}\hat{b}_{2}^{\dagger}\rangle -\gamma_{1}\langle \hat{b}_{1}^{\dagger}\hat{b}_{1}\rangle +\gamma_{1}\bar{n}_{\text{th},1},   \\
\frac{d}{dt}\langle \hat{b}_{1}^{\dagger}\hat{b}_{2}\rangle &=\left[i(\Delta_{b,1}-\Delta_{b,2}) -\left(\frac{\gamma _{1}}{2}+\frac{%
\gamma_{2}}{2}\right)\right] \langle\hat{b}_{1}^{\dagger}\hat{b}_{2}\rangle -i\xi\langle\hat{b}_{1}^{\dagger}\hat{b}%
_{1}\rangle +i\xi \langle \hat{b}_{2}^{\dagger}\hat{b}_{2}\rangle,       \\
\frac{d}{dt}\langle \hat{b}_{1}\hat{b}_{2}^{\dagger}\rangle &=\left[-i(\Delta_{b,1}-\Delta_{b,2})-\frac{\gamma_{1}}{2}-\frac{%
\gamma_{2}}{2}\right] \langle\hat{b}_{1}\hat{b}_{2}^{\dagger}\rangle +i\xi \langle\hat{b}_{1}^{\dagger}\hat{b}%
_{1}\rangle -i\xi \langle\hat{b}_{2}^{\dagger}\hat{b}_{2}\rangle,   \\
\frac{d}{dt}\langle\hat{b}_{2}^{\dagger}\hat{b}_{2}\rangle &=i\xi\langle \hat{b}_{1}^{\dagger}\hat{b}_{2}\rangle -i\xi
\langle\hat{b}_{1}\hat{b}_{2}^{\dagger}\rangle -\gamma_{2}\langle\hat{b}_{2}^{\dagger}\hat{b}_{2}\rangle +\gamma_{2}\bar{n}_{\text{th},2}.
\end{align}
\end{subequations}
In the long-time limit, we can obtain the steady-state average phonon numbers as
\begin{subequations}
\label{phononavergequ}
\begin{align}
\langle\hat{b}_{1}^{\dagger}\hat{b}_{1}\rangle_{ss}& =\frac{4\xi^{2}(\gamma_{1}+\gamma_{2})(\gamma_{1}\bar{n}_{\text{th},1}+\gamma_{2}\bar{n}
_{\text{th},2})+\gamma_{1}\gamma_{2}\bar{n}_{\text{th},1}[4(\Delta_{b,1}-\Delta_{b,2})^{2}+(\gamma_{1}+\gamma_{2})^{2}]}{4\xi^{2}(\gamma_{1}+\gamma
_{2})^{2}+\gamma_{1}\gamma_{2}[4(\Delta_{b,1}-\Delta_{b,2})^{2}+(\gamma_{1}+\gamma_{2})^{2}]},\\
\langle\hat{b}_{2}^{\dagger}\hat{b}_{2}\rangle_{ss}& =\frac{4\xi^{2}(\gamma_{1}+\gamma_{2})(\gamma_{1}\bar{n}_{\text{th},1}+\gamma_{2}\bar{n}
_{\text{th},2})+\gamma_{1}\gamma_{2}\bar{n}_{\text{th},2}[4(\Delta_{b,1}-\Delta_{b,2})^{2}+(\gamma_{1}+\gamma_{2})^{2}]}{4\xi^{2}(\gamma_{1}+\gamma
_{2})^{2}+\gamma_{1}\gamma_{2}[4(\Delta_{b,1}-\Delta_{b,2})^{2}+(\gamma_{1}+\gamma_{2})^{2}]}.
\end{align}
\end{subequations}
In particular, to define the mode temperature of the two mechanical modes, we need to show that the two mirrors will be in the thermal state in the steady state. We have numerically checked the thermal state of system in the steady state. Below, we will analytically show that the steady states of the two moving mirrors are thermal states. To this end, we derive the equations of motion for the four-order moments as follows,
\begin{subequations}
\label{avergequ}
\begin{align}
\frac{d}{dt}\langle\hat{b}_{1}^{\dagger}\hat{b}_{1}^{\dagger}\hat{b}_{1}\hat{b}_{1}\rangle =&-2i\xi\langle\hat{b}_{1}^{\dagger}\hat{b}%
_{1}^{\dagger}\hat{b}_{1}\hat{b}_{2}\rangle+2i\xi\langle\hat{b}_{1}^{\dagger}\hat{b}_{1}\hat{b}_{1}\hat{b}_{2}^{\dagger
}\rangle-2\gamma_{1}\langle\hat{b}_{1}^{\dagger}\hat{b}_{1}^{\dagger}\hat{b}_{1}\hat{b}_{1}\rangle+4\gamma_{1}\bar{n}%
_{\text{th},1}\langle\hat{b}_{1}^{\dagger}\hat{b}_{1}\rangle,                                                                              \\
\frac{d}{dt}\langle\hat{b}_{1}^{\dagger}\hat{b}_{1}^{\dagger}\hat{b}_{1}\hat{b}_{2}\rangle =&\left[i(\Delta_{b,1}-\Delta_{b,2})-%
\frac{3\gamma_{1}}{2}-\frac{\gamma_{2}}{2}\right]\langle\hat{b}_{1}^{\dagger}\hat{b}_{1}^{\dagger}\hat{b}_{1}\hat{b}_{2}\rangle
-i\xi\langle\hat{b}_{1}^{\dagger}\hat{b}_{1}^{\dagger}\hat{b}_{1}\hat{b}_{1}\rangle      \nonumber \\
&-i\xi\langle\hat{b}_{1}^{\dagger}\hat{b}_{1}^{\dagger}\hat{b}_{2}\hat{b}_{2}\rangle+2i\xi\langle\hat{b}_{1}^{\dagger}%
\hat{b}_{1}\hat{b}_{2}^{\dagger}\hat{b}_{2}\rangle+2\gamma_{1}\bar{n}_{\text{th},1}\langle\hat{b}_{1}^{\dagger}\hat{b}_{2}\rangle,    \\
\frac{d}{dt}\langle\hat{b}_{1}^{\dagger}\hat{b}_{1}\hat{b}_{1}\hat{b}_{2}^{\dagger}\rangle =&\left[-i(\Delta_{b,1}-\Delta_{b,2}) -\frac{3\gamma _{1}}{2}-\frac{\gamma_{2}}{2}\right]\langle\hat{b}_{1}^{\dagger}\hat{b}_{1}\hat{b}_{1}\hat{b}%
_{2}^{\dagger}\rangle+i\xi\langle\hat{b}_{1}^{\dagger}\hat{b}_{1}^{\dagger}\hat{b}_{1}\hat{b}_{1}\rangle          \nonumber \\
&+i\xi\langle\hat{b}_{1}\hat{b}_{1}\hat{b}_{2}^{\dagger}\hat{b}_{2}^{\dagger}\rangle-2i\xi\langle\hat{b}_{1}^{\dagger}%
\hat{b}_{1}\hat{b}_{2}^{\dagger}\hat{b}_{2}\rangle+2\gamma_{1}\bar{n}_{\text{th},1}\langle\hat{b}_{1}\hat{b}_{2}^{\dagger}\rangle,     \\
\frac{d}{dt}\langle\hat{b}_{1}^{\dagger}\hat{b}_{1}^{\dagger}\hat{b}_{2}\hat{b}_{2}\rangle &=\left[2i(\Delta_{b,1}-\Delta_{b,2})
-\gamma_{1}-\gamma_{2}\right]\langle\hat{b}_{1}^{\dagger}\hat{b}_{1}^{\dagger}\hat{b}_{2}\hat{b}_{2}\rangle-2i\xi\langle
\hat{b}_{1}^{\dagger}\hat{b}_{1}^{\dagger}\hat{b}_{1}\hat{b}_{2}\rangle+2i\xi\langle\hat{b}_{1}^{\dagger}\hat{b}%
_{2}^{\dagger}\hat{b}_{2}\hat{b}_{2}\rangle,    \\
\frac{d}{dt}\langle\hat{b}_{1}^{\dagger}\hat{b}_{1}\hat{b}_{2}^{\dagger}\hat{b}_{2}\rangle =&i\xi\langle\hat{b}_{1}^{\dagger}\hat{b}%
_{1}^{\dagger}\hat{b}_{1}\hat{b}_{2}\rangle-i\xi\langle\hat{b}_{1}^{\dagger}\hat{b}_{1}\hat{b}_{1}\hat{b}_{2}^{\dagger
}\rangle-(\gamma_{1}+\gamma_{2})\langle\hat{b}_{1}^{\dagger}\hat{b}_{1}\hat{b}_{2}^{\dagger}\hat{b}_{2}\rangle
-i\xi\langle\hat{b}_{1}^{\dagger}\hat{b}_{2}^{\dagger}\hat{b}_{2}\hat{b}_{2}\rangle       \nonumber \\
&+i\xi\langle\hat{b}_{1}\hat{b}_{2}^{\dagger}\hat{b}_{2}^{\dagger}\hat{b}_{2}\rangle+\gamma_{2}\bar{n}_{\text{th},2}\langle
\hat{b}_{1}^{\dagger}\hat{b}_{1}\rangle+\gamma_{1}\bar{n}_{\text{th},1}\langle\hat{b}_{2}^{\dagger}\hat{b}_{2}\rangle,  \\
\frac{d}{dt}\langle\hat{b}_{1}\hat{b}_{1}\hat{b}_{2}^{\dagger}\hat{b}_{2}^{\dagger}\rangle =&\left[-2i(\Delta_{b,1}-\Delta_{b,2})
-(\gamma_{1}+\gamma_{2})\right]\langle\hat{b}_{1}\hat{b}_{1}\hat{b}_{2}^{\dagger}\hat{b}_{2}^{\dagger}\rangle-2i\xi
\langle\hat{b}_{1}\hat{b}_{2}^{\dagger}\hat{b}_{2}^{\dagger}\hat{b}_{2}\rangle+2i\xi\langle\hat{b}_{1}^{\dagger}\hat{b}_{1}%
\hat{b}_{1}\hat{b}_{2}^{\dagger}\rangle,      \\
\frac{d}{dt}\langle\hat{b}_{1}^{\dagger}\hat{b}_{2}^{\dagger}\hat{b}_{2}\hat{b}_{2}\rangle  =&\left[i(\Delta_{b,1}-\Delta_{b,2}) -\frac{\gamma_{1}}{2}-\frac{3\gamma_{2}}{2}\right]\langle\hat{b}%
_{1}^{\dagger}\hat{b}_{2}^{\dagger}\hat{b}_{2}\hat{b}_{2}\rangle+i\xi\langle\hat{b}_{2}^{\dagger}\hat{b}_{2}^{\dagger}\hat{b}%
_{2}\hat{b}_{2}\rangle  \nonumber \\
&+i\xi\langle\hat{b}_{1}^{\dagger}\hat{b}_{1}^{\dagger}\hat{b}_{2}\hat{b}_{2}\rangle-2i\xi\langle\hat{b}_{1}^{\dagger}%
\hat{b}_{1}\hat{b}_{2}^{\dagger}\hat{b}_{2}\rangle+2\gamma_{2}\bar{n}_{\text{th},2}\langle\hat{b}_{1}^{\dagger}\hat{b}_{2}\rangle,    \\
\frac{d}{dt}\langle\hat{b}_{1}\hat{b}_{2}^{\dagger}\hat{b}_{2}^{\dagger}\hat{b}_{2}\rangle  =&\left[-i(\Delta_{b,1}-\Delta_{b,2}) -%
\frac{\gamma_{1}}{2}-\frac{3\gamma_{2}}{2}\right]\langle\hat{b}_{1}\hat{b}_{2}^{\dagger}\hat{b}_{2}^{\dagger}\hat{b}_{2}\rangle
+2i\xi\langle\hat{b}_{1}^{\dagger}\hat{b}_{1}\hat{b}_{2}^{\dagger}\hat{b}_{2}\rangle      \nonumber \\
&-i\xi\langle\hat{b}_{2}^{\dagger}\hat{b}_{2}^{\dagger}\hat{b}_{2}\hat{b}_{2}\rangle-i\xi\langle\hat{b}_{1}\hat{b}_{1}%
\hat{b}_{2}^{\dagger}\hat{b}_{2}^{\dagger}\rangle+2\gamma_{2}\bar{n}_{\text{th},2}\langle\hat{b}_{1}\hat{b}_{2}^{\dagger}\rangle,     \\
\frac{d}{dt}\langle\hat{b}_{2}^{\dagger}\hat{b}_{2}^{\dagger}\hat{b}_{2}\hat{b}_{2}\rangle =&2i\xi\langle\hat{b}_{1}^{\dagger}\hat{b}%
_{2}^{\dagger}\hat{b}_{2}\hat{b}_{2}\rangle-2i\xi\langle\hat{b}_{1}\hat{b}_{2}^{\dagger}\hat{b}_{2}^{\dagger}\hat{b}%
_{2}\rangle-2\gamma_{2}\langle\hat{b}_{2}^{\dagger}\hat{b}_{2}^{\dagger}\hat{b}_{2}\hat{b}_{2}\rangle+4\gamma_{2}\bar{n}%
_{\text{th},2}\langle\hat{b}_{2}^{\dagger}\hat{b}_{2}\rangle.
\end{align}
\end{subequations}
By setting the derivatives at the left-hand side of the above equations as zero, we can further obtain the steady-state solution of these equations. Here we only show the results of $\langle b_{1}^{\dag}b_{1}^{\dag}b_{1}b_{1}\rangle_{ss}$ and $\langle b_{2}^{\dag}b_{2}^{\dag}b_{2}b_{2}\rangle_{ss}$, which will be used in the calculation of the second-order correlation function,
\begin{subequations}
\label{solutionss}
\begin{align}
\langle b_{1}^{\dag}b_{1}^{\dag}b_{1}b_{1}\rangle_{ss}=&\frac{2\{\bar{n}_{\text{th,1}}\gamma_{1}\gamma_{2}[(\gamma_{1}+\gamma_{2})^{2}+4(\Delta_{b,1}-\Delta_{b,2})^{2}]
+4(\gamma_{1}+\gamma_{2})(\bar{n}_{\text{th,1}}\gamma_{1}+\bar{n}_{\text{th,2}}\gamma_{2})\xi^{2}\}^{2}}{\{\gamma_{1}\gamma_{2}[(\gamma_{1}+\gamma_{2})^{2}+4(\Delta_{b,1}-\Delta_{b,2})^{2}]+4(\gamma_{1}+\gamma_{2})^{2}\xi^{2}\}^{2}},\\
\langle b_{2}^{\dag}b_{2}^{\dag}b_{2}b_{2}\rangle_{ss}=&\frac{2\{\bar{n}_{\text{th,2}}\gamma_{1}\gamma_{2}[(\gamma_{1}+\gamma_{2})^{2}+4(\Delta_{b,1}-\Delta_{b,2})^{2}]
+4(\gamma_{1}+\gamma_{2})(\bar{n}_{\text{th,1}}\gamma_{1}+\bar{n}_{\text{th,2}}\gamma_{2})\xi^{2}\}^{2}}{\{\gamma_{1}\gamma_{2}[(\gamma_{1}+\gamma_{2})
^{2}+4(\Delta_{b,1}-\Delta_{b,2})^{2}]+4(\gamma_{1}+\gamma_{2})^{2}\xi^{2}\}^{2}}.
\end{align}
\end{subequations}
Based on Eqs.~(\ref{solutionss}), we can prove that the second-order correlation functions in the steady state are
\begin{subequations}
\label{socorr}
\begin{align}
g^{(2)}(0)& =\frac{\langle\hat{b}_{1}^{\dagger}\hat{b}_{1}^{\dagger}\hat{b
}_{1}\hat{b}_{1}\rangle_{ss}}{\langle\hat{b}_{1}^{\dagger}\hat{b}_{1}\rangle_{ss}^{2}}=2, \\
g^{(2)}(0)& =\frac{\langle\hat{b}_{2}^{\dagger}\hat{b}_{2}^{\dagger}\hat{b
}_{2}\hat{b}_{2}\rangle_{ss}}{\langle\hat{b}_{2}^{\dagger}\hat{b}_{2}\rangle_{ss}^{2}}=2.
\end{align}
\end{subequations}

When the two moving end mirrors are in thermal states, we can introduce the mode temperatures $T_{m1}$ and $T_{m2}$ of the two moving mirrors by
\begin{subequations}
\label{thphonaverq1}
    \begin{align}
\langle\hat{b}_{1}^{\dagger}\hat{b}_{1}\rangle_{ss} &=\frac{1}{e^{\hbar\omega_{M}/k_{B}T_{m1}}-1},   \\
\langle \hat{b}_{2}^{\dagger}\hat{b}_{2}\rangle_{ss} &=\frac{1}{e^{\hbar\omega_{M}/k_{B}T_{m2}}-1}.
\end{align}
\end{subequations}
According to Eq.~(\ref{thphonaverq1}), the two mode temperatures can be written as
\begin{subequations}
    \begin{align}
T_{m1}&=\frac{\hbar\omega_{M}}{k_{B}\ln \left(\frac{1}{\langle\hat{b}_{1}^{\dagger}\hat{b}_{1}\rangle_{ss}}+1\right)},     \\
T_{m2}&=\frac{\hbar\omega_{M}}{k_{B}\ln \left(\frac{1}{\langle\hat{b}_{2}^{\dagger}\hat{b}_{2}\rangle_{ss}}+1\right)}.
\end{align}
\end{subequations}
In the high-temperature limit ($\hbar\omega_{M}\ll k_{B}T_{1}$ and $\hbar\omega_{M}\ll k_{B}T_{2}$), we have the relations
\begin{subequations}
\label{avergequhighlimit}
    \begin{align}
\langle\hat{b}_{1}^{\dagger}\hat{b}_{1}\rangle_{ss} &\simeq  \frac{k_{B}T_{m1}}{\hbar\omega_{M}},   \nonumber \\
\langle\hat{b}_{2}^{\dagger}\hat{b}_{2}\rangle_{ss} &\simeq  \frac{k_{B}T_{m2}}{\hbar\omega_{M}}.
\end{align}
\end{subequations}
According to Eqs.~(\ref{phononavergequ}) and (\ref{avergequhighlimit}), we can obtain the expression of the mode temperatures as
\begin{subequations}
    \begin{align}
T_{m1}&=\frac{\hbar\omega_{M}}{k_{B}}\frac{\xi^{2}(\gamma_{1}+\gamma_{2})(\gamma_{1}\bar{n}_{\text{th},1}+\gamma_{2}\bar{n}_{\text{th},2})+\gamma_{1}\gamma _{2}\bar{n}_{\text{th},1}\left[(\Delta_{b,1}-\Delta_{b,2})^{2}+(\gamma_{1}+\gamma_{2})^{2}/4\right]}{\xi^{2}(\gamma_{1}+\gamma_{2})^{2}+\gamma_{1}\gamma
_{2}\left[(\Delta_{b,1}-\Delta_{b,2})^{2}+(\gamma_{1}+\gamma_{2})^{2}/4\right]},                  \\
T_{m2}&=\frac{\hbar\omega_{M}}{k_{B}}\frac{\xi^{2}(\gamma_{1}+\gamma_{2})(\gamma_{1}\bar{n}_{\text{th},1}+\gamma_{2}\bar{n}_{\text{th},2})+\gamma_{1}\gamma_{2}%
\bar{n}_{\text{th},2}\left[(\Delta_{b,1}-\Delta_{b,2})^{2}+(\gamma_{1}+\gamma_{2})^{2}/4\right]}{\xi^{2}(\gamma_{1}+\gamma_{2})^{2}+\gamma_{1}\gamma
_{2}\left[(\Delta_{b,1}-\Delta_{b,2})^{2}+(\gamma_{1}+\gamma_{2})^{2}/4\right]}.
\end{align}
\end{subequations}
When the masses of the two mechanical oscillators are equal, the expression of the mode temperatures can be further simplified as follows,
\begin{subequations}
\label{modetempera}
\begin{align}
T_{m1}=&T_{1}+\frac{4\xi^{2}\gamma_{2}(T_{2}-T_{1})}{(\gamma_{1}+\gamma_{2})(4\xi^{2}+\gamma_{1}\gamma_{2})}, \\
T_{m2}=&T_{2}+\frac{4\xi^{2}\gamma_{1}(T_{1}-T_{2})}{(\gamma_{1}+\gamma_{2})(4\xi^{2}+\gamma_{1}\gamma_{2})}.
\end{align}%
\end{subequations}

\section{VII. Analyses of the system parameters}  \label{atep}

In this section, we present some detailed analyses on the system parameters. In our system, there are two moving mirrors and many cavity field modes. Theoretically speaking, there are infinite cavity-field modes in the cavity. In practice, the number of the resonance modes in the cavity is determined by the cut-off frequency of the cavity mirrors and the static cavity length. The two moving mirrors experience the harmonic oscillations, and we assume that the two moving mirrors have the same parameters, namely, the same resonance frequency $\omega_{M}$, mass $m_{M}$, and decay rate $\gamma_{M}$ (the quantity factor $Q=\omega_{M}/\gamma_{M}$). Based on the mass and resonance frequency, we can calculate the zero-point fluctuation $x_{\text{zpf}}=\sqrt{\hbar/(2m_{M}\omega_{M})}$ of the moving mirrors. In Table 1, we show the zero-point fluctuation when the mass takes $m_{M}=10^{-19}-10^{-13}$ kg, and the resonance frequency $\omega_{M}=2\pi\times10^{4}-10^{9}$ Hz. We can see that the zero-point fluctuation can take $10^{-16}-10^{-11}$ m.
\begin{table}[h!]
  \begin{center}
    \caption{The zero-point fluctuation $x_{\text{zpf}}$ of the moving mirrors with different values of the mass $m_{M}$  and resonance frequency $\omega_{M}$.}
    \label{tab:table1}
    \begin{tabular}{|c|c|c|c|c|c|c|} 
      \hline
      \diagbox{$m_{M}$ [kg]}{$x_{\text{zpf}}=\sqrt{\frac{\hbar}{2m_{M}\omega_{M}}}$ [m]}{$\omega_{M}/(2\pi)$ [Hz]} & $10^{4}$  & $10^{5}$  & $10^{6}$  & $10^{7}$  & $10^{8}$  &  $10^{9}$   \\
      \hline
      $10^{-13}$ & $9.161\times10^{-14}$ & $2.897\times10^{-14}$ & $9.161\times10^{-15}$ & $2.897\times10^{-15}$ & $9.161\times10^{-16}$ & $2.897\times10^{-16}$\\
      \hline
      $10^{-14}$ & $2.897\times10^{-13}$ & $9.161\times10^{-14}$ & $2.897\times10^{-14}$ & $9.161\times10^{-15}$ & $2.897\times10^{-15}$ & $9.161\times10^{-16}$\\
      \hline
      $10^{-15}$ & $9.161\times10^{-13}$ & $2.897\times10^{-13}$ & $9.161\times10^{-14}$ & $2.897\times10^{-14}$ & $9.171\times10^{-15}$ & $2.897\times10^{-15}$\\
      \hline
      $10^{-16}$ & $2.897\times10^{-12}$ & $9.161\times10^{-13}$ & $2.897\times10^{-13}$ & $9.161\times10^{-14}$ & $2.897\times10^{-14}$ & $9.161\times10^{-15}$\\
      \hline
      $10^{-17}$ & $9.161\times10^{-12}$ & $2.897\times10^{-12}$ & $9.161\times10^{-13}$ & $2.897\times10^{-13}$ & $9.161\times10^{-14}$ & $2.897\times10^{-14}$\\
      \hline
      $10^{-18}$ & $2.897\times10^{-11}$ & $9.161\times10^{-12}$ & $2.897\times10^{-12}$ & $9.161\times10^{-13}$ & $2.897\times10^{-13}$ & $9.161\times10^{-14}$\\
      \hline
      $10^{-19}$ & $9.161\times10^{-11}$ & $2.897\times10^{-11}$ & $9.161\times10^{-12}$ & $2.897\times10^{-12}$ & $9.161\times10^{-13}$ & $2.897\times10^{-13}$\\
      \hline
    \end{tabular}
  \end{center}
\end{table}

For the cavity, when the two moving mirrors are located at their equilibrium points, the frequencies of these resonant cavity modes take the form as
\begin{equation}
\omega_{k0}=k\cdot\Delta\omega_{c}=k\cdot\frac{\pi c}{l_{0}},\hspace{1 cm} k=1,2,3,\cdots, \label{atepfwc}
\end{equation}%
where $\Delta\omega_{c}=\pi c/l_{0}$ is the free spectrum range of the cavity field. Note that here $c$ is introduced to denote the velocity of light in vacuum. We see that $\Delta\omega_{c}$ is a function of the rest cavity length $l_{0}$. In a realistic case, there exists a cut-off frequency $\omega_{\text{cut}}$ of the cavity field modes. Therefore, the number of the existing resonant cavity modes is a finite number $N_{\text{cut}}=$Floor$[\omega_{\text{cut}}/\Delta\omega_{c}]$. It is obvious that the resonant cavity mode number $N_{\text{cut}}$ will be changed with the change of the cavity length $l_{0}$ under a given cut-off frequency $\omega_{\text{cut}}$.

In our system, both the frequency shift $\Delta_{b}$ and the effective coupling strength $\xi$ are important physical quantities, which are detectable in experiments. These two quantities are also very important to determine the mode temperatures and the heat flux. Below, we present detailed analyses on these two quantities. In particular, we introduce the dimensionless frequency shift and dimensionless coupling strength,
\begin{subequations}
\label{dimfrshifandcoup}
\begin{align}
\frac{\Delta_{b}}{\omega_{M}}& =\frac{\pi c}{\omega_{M}}\frac{x_{\text{zpf%
}}^{2}}{l_{0}^{3}}\sum_{k,j=1}^{N_{\text{cut}}}\frac{jk(j+k)}{\left(\frac{%
\omega_{M}}{\Delta\omega_{c}}\right)^{2}-(j+k)^{2}},     \label{dimfrshifandcoupa} \\
\frac{\xi}{\omega_{M}}& =-\frac{\pi c}{\omega_{M}}\frac{x_{\text{zpf}}^{2}%
}{l_{0}^{3}}\sum_{k,j=1}^{N_{\text{cut}}}\frac{(-1)^{j+k}jk(j+k)}{\left(
\frac{\omega_{M}}{\Delta\omega_{c}}\right)^{2}-(j+k)^{2}}.   \label{dimfrshifandcoupb}
\end{align}
\end{subequations}
The quantity $\Delta_{b}/\omega_{M}$ can be used to characterize the deviation of the frequency shift away from the mechanical resonance frequency. The quantity $\xi/\omega_{M}$ can be used to characterize if the system can enter the ultrastrong even deep-strong coupling regimes.
In this two-moving-mirror system, the effective Hamiltonian describes a phonon-exchange coupling, and hence we need to analyze if the system works in either the strong-coupling regime or the weak-coupling regime. To this end, we introduce the ratio
\begin{equation}
\frac{\xi}{\gamma_{M}}=-\frac{\pi c}{\gamma_{M}}\frac{x_{\text{zpf}}^{2}}{%
l_{0}^{3}}\sum_{k,j=1}^{N_{\text{cut}}}\frac{(-1)^{j+k}jk(j+k)}{\left(\frac{\omega_{M}}{\Delta\omega_{c}}\right)^{2}-(j+k)^{2}}.  \label{dimcoupre}
\end{equation}
Under the condition $\omega_{M}\ll \Delta\omega_{c}$, Eqs.~(\ref{dimfrshifandcoup}) and (\ref{dimcoupre}) can be approximately reduced to
\begin{subequations}
\label{dimfrshndcoup1}
\begin{align}
\frac{\Delta_{b}}{\omega_{M}}& \simeq -\frac{\pi c}{\omega_{M}}\frac{x_{%
\text{zpf}}^{2}}{l_{0}^{3}}\sum_{k,j=1}^{N_{\text{cut}}}\frac{jk}{j+k},   \label{dimfrshndcoup1a} \\
\frac{\xi}{\omega_{M}}& \simeq \frac{\pi c}{\omega_{M}}\frac{x_{\text{zpf}%
}^{2}}{l_{0}^{3}}\sum_{k,j=1}^{N_{\text{cut}}}\frac{(-1)^{j+k}jk}{j+k},    \label{dimfrshndcoup1b} \\
\frac{\xi}{\gamma_{M}}& \simeq \frac{\pi c}{\gamma_{M}}\frac{x_{\text{zpf}%
}^{2}}{l_{0}^{3}}\sum_{k,j=1}^{N_{\text{cut}}}\frac{(-1)^{j+k}jk}{j+k}.    \label{dimfrshndcoup1c}
\end{align}
\end{subequations}
The dependence of the ratios $\Delta_{b}/\omega_{M}$ and $\xi/\omega_{M}$ on the static cavity length $l_{0}$ has been discussed in the main text.
\begin{table}[h!]
  \begin{center}
    \caption{The values of the static cavity length $l_{0}$, the free spectrum range $\Delta\omega_{c}$, and the summations $\Sigma_{1}$ and $\Sigma _{2}$ under the cut-off mode numbers for Au and Ag.}
    \label{tab:table2}
    \begin{tabular}{|c|c|c|c|c|c|} 
      \hline
      $l_{0} $ [m] & $10^{-7}$  & $10^{-6}$  & $10^{-5}$  & $10^{-4}$  &  $10^{-3}$   \\
      \hline
      $\Delta\omega_{c}/(2\pi\times1.5)$ [Hz] & $10^{15}$ & $10^{14}$ & $10^{13}$ & $10^{12}$ & $10^{11}$\\
      \hline
      \hline
      $n_{\text{cut}}$ for Au with $\omega_{\text{cut}}=2\pi\times2.196\times10^{15}$  & $1$ & $14$ & $146$ & $1464$ & $14640$\\
      \hline
      $\sum_{1}$ & $0.5$ & $621.181$ & $643182.716$ & $6.4255\times10^{8}$ & $6.419\times10^{11}$\\
      \hline
      $\sum_{2}$ & $0.5$ & $1.6475$ & $18.1459$ & $182.896$ & $1829.9$\\
      \hline
      \hline
      $n_{\text{cut}}$ for Ag with $\omega_{\text{cut}}=2\pi\times2.18\times10^{15} $  & $1$ & $14$ & $145$ & $1453$ & $14533$\\
      \hline
      $\sum_{1}$ & $0.5$ & $621.181$ & $630101.092$ & $6.2818\times10^{8}$ & $6.2798\times10^{11}$\\
      \hline
      $\sum_{2}$ & $0.5$ & $1.6475$ & $18.5209$ & $182.021$ & $1817.02$\\
      \hline
    \end{tabular}
  \end{center}
\end{table}

In Eq.~(\ref{dimfrshndcoup1a}), the summation is independent of the system parameters. However, when considering the cut-off frequency, the upper bound of the summation indexes $j$ and $k$ is given by $N_{\text{cut}}$. Therefore, the values of Eqs.~(\ref{dimfrshifandcoupa}) and (\ref{dimfrshndcoup1b}) are determined by the ratios $x_{\text{zpf}}/l_{0}$, $\Delta\omega_{c}/\omega_{M}$, and the summations
\begin{subequations}
\label{Sigma12defs}
\begin{align}
\Sigma_{1}&=\sum_{k,j=1}^{N_{\text{cut}}}\frac{jk}{j+k},\\
\Sigma_{2}&=\sum_{k,j=1}^{N_{\text{cut}}}\frac{(-1)^{j+k}jk}{j+k}.
\end{align}
\end{subequations}
We note that the two summations $\Sigma_{1}$ and $\Sigma_{2}$ can determine the multimode cavity field effect. We have discussed the multimode cavity field effect in the main text.

Based on the relations $N_{\text{cut}}=$Floor$[\omega_{\text{cut}}/\Delta\omega_{c}]$ and $\Delta\omega_{c}=\pi c/l_{0}$, we know that the values of Eqs.~(\ref{dimfrshndcoup1a}) and~(\ref{dimfrshndcoup1b}) are determined by $m_{\mu}$, $\omega_{M}$, $l_{0}$, and $\omega_{\text{cut}}$. When $\omega_{\text{cut}}$, $m_{\mu}$, and $\omega_{M}$ are given, proper values of $\Delta_{b}/\omega_{M}$ and $\xi/\omega_{M}$ can be obtained by choosing the proper value of $l_{0}$. Based on Eqs.~(\ref{dimfrshndcoup1a}) and~(\ref{dimfrshndcoup1b}), we can see that $|\Delta_{b}/\xi |=|\Sigma _{1}/\Sigma_{2}|$, which is determined by the cut-off dimension $N_{\text{cut}}$. In addition, the ratio $\xi/\gamma_{M}=(\xi/\omega_{M})/(\gamma_{M}/\omega_{M})$ can be analyzed by comparing the values of $\xi/\omega
_{M}$ and $\gamma_{M}/\omega_{M}$. The strong coupling between the two mechanical modes will be helpful to the enhancement of the heat transfer between the two moving mirrors. Meanwhile, there exist sufficient independently tunable parameters such that proper coupling regime can be achieved. These analyses are also useful to the design of the experimental system and parameters. In Table II, we show the statistic cavity length $l_{0}$, the free spectrum range $\Delta\omega_{c}$, and the summations $\Sigma_{1}$ and $\Sigma_{2}$. Here, we can see that $N_{\text{cut}}$ need to be increased with the increase of the rest cavity length $l_{0}$.

We also want to analyze how to check the correctness of the present theory in experiments. According to the above theoretical analyses, we know that the theory can be checked via analyzing the frequency shift of the moving mirrors and the coupling strength between the two moving mirrors. According to the expression of $\Delta_{b}$ and $\xi$, we can see that
\begin{equation}
\frac{\Delta_{b}}{\xi}=-\frac{\Sigma_{1}}{\Sigma_{2}},
\end{equation}
which is a constant determined by the cut-off mode number $N_{\text{cut}}$. In addition, we can check the correctness of the theory by only analyzing the frequency shift. Since the frequency shift depends on the static cavity length, we can choose two different values of the static cavity length, i.e., $l_{0}$ and $l_{0}^{\prime}$. The frequency shifts corresponding to $l_{0}$ and $l_{0}^{\prime}$ are denoted by $\Delta_{b}(l_{0})$ and $\Delta_{b}(l_{0}^{\prime})$, respectively. We assume that, for the two values $l_{0}$ and $l_{0}^{\prime}$, the cut-off mode number does not change. Then we have
\begin{equation}
\frac{\Delta_{b}(l_{0}^{\prime})}{\Delta_{b}(l_{0})}=\left(\frac{l_{0}}{l_{0}^{\prime}}\right)^{3}.
\end{equation}
Therefore, if the relation $\Delta_{b}(l_{0}^{\prime})/\Delta_{b}(l_{0})=\left(l_{0}/l_{0}^{\prime}\right)^{3}$ is satisfied, then our theory works for the present system. When $l_{0}$ and $l_{0}^{\prime}$ correspond to the same cut-off mode number, then $|l_{0}^{\prime}-l_{0}|$
should be within a finite range. Based on the relation $\omega_{\text{cut}}/\Delta\omega_{c}=\omega_{\text{cut}}l_{0}/(\pi c)=N_{\text{cut}}$, we
have $\omega_{\text{cut}}\delta l_{0}/(\pi c)=\delta N_{\text{cut}}$ and then $\delta l_{0}=\pi c\delta N_{\text{cut}}/\omega_{\text{cut}}$. Here $\delta l_{0}$ and $\delta N_{\text{cut}}$ denote the variances of the static cavity length and the cut-off mode number, respectively. To make
sure that $l_{0}^{\prime}$ and $l_{0}$ are associated with the same $N_{\text{cut}}$, then the relation $|l_{0}^{\prime}-l_{0}|<\pi c/\omega_{\text{cut}}$ should be satisfied.

\section{VIII. Comparison of the phonon-exchange coupling strengths between the virtual-process mechanism and the Casimir-coupling mechanism}  \label{ph}

In our derivation of the effective coupling between the two mechanical resonators, we have neglected the zero-point energy in the Hamiltonian. Therefore, our consideration does not include the Casimir interaction between the two mechanical resonators. To evaluate the magnitude of the Casimir interaction, we present a phenomenological estimation based on the expression of the Casimir energy for a one-dimension two-plate system. Further, we make a comparison of the effective phonon-exchange coupling strengths between the virtual process mechanism and the Casimir coupling mechanism.

In this section, we estimate the effective coupling strength between the two mirrors caused by the Casimir interaction following the method in Ref.~\cite{FongNature2019}. The Hamiltonian of the two mechanical resonators can be expressed as
\begin{equation}
\hat{H}_{S}=\hbar\omega_{M}\hat{b}_{1}^{\dagger}\hat{b}_{1}+\hbar\omega_{M}\hat{b}_{2}^{\dagger}\hat{b}_{2}-\frac{\hbar c\pi}{24(l_{0}+\hat{x}%
_{1}-\hat{x}_{2})},  \label{phwice}
\end{equation}%
where the third term is the Casimir energy. Next, we perform the Taylor expansion of the Casimir energy and keep the terms upto the second order of a
small amount $(\hat{x}_{1}-\hat{x}_{2})^{2}/l_{0}$, then the Hamiltonian~(\ref{phwice}) is reduced to
\begin{equation}
\hat{H}_{S}=\hbar\omega_{M}\hat{b}_{1}^{\dagger}\hat{b}_{1}+\hbar\omega_{M}\hat{b}_{2}^{\dagger}\hat{b}_{2}-\frac{\hbar c\pi}{24l_{0}}+\frac{%
\hbar c\pi}{24l_{0}^{2}}(\hat{x}_{1}-\hat{x}_{2})-\frac{\hbar c\pi}{24l_{0}^{3}}\hat{x}_{1}^{2}-\frac{\hbar c\pi}{24l_{0}^{3}}\hat{x}_{2}^{2}+%
\frac{\hbar c\pi}{12l_{0}^{3}}\hat{x}_{1}\hat{x}_{2}.   \label{phwicewihex}
\end{equation}%
Using the creation and annihilation operators introduced in Eq.~(\ref{phcadefined}), the Hamiltonian~(\ref{phwicewihex}) becomes
\begin{eqnarray}
\hat{H}_{S} &=&\hbar\omega_{M}\hat{b}_{1}^{\dagger}\hat{b}_{1}+\hbar\omega_{M}\hat{b}_{2}^{\dagger}\hat{b}_{2}-\frac{\hbar c\pi}{24l_{0}}+%
\frac{\hbar c\pi}{24l_{0}^{2}}x_{1,\text{zpf}}(\hat{b}_{1}^{\dagger}+\hat{b}_{1})-\frac{\hbar c\pi}{24l_{0}^{2}}x_{2,\text{zpf}}(\hat{b}_{2}^{\dagger
}+\hat{b}_{2}) \nonumber\\
&&-\frac{\hbar c\pi}{24l_{0}^{3}}x_{1,\text{zpf}}^{2}(\hat{b}_{1}^{\dagger}+\hat{b}_{1})^{2}-\frac{\hbar c\pi}{24l_{0}^{3}}x_{2,\text{zpf}}^{2}(\hat{b}_{2}^{\dagger}+\hat{b}_{2})^{2}+\frac{\hbar c\pi}{12l_{0}^{3}}x_{1,\text{%
zpf}}x_{2,\text{zpf}}(\hat{b}_{1}^{\dagger}+\hat{b}_{1})(\hat{b}_{2}^{\dagger}+\hat{b}_{2}).  \label{phwicewihexca}
\end{eqnarray}%
Based on the expressions defined in Eq.~(\ref{couplingstrong}),
\begin{equation}
g_{1,1}=\frac{\omega_{10}}{l_{0}}x_{1,\text{zpf}},\hspace{1.0 cm} g_{1,2}=\frac{\omega_{10}}{l_{0}}x_{2,\text{zpf}},
\end{equation}%
Eq.~(\ref{phwicewihexca}) can be expressed as
\begin{eqnarray}
\hat{H}_{S} &=&\hbar\omega_{M}\hat{b}_{1}^{\dagger}\hat{b}_{1}+\hbar\omega_{M}\hat{b}_{2}^{\dagger}\hat{b}_{2}-\hbar\frac{\omega_{10}}{24}%
+\hbar\frac{g_{1,1}}{24}(\hat{b}_{1}^{\dagger}+\hat{b}_{1})-\hbar\frac{g_{1,2}}{24}(\hat{b}_{2}^{\dagger}+\hat{b}_{2}) \nonumber\\
&&-\hbar\frac{g_{1,1}^{2}}{24\omega_{10}}(\hat{b}_{1}^{\dagger}+\hat{b}_{1})^{2}-\hbar\frac{g_{1,2}^{2}}{24\omega_{10}}(\hat{b}_{2}^{\dagger}+%
\hat{b}_{2})^{2}+\hbar\frac{g_{1,1}g_{1,2}}{12\omega_{10}}(\hat{b}_{1}^{\dagger}+\hat{b}_{1})(\hat{b}_{2}^{\dagger}+\hat{b}_{2}).
\end{eqnarray}%
Under the condition $\pi cx_{1,\text{zpf}}^{2}/(48l_{0}^{3}\omega_{M})\ll 1$, $\pi cx_{2,\text{zpf}}^{2}/(48l_{0}^{3}\omega_{M})\ll 1$, and $\pi cx_{1,\text{zpf}}x_{2,\text{zpf}}/(24l_{0}^{3}\omega_{M})\ll 1$, which can also be expressed as $g_{1,1}^{2}/(48\omega_{10}\omega_{M})\ll 1$, $%
g_{1,2}^{2}/(48\omega_{10}\omega_{M})\ll 1$, and $g_{1,1}g_{1,2}/(24\omega_{10}\omega_{M})\ll 1$, we can perform the rotating-wave approximation, then Eq.~(\ref{phwicewihexca}) is reduced to
\begin{eqnarray}
\hat{H}_{S} &=&\hbar(\omega_{M}-\delta_{b,1})\hat{b}_{1}^{\dagger}\hat{b}_{1}+\hbar(\omega_{M}-\delta_{b,2})\hat{b}_{2}^{\dagger}\hat{b}_{2}-%
\frac{\hbar \pi c}{24l_{0}} \nonumber\\
&&+\frac{\hbar\pi c}{24l_{0}^{2}}x_{1,\text{zpf}}(\hat{b}_{1}^{\dagger}+\hat{b}_{1})-\frac{\hbar\pi c}{24l_{0}^{2}}x_{2,\text{zpf}}(\hat{b}%
_{2}^{\dagger}+\hat{b}_{2})+\hbar\zeta(\hat{b}_{1}^{\dagger}\hat{b}_{2}+\hat{b}_{1}\hat{b}_{2}^{\dagger}),
\end{eqnarray}%
where we introduce the paramerters
\begin{subequations}
\label{freandcouplscs}
\begin{align}
\delta_{b,n} =&\pi c\frac{x_{n,\text{zpf}}^{2}}{12l_{0}^{3}}, \\
\zeta  =&\pi c\frac{x_{1,\text{zpf}}x_{2,\text{zpf}}}{12l_{0}^{3}}. \label{couplscs}
\end{align}
\end{subequations}
In Eq.~(\ref{freandcouplscs}), $\delta_{b,n}$ is the frequency shift of the $n$th ($n=1,2$) moving end mirror induced by the Casimir force, and $\zeta$ is the effective coupling strength between the two moving end mirrors induced by the Casimir force. By comparing Eqs.~(\ref{couplsva}) with~(\ref{couplscs}), we can obtain
\begin{equation}
\frac{\xi}{\zeta}=12\Sigma_{2}.
\end{equation}%
In the multimode-cavity-field framework, we have $\Sigma_{2}\gg1$. Therefore, the coupling strength of the resonant second-order effective phonon exchange coupling between the two moving mirrors induced by the quantum vacuum is much greater than that induced by the Casimir force.

\section{IX. The quantized Hamiltonian of the three-cavity optomechanical system}  \label{htcss}

In this section, we present the detailed derivation of the Hamiltonian of the system in a consistent way. This is achieved by introducing two auxiliary cavities to hold the two-moving-mirror optomechanical cavity. The three-cavity system forms a closed quantum system. Our motivation of this section is not to perform all the calculations within the three-cavity framework, but to show that the two-moving-mirror optomechanical cavity system is valid to support the studies presented in this work.

\subsection{A. The equations of motion of the three-cavity optomechanical system}

\begin{figure}[tbp]
\center
\includegraphics[bb=2 230 608 453, width=0.6 \textwidth]{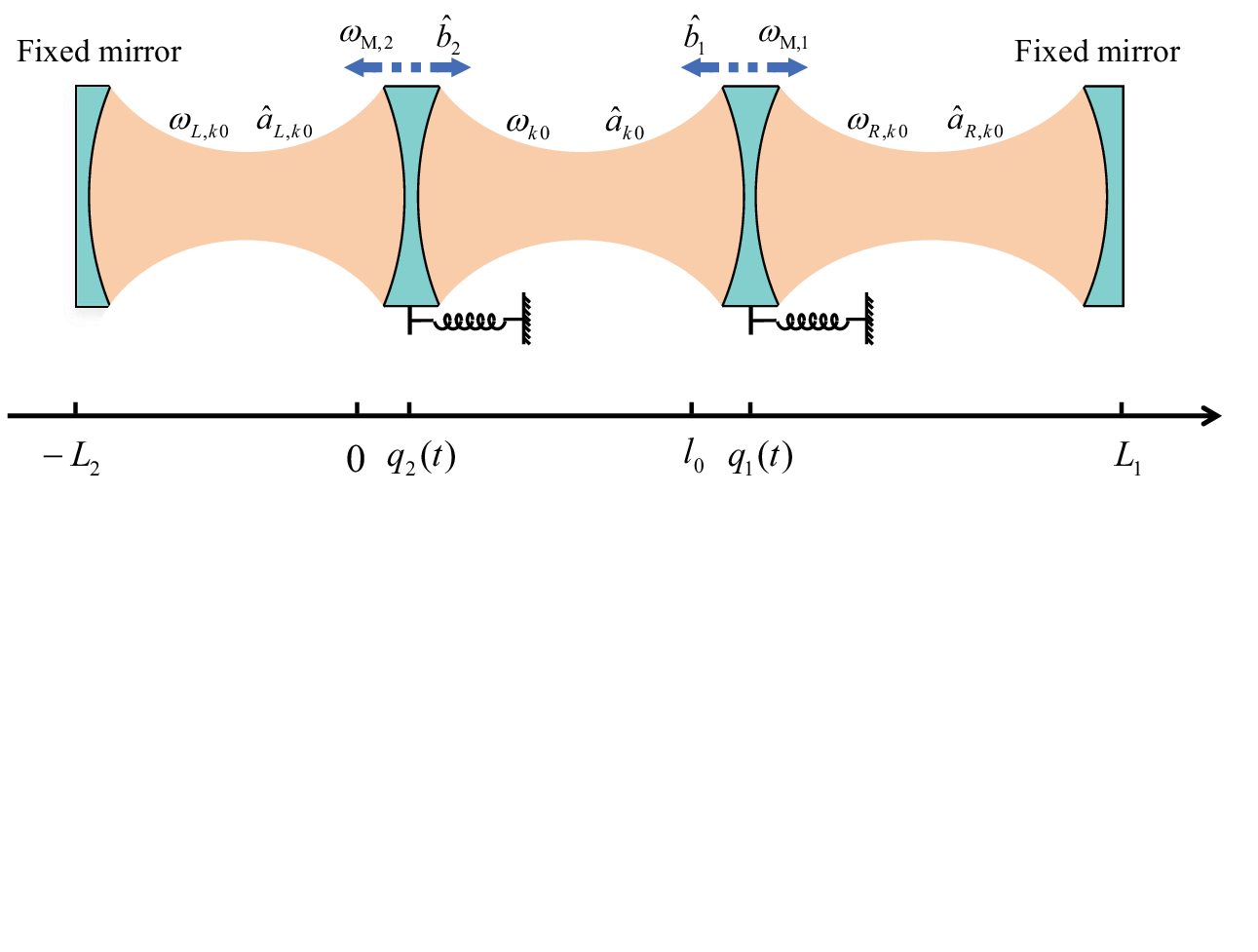}
\caption{(Color online) Schematic of the three-cavity optomechanical system. The second cavity with a bare length $l_{0}$ are composed of two moving end mirrors confined in the harmonic potential wells. The equilibrium position of the left (right) moving mirror is given by $x=0$ ($x=l_{0}$). The first (third) cavity with a bare length $L_{2}$ ($L_{1}-l_{0}$) has only one moving end mirror. The $\hat{a}_{k0}$ represents the annihilation operator of the field in the second cavity with resonance frequency $\omega_{k0}$, the $\hat{a}_{L,k0}$ ($\hat{a}_{R,k0}$) represents the annihilation operator of the field in the first (third) cavity field with resonance frequency $\omega_{L,k0}$ ($\omega_{R,k0}$), and the operator $\hat{b}_{1}$ ($\hat{b}_{2}$) is the annihilation operator of the right (left) moving mirrors with resonance frequencies $\omega_{\text{M,1}}$ ($\omega_{\text{M,2}}$).}\label{mode2}
\end{figure}

In our previous considerations, we only consider the fields inside the cavity. In fact, the external fields outside the cavity can also affect the dynamics of the moving mirrors and the internal cavity fields. On the contrary, the dynamic process of the external fields can also be influenced by the oscillation of the moving mirrors and the fields inside the cavity. As a result, to completely describe the physics in the present model, we need to consider a three-cavity optomechanical system. Concretely, we consider a one-dimensional cavity with two perfectly reflecting fixed end mirrors (see Fig.~\ref{mode2}). This cavity can be considered as a whole one-dimension universe according to the quasi-normal mode theory. We also assume that two perfectly reflecting moving mirrors are placed inside the cavity. The two mirrors move in the potential wells $V_{1}(q_{1})$ and $V_{2}(q_{2})$. In this way, all the fields inside and outside of the internal cavity have been considered. The motion of the moving mirrors will also be influenced by the radiation pressure of the fields at the two sides of each moving mirrors.

In this system, we denote the position (mass) of the $j$th moving mirror as $q_{j}(t)$ ($m_{j}$), where $j=1$ ($j=2$) denotes the right (left) end mirror. The whole one-dimension space is divided into three parts: the first cavity region $-L_{2}\leq x\leq q_{2}(t)$, the second cavity region $q_{2}(t) \leq x\leq q_{1}(t)$, and the third cavity region $q_{1}(t) \leq x\leq L_{1}$. We mark the vector potentials of the first, second, and third cavity fields as $\vec{A}_{L}(x,t)$, $\vec{A}(x,t)$, and $\vec{A}_{R}(x,t)$, respectively. The wave equations of the vector potentials $\vec{A}_{L}(x,t)$, $\vec{A}(x,t)$, and $\vec{A}_{R}(x,t)$ are given by
\begin{subequations}
\label{equationv3}
    \begin{align}
\frac{\partial ^{2}\vec{A}_{L}(x,t)}{\partial x^{2}} &=\frac{\partial ^{2}%
\vec{A}_{L}(x,t)}{\partial t^{2}}, \hspace{1 cm} -L_{2}\leq x\leq q_{2}(t),  \\
\frac{\partial ^{2}\vec{A}(x,t)}{\partial x^{2}} &=\frac{\partial ^{2}\vec{A%
}(x,t)}{\partial t^{2}}, \hspace{1 cm}  q_{2}(t) \leq x\leq q_{1}(t),  \\
\frac{\partial ^{2}\vec{A}_{R}(x,t)}{\partial x^{2}} &=\frac{\partial ^{2}%
\vec{A}_{R}(x,t)}{\partial t^{2}}, \hspace{1 cm}  q_{1}(t) \leq x\leq L_{1}.
\end{align}
\end{subequations}
Since the electric fields are always equal to zero in the rest frame of the mirror surface~\cite{MooreJMPs1970}, the time-dependent boundary conditions of the fields in these three cavities can be expressed as
\begin{subequations}
\label{boundcond3}
    \begin{align}
\vec{A}_{L}(-L_{2},t)&=\vec{A}_{L}(q_{2}(t),t)=0,   \\
\vec{A}(q_{2}(t),t)&=\vec{A}(q_{1}(t),t)=0,   \\
\vec{A}_{R}(q_{1}(t),t)&=\vec{A}_{R}(L_{1},t)=0.
\end{align}
\end{subequations}
Similarly, we consider the linear polarization of the fields, then the vector potentials $\vec{A}_{L}(x,t)$, $\vec{A}(x,t)$, and $\vec{A}_{R}(x,t)$ can be regarded as the scalar quantities $A_{L}(x,t)$, $A(x,t)$, and $A_{R}(x,t)$, respectively.

According to Newton's second law, the equations of motion for the moving mirrors can be written as
\begin{subequations}
\label{miequationv3}
    \begin{align}
m_{1}\ddot{q}_{1} &=-\partial_{q_{1}}V_{1}(q_{1})+\frac{1}{2}[\partial_{x}A(x,t)]^{2}|_{x=q_{1}(t)}-\frac{1}{2}[\partial
_{x}A_{R}(x,t)]^{2}|_{x=q_{1}(t)},  \\
m_{2}\ddot{q}_{2} &=-\partial_{q_{2}}V_{2}\left(q_{2}\right) -\frac{1}{2}[\partial _{x}A(x,t)]^{2}|_{x=q_{2}(t)}+\frac{1}{2}[\partial
_{x}A_{L}(x,t)]^{2}|_{x=q_{2}(t)},
\end{align}
\end{subequations}
where the first term on the right-hand side of Eq.~(\ref{miequationv3}) is called conservative force, both the second and third terms are the radiation-pressure forces of the cavity fields acting on the moving mirrors. We can introduce three sets of generalized coordinates $\{Q_{L,k}(t)\}$, $\{Q_{k}(t)\}$, and $\{Q_{R,k}(t)\}$, which are defined by
\begin{subequations}
\label{gcoord3}
    \begin{align}
Q_{L,k}(t)&=\int_{-L_{2}}^{q_{2}(t)}\varphi_{L,k}(x)A_{L}(x,t)dx,  \\
Q_{k}(t)&=\int_{q_{2}(t)}^{q_{1}(t)}\varphi_{k}(x)A(x,t)dx,  \\
Q_{R,k}(t)&=\int_{q_{1}(t)}^{L_{1}}\varphi_{R,k}(x)A_{R}(x,t)dx,
\end{align}
\end{subequations}
where the function $\varphi_{k}(x)$ is defined by Eq.~(\ref{function}), and the functions $\varphi_{L,k}(x)$ and $\varphi_{R,k}(x)$ are, respectively, defined by
\begin{subequations}
\label{functi3}
    \begin{align}
\varphi_{L,k}(x)&=\sqrt{\frac{2}{L_{2}+q_{2}}}\sin[\omega_{L,k}(x+L_{2})],  \\
\varphi_{R,k}(x)&=\sqrt{\frac{2}{L_{1}-q_{1}}}\sin[\omega_{R,k}(L_{1}-x)],
\end{align}
\end{subequations}
with the position-dependent resonance frequencies
\begin{subequations}
\label{dfren3}
    \begin{align}
\omega_{L,k}&=\frac{k\pi}{L_{2}+q_{2}},   \\
\omega_{R,k}&=\frac{k\pi}{L_{1}-q_{1}}.
\end{align}
\end{subequations}
The scalar quantity $A(x,t)$ is given by Eq.~(\ref{vector potential}), and the scalar quantities $A_{L}(x,t)$ and $A_{R}(x,t)$ can be written as
\begin{subequations}
\label{vecpt3}
    \begin{align}
A_{L}(x,t)&=\sum_{k=1}^{\infty}Q_{L,k}(t)\varphi_{L,k}(x),  \\
A_{R}(x,t)&=\sum_{k=1}^{\infty}Q_{R,k}(t)\varphi_{R,k}(x).
\end{align}
\end{subequations}
Considering the boundary conditions given in Eq.~(\ref{boundcond3}), we can also obtain Eq.~(\ref{vecpt3}) using the Fourier series
expansion, with the generalized coordinates $\{Q_{L,k}(t)\}$ and $\{Q_{R,k}(t)\}$ being the expansion coefficients. Without causing confusion and for keeping the equations concise, hereafter we denote $Q_{L,k}(t)$, $Q_{k}(t)$, and $Q_{R,k}(t)$ as $Q_{L,k}$, $Q_{k}$, and $Q_{R,k}$, respectively. We also denote $\varphi_{L,k}(x)$, $\varphi_{k}(x)$, and $\varphi_{R,k}(x)$ as $\varphi_{L,k}$, $\varphi_{k}$, and $\varphi_{R,k}$, respectively. Next, we can calculate the first and second derivatives of $A_{L}(x,t)$ and $A_{R}(x,t)$ with respect to the position $x$ and the time $t$ as follows,
\begin{subequations}
\label{threevpdos}
    \begin{align}
\frac{\partial A_{\alpha}(x,t)}{\partial x}&=\sum_{k=1}^{\infty}Q_{\alpha,k}\frac{\partial\varphi_{\alpha,k}}{\partial x},  \\
\frac{\partial^{2}A_{\alpha}(x,t)}{\partial x^{2}}&=-\sum_{k=1}^{\infty}\omega_{\alpha,k}^{2}Q_{\alpha,k}\varphi_{\alpha,k},   \\
\frac{\partial A_{L}(x,t)}{\partial t}&=\sum_{k=1}^{\infty}\dot{Q}_{L,k}\varphi_{L,k}+\sum_{k=1}^{\infty}Q_{L,k}
\frac{\partial\varphi_{L,k}}{\partial q_{2}}\dot{q}_{2},      \\
\frac{\partial^{2}A_{L}(x,t)}{\partial t^{2}}&=\sum_{k=1}^{\infty}\ddot{Q}_{L,k}\varphi_{L,k}
+2\sum_{k=1}^{\infty}\dot{Q}_{L,k}\frac{\partial\varphi_{L,k}}{\partial q_{2}}\dot{q}_{2}
+\sum_{k=1}^{\infty}Q_{L,k}\frac{\partial^{2}\varphi_{L,k}}{\partial q_{2}^{2}}\dot{q}_{2}^{2}+\sum_{k=1}^{\infty}Q_{L,k}
\frac{\partial\varphi_{L,k}}{\partial q_{2}}\ddot{q}_{2},          \\
\frac{\partial A_{R}(x,t)}{\partial t}&=\sum_{k=1}^{\infty}\dot{Q}_{R,k} \varphi_{R,k}+\sum_{k=1}^{\infty}Q_{R,k}
\frac{\partial \varphi_{R,k}}{\partial q_{1}}\dot{q}_{1},      \\
\frac{\partial^{2}A_{R}(x,t)}{\partial t^{2}}&=\sum_{k=1}^{\infty}\ddot{Q}_{R,k}\varphi_{R,k}
+2\sum_{k=1}^{\infty}\dot{Q}_{R,k}\frac{\partial\varphi_{R,k}}{\partial q_{1}}\dot{q}_{1}
+\sum_{k=1}^{\infty}Q_{R,k}(t)\frac{\partial^{2}\varphi_{R,k}}{\partial q_{1}^{2}}\dot{q}_{1}^{2}+\sum_{k=1}^{\infty}Q_{R,k}
\frac{\partial\varphi_{R,k}}{\partial q_{1}}\ddot{q}_{1},
\end{align}
\end{subequations}
where $\alpha =L$, $R$.

Based on Eq.~(\ref{threevpdos}), the wave equations of the scalar quantities $A_{L}(x,t)$ and $A_{R}(x,t)$ can be expressed as
\begin{subequations}
\label{threevecgcpt3}
    \begin{align}
-\sum_{k=1}^{\infty}\omega_{L,k}^{2}Q_{L,k}\varphi_{L,k}&=\sum_{k=1}^{\infty}\ddot{Q}_{L,k} \varphi_{L,k}
+2\sum_{k=1}^{\infty}\dot{Q}_{L,k}\frac{\partial\varphi_{L,k}}{\partial q_{2}}\dot{q}_{2}
+\sum_{k=1}^{\infty}Q_{L,k}\frac{\partial^{2}\varphi_{L,k}}{\partial q_{2}^{2}}\dot{q}_{2}^{2}
+\sum_{k=1}^{\infty}Q_{L,k}\frac{\partial\varphi_{L,k}}{\partial q_{2}}\ddot{q}_{2},                   \\
-\sum_{k=1}^{\infty}\omega_{R,k}^{2}Q_{R,k}\varphi_{R,k}&=\sum_{k=1}^{\infty}\ddot{Q}_{R,k} \varphi_{R,k}
+2\sum_{k=1}^{\infty}\dot{Q}_{R,k}\frac{\partial\varphi_{R,k}}{\partial q_{1}}\dot{q}_{1}
+\sum_{k=1}^{\infty}Q_{R,k}\frac{\partial^{2}\varphi_{R,k}}{\partial q_{1}^{2}}\dot{q}_{1}^{2}
+\sum_{k=1}^{\infty }Q_{R,k}\frac{\partial\varphi_{R,k}}{\partial q_{1}}\ddot{q}_{1}.
\end{align}
\end{subequations}
Multiply both sides of Eq.~(\ref{threevecgcpt3}) by the functions $\varphi_{L,s}$ and $\varphi_{R,s}$, respectively, and perform the integration with respect to $x$, we obtain
\begin{subequations}
\label{vecgcpt3}
    \begin{align}
\ddot{Q}_{L,k} &=-\omega_{L,k}^{2}Q_{L,k}+\frac{2\dot{q}_{2}}{L_{2}+q_{2}}\sum_{j=1}^{\infty}\dot{Q}_{L,j}g_{L,kj}+\frac{(L_{2}+q_{2})
\ddot{q}_{2}-\dot{q}_{2}^{2}}{(L_{2}+q_{2})^{2}}\sum_{j=1}^{\infty}Q_{L,j}g_{L,kj}+\frac{\dot{q}_{2}^{2}}{(
L_{2}+q_{2})^{2}}\sum_{l,j=1}^{\infty}Q_{L,l}g_{L,jk}g_{L,jl},                                                     \\
\ddot{Q}_{R,k} &=-\omega_{R,k}^{2}Q_{R,k}-\frac{2\dot{q}_{1}}{L_{1}-q_{1}}\sum_{j=1}^{\infty}\dot{Q}_{R,j}g_{R,jk}-\frac{\dot{q}_{1}^{2}+(
L_{1}-q_{1})\ddot{q}_{1}}{(L_{1}-q_{1})^{2}}\sum_{j=1}^{\infty}Q_{R,j}g_{R,jk}+\frac{\dot{q}_{1}^{2}}{(
L_{1}-q_{1})^{2}}\sum_{l,j=1}^{\infty}Q_{R,l}g_{R,kj}g_{R,lj},
\end{align}
\end{subequations}
where the parameters $g_{L,kj}$ and $g_{R,kj}$ are introduced by
\begin{subequations}
\label{parametggt3}
    \begin{align}
g_{L,kj}&=\left\{
\begin{array}{c}
\frac{2jk}{j^{2}-k^{2}}(-1)^{j+k},\hspace{1 cm} j\neq k, \\
0,\hspace{2.9 cm} j=k,%
\end{array}%
\right.    \\
g_{R,kj}&=\left\{
\begin{array}{c}
-\frac{2jk}{j^{2}-k^{2}}(-1)^{j+k},\hspace{1 cm} j\neq k, \\
0,\hspace{3.2 cm} j=k.%
\end{array}%
\right.
\end{align}
\end{subequations}
In the derivation of Eqs.~(\ref{vecgcpt3}), we have used the following relations
\begin{subequations}
\label{usedpart3}
    \begin{align}
\int_{-L_{2}}^{q_{2}}\varphi_{L,j}\varphi_{L,k}dx &=\delta_{j,k},  \\
\int_{q_{1}}^{L_{1}}\varphi_{R,j}\varphi_{R,k}dx &=\delta_{j,k},  \\
\int_{-L_{2}}^{q_{2}}\varphi_{L,j}\frac{\partial\varphi_{L,k}}{\partial q_{2}}dx &=\frac{g_{L,kj}}{L_{2}+q_{2}},  \\ \int_{q_{1}}^{L_{1}}\varphi_{R,j}\frac{\partial \varphi_{R,k}}{\partial q_{1}}dx &=\frac{g_{R,kj}}{L_{1}-q_{1}},      \\
\int_{-L_{2}}^{q_{2}}\frac{\partial\varphi_{L,j}}{\partial q_{2}}\frac{\partial\varphi_{L,l}}{\partial q_{2}}dx &=\frac{\sum_{k=1}^{\infty
}g_{L,jk}g_{L,lk}}{(L_{2}+q_{2})^{2}},                                               \\
\int_{q_{1}}^{L_{1}}\frac{\partial\varphi_{R,j}}{\partial q_{1}}\frac{\partial\varphi_{R,k}}{\partial q_{1}}dx &=\frac{\sum_{l=1}^{\infty}g_{R,jl}g_{R,kl}}{(L_{1}-q_{1})^{2}},              \\
\int_{-L_{2}}^{q_{2}}\varphi_{L,j}\frac{\partial^{2}\varphi_{L,k}}{%
\partial q_{2}^{2}}dx &=-\frac{g_{L,kj}}{(L_{2}+q_{2})^{2}}-\frac{1}{(L_{2}+q_{2})^{2}}\sum_{l=1}^{\infty }g_{L,jl}g_{L,kl},     \\
\int_{q_{1}}^{L_{1}}\varphi_{R,j}\frac{\partial^{2}\varphi_{R,k}}{\partial q_{1}^{2}}dx &=\frac{g_{R,kj}}{(L_{1}-q_{1})^{2}}-%
\frac{1}{(L_{1}-q_{1})^{2}}\sum_{l=1}^{\infty}g_{R,jl}g_{R,kl}.
\end{align}
\end{subequations}
Note that the above relations in Eqs.~(\ref{usedpart3}) can be proved with the same method used in Eqs.~(\ref{matherela}), here we do not present the detailed derivations for keeping conciseness. The wave equations of the scalar quantity $A(x,t)$ can be expressed as Eq.~(\ref{dynaequationsa}). In terms of the generalized coordinates, the equations of motion for the two moving mirrors can be expressed as
\begin{subequations}
\label{mirroreqution3}
    \begin{align}
m_{1}\ddot{q}_{1} &=-\frac{\partial V_{1}(q_{1})}{\partial q_{1}}+\frac{1}{q_{1}-q_{2}}\sum_{j,k=1}^{\infty}(-1)^{j+k}\omega_{k}\omega
_{j}Q_{k}Q_{j}-\frac{1}{L_{1}-q_{1}}\sum_{j,k=1}^{\infty}(-1)^{j+k}\omega_{R,k}\omega_{R,j}Q_{R,k}Q_{R,j},                        \\
m_{2}\ddot{q}_{2} &=-\frac{\partial V_{2}(q_{2})}{\partial q_{2}}-\frac{1}{q_{1}-q_{2}}\sum_{j,k=1}^{\infty}\omega_{k}\omega_{j}Q_{k}Q_{j}+\frac{1}{%
L_{2}+q_{2}}\sum_{j,k=1}^{\infty}(-1)^{j+k}\omega_{L,k}\omega_{L,j}Q_{L,k}Q_{L,j}.
\end{align}
\end{subequations}
Based on Eqs.~(\ref{dynaequationsa}),~(\ref{vecgcpt3}), and~(\ref{mirroreqution3}), we can construct the Lagrange function and Hamiltonian of the three-cavity system.

\subsection{B. The Lagrange function of the three-cavity optomechanical system}

For the three-cavity system, the Lagrange function $L$ can be written as
\begin{equation}
L=L_{\text{field}}+L_{L,\text{field}}+L_{R,\text{field}}+\frac{1}{2}m_{1}\dot{q}_{1}^{2}+\frac{1}{2}m_{2}\dot{q}_{2}^{2}-V_{1}(q_{1})-V_{2}(q_{2}), \label{threetotalLang}
\end{equation}
where $L_{\text{field}}$ is the Lagrange function of the second cavity field which could be defined as Eq.~(\ref{filedL}), $L_{L,\text{field}}$ and $L_{R,\text{field}}$ are, respectively, the Lagrange functions of the first and third cavity fields,
\begin{subequations}
\label{threefiledL}
    \begin{align}
L_{L,\text{field}}&=\frac{1}{2}\int_{-L_{2}}^{q_{2}}\left(\frac{\partial A_{L}(x,t)}{\partial t}\right)^{2}dx-\frac{1}{2}%
\int_{-L_{2}}^{q_{2}}\left(\frac{\partial A_{L}(x,t)}{\partial x}\right)^{2}dx,                       \\
L_{R,\text{field}}&=\frac{1}{2}\int_{q_{1}}^{L_{1}}\left(\frac{\partial A_{R}(x,t)}{\partial t}\right)^{2}dx-\frac{1}{2}%
\int_{q_{1}}^{L_{1}}\left(\frac{\partial A_{R}(x,t) }{\partial x}\right)^{2}dx.
\end{align}
\end{subequations}
In terms of Eqs.~(\ref{vpdos}), (\ref{matherela}), (\ref{filedL}), (\ref{threevpdos}), (\ref{usedpart3}), and (\ref{threefiledL}), the assumed Lagrange function $L$ given in Eq.~(\ref{threetotalLang}) can be expressed as
\begin{eqnarray}
L &=&\frac{1}{2}\sum_{k=1}^{\infty}(\dot{Q}_{L,k}^{2}-\omega_{L,k}^{2}Q_{L,k}^{2})+\sum_{j,k=1}^{\infty}\frac{\dot{q}_{2}g_{L,kj}}{%
L_{2}+q_{2}}\dot{Q}_{L,j}Q_{L,k}+\sum_{j,k,l=1}^{\infty}\frac{\dot{q}_{2}^{2}g_{L,jl}g_{L,kl}}{2(L_{2}+q_{2})^{2}}Q_{L,k}Q_{L,j} \nonumber \\
&&+\frac{1}{2}\sum_{k=1}^{\infty}(\dot{Q}_{k}^{2}-\omega_{k}^{2}Q_{k}^{2})+\sum_{n=1,2}\sum_{j,k=1}^{\infty}\frac{\dot{q}%
_{n}g_{jk}^{(n)}}{q_{1}-q_{2}}\dot{Q}_{k}Q_{j}+\sum_{n,n^{\prime}=1,2}\sum_{j,k,l=1}^{\infty}\frac{\dot{q}_{n}\dot{q}_{n^{\prime
}}g_{jl}^{(n)}g_{kl}^{(n^{\prime})}}{2((q_{1}-q_{2})^{2}}Q_{k}Q_{j}                                                              \nonumber \\
&&+\frac{1}{2}\sum_{k=1}^{\infty}(\dot{Q}_{R,k}^{2}-\omega_{R,k}^{2}Q_{R,k}^{2})+\sum_{j,k=1}^{\infty}\frac{\dot{q}_{1}g_{R,kj}}{%
L_{1}-q_{1}}\dot{Q}_{R,j}Q_{R,k}+\sum_{j,k,l=1}^{\infty}\frac{\dot{q}_{1}^{2}g_{R,jl}g_{R,kl}}{2(L_{1}-q_{1})^{2}}Q_{R,k}Q_{R,j} \nonumber \\
&&+\frac{1}{2}m_{1}\dot{q}_{1}^{2}+\frac{1}{2}m_{2}\dot{q}_{2}^{2}-V_{1}(q_{1})-V_{2}(q_{2}).     \label{lagfunc3}
\end{eqnarray}
In principle, the correctness of the Lagrange function~(\ref{lagfunc3}) could be confirmed by checking the equations of motion for the cavity fields and the moving mirrors. Concretely, the Euler-Lagrangian equations of the $k$th cavity mode in the first and third cavities are, respectively, given by
\begin{subequations}
\label{EulerLagrangiankfc}
    \begin{align}
\frac{d}{dt}\left(\frac{\partial L}{\partial\dot{Q}_{L,k}}\right) -\frac{\partial L}{\partial Q_{L,k}}&=0,   \\
\frac{d}{dt}\left(\frac{\partial L}{\partial\dot{Q}_{R,k}}\right) -\frac{\partial L}{\partial Q_{R,k}}&=0.
\end{align}
\end{subequations}
By substituting Eq.~(\ref{lagfunc3}) into Eqs.~(\ref{EulerLagrangiankfc}), we can obtain the equations of motion
\begin{subequations}
\label{thefirstthirdcavitymodek}
\begin{align}
\ddot{Q}_{L,k}+\frac{\ddot{q}_{2}(L_{2}+q_{2})-\dot{q}_{2}^{2}}{(L_{2}+q_{2})^{2}}\sum_{s=1}^{\infty}Q_{L,s}g_{L,sk}+\frac{
\dot{q}_{2}}{L_{2}+q_{2}}\sum_{s=1}^{\infty}\dot{Q}_{L,s}g_{L,sk} =&-\omega_{L,k}^{2}Q_{L,k}+\frac{\dot{q}_{2}}{L_{2}+q_{2}}\sum_{j=1}^{\infty}\dot{Q}_{L,j}g_{L,kj}\nonumber \\&+\frac{\dot{q}_{2}^{2}}{(L_{2}+q_{2})^{2}}\sum_{j,l=1}^{\infty}Q_{L,j}g_{L,jl}g_{L,kl},\\
\ddot{Q}_{R,k}+\frac{\ddot{q}_{1}(L_{1}-q_{1})+\dot{q}_{1}^{2}}{(L_{1}-q_{1})^{2}}\sum_{s=1}^{\infty}Q_{R,s}g_{R,sk}+\frac{
\dot{q}_{1}}{L_{1}-q_{1}}\sum_{s=1}^{\infty}\dot{Q}_{R,s}g_{R,sk}
=&-\omega_{R,k}^{2}Q_{R,k}+\frac{\dot{q}_{1}}{L_{1}-q_{1}}\sum_{j=1}^{\infty}\dot{Q}_{R,j}g_{R,kj} \nonumber \\
&+\frac{\dot{q}_{1}^{2}}{(L_{1}-q_{1})^{2}}\sum_{j,l=1}^{\infty}Q_{R,j}g_{R,jl}g_{R,kl},
\end{align}
\end{subequations}
which can be further rewritten as Eq.~(\ref{vecgcpt3}). Therefore, we can recover Eq.~(\ref{vecgcpt3}) based on the Euler-Lagrange equation and the Lagrange function in Eq.~(\ref{lagfunc3}).

For the second cavity, the Euler-Lagrangian equation of the $k$th cavity mode is given by Eq.~(\ref{Euler-Lagrangiank}). By substituting Eq.~(\ref{lagfunc3}) into Eq.~(\ref{Euler-Lagrangiank}), we can obtain Eq.~(\ref{cavitymodek}). Finally, we can obtain Eq.~(\ref{dynaequationsa}) from Eq.~(\ref{cavitymodek}). Therefore, we can obtain Eq.~(\ref{dynaequationsa}) based on the Euler-Lagrange equation and the Lagrange function~(\ref{lagfunc3}).

The Euler-Lagrangian equation of the right moving mirror is given by Eq.~(\ref{Euler-Lagrangianrighimoving}). Substituting Eq.~(\ref{lagfunc3}) into Eq.~(\ref{Euler-Lagrangianrighimoving}), we can obtain the equation
\begin{eqnarray}
&&-\frac{\dot{q}_{1}-\dot{q}_{2}}{(q_{1}-q_{2})^{2}}\sum_{j,k=1}^{\infty}\dot{Q}_{k}Q_{j}g_{jk}^{(1)}+\frac{1}{%
q_{1}-q_{2}}\sum_{j,k=1}^{\infty}\ddot{Q}_{k}Q_{j}g_{jk}^{(1)}+\frac{1}{q_{1}-q_{2}}\sum_{j,k=1}^{\infty}\dot{Q}_{k}\dot{Q}%
_{j}g_{jk}^{(1)}                                                                  \nonumber \\
&&+\sum_{n=1,2}\frac{\ddot{q}_{n}(q_{1}-q_{2}) -2(\dot{q}_{1}-\dot{q}_{2})\dot{q}_{n}}{(q_{1}-q_{2})^{3}}%
\sum_{j,k,l=1}^{\infty}Q_{k}Q_{j}g_{jl}^{(n)}g_{kl}^{(1)}                          \nonumber \\
&&+\sum_{n=1,2}\frac{\dot{q}_{n}}{(q_{1}-q_{2})^{2}}\sum_{j,k,l=1}^{\infty}(\dot{Q}_{k}Q_{j}+Q_{k}\dot{Q}_{j})g_{jl}^{(
n)}g_{kl}^{(1)}+\frac{\dot{q}_{1}}{(L_{1}-q_{1})^{2}}\sum_{j,k=1}^{\infty}\dot{Q}_{R,j}Q_{R,k}g_{R,kj}      \nonumber \\
&&+\frac{1}{L_{1}-q_{1}}\sum_{j,k=1}^{\infty}\ddot{Q}_{R,j}Q_{R,k}g_{R,kj}+%
\frac{1}{L_{1}-q_{1}}\sum_{j,k=1}^{\infty}\dot{Q}_{R,j}\dot{Q}_{R,k}g_{R,kj}    \nonumber \\
&&+\frac{\ddot{q}_{1}(L_{1}-q_{1}) +2\dot{q}_{1}^{2}}{(L_{1}-q_{1})^{3}}\sum_{j,k,l=1}^{\infty
}Q_{R,k}Q_{R,j}g_{R,jl}g_{R,kl}+\frac{2\dot{q}_{1}}{(L_{1}-q_{1})^{2}}\sum_{j,k,l=1}^{\infty}\dot{Q}%
_{R,k}Q_{R,j}g_{R,jl}g_{R,kl}+m_{1}\ddot{q}_{1}             \nonumber \\
&=&\frac{1}{q_{1}-q_{2}}\sum_{k=1}^{\infty}\omega_{k}^{2}Q_{k}^{2}-\sum_{n=1,2}\frac{\dot{q}_{n}}{( q_{1}-q_{2})
^{2}}\sum_{j,k=1}^{\infty}\dot{Q}_{k}Q_{j}g_{jk}^{(n)}     \nonumber \\
&&-\sum_{n,n^{\prime}=1,2}\sum_{j,k,l=1}^{\infty}\frac{\dot{q}_{n}\dot{q}_{n^{\prime}}}{(q_{1}-q_{2})^{3}}Q_{k}Q_{j}g_{jl}^{(
n)}g_{kl}^{(n^{\prime})}-\frac{1}{L_{1}-q_{1}}\sum_{k=1}^{\infty}\omega_{R,k}^{2}Q_{R,k}^{2}-\frac{\partial V(q_{1})}{\partial q_{1}} \nonumber \\
&&+\frac{\dot{q}_{1}}{(L_{1}-q_{1})^{2}}\sum_{j,k=1}^{\infty}\dot{Q}_{R,j}Q_{R,k}g_{R,kj}+\frac{2\dot{q}_{1}^{2}}{2(
L_{1}-q_{1})^{3}}\sum_{j,k,l=1}^{\infty}Q_{R,k}Q_{R,j}g_{R,jl}g_{R,kl}.  \label{equationrightmovingmirrors}
\end{eqnarray}
In terms of the relations
\begin{subequations}
\label{zhnemgxuy}
\begin{align}
\sum_{j,k=1}^{\infty}\dot{Q}_{R,j}\dot{Q}_{R,k}g_{R,kj}& =0,     \label{zhnxuy:1a} \\
\sum_{j,k,l,s=1}^{\infty}Q_{R,l}Q_{R,k}g_{R,ls}g_{R,js}g_{R,kj}& =0,              \label{zhnxuy:1b}
\end{align}%
\end{subequations}
Eq.~(\ref{equationrightmovingmirrors}) can be simplified as
\begin{eqnarray}
m_{1}\ddot{q}_{1} &=&\frac{1}{q_{1}-q_{2}}\sum_{k}^{\infty}\omega_{k}^{2}Q_{k}^{2}-\frac{1}{L_{1}-q_{1}}\sum_{k=1}^{\infty}\omega
_{R,k}^{2}Q_{R,k}^{2}-\frac{\partial V(q_{1})}{\partial q_{1}}+\frac{1}{L_{1}-q_{1}}\sum_{j,k=1}^{\infty}\omega_{R,j}^{2}Q_{R,j}Q_{R,k}g_{R,kj}\nonumber \\
&&-\frac{1}{q_{1}-q_{2}}\sum_{j,k}^{\infty}\omega_{k}^{2}Q_{k}Q_{j}g_{kj}.\label{threefangc2}
\end{eqnarray}
After some algebra, Eq.~(\ref{threefangc2}) can be reduced to Eq.~(\ref{mirroreqution3}).

With the similar calculations, we can derive the second equation in Eq.~(\ref{mirroreqution3}) for the left moving mirror based on the Euler-Lagrangian equation. Therefore, the assumed Lagrange function $L$ given in Eq.~(\ref{lagfunc3}) is valid for the three-cavity system.

In the rest part of this section, we present the detailed derivation of the relations given in Eqs.~(\ref{zhnemgxuy}). To prove Eq.~(\ref{zhnxuy:1a}), we make the following simplification based on Eqs.~(\ref{parametggt3}),
\begin{eqnarray}
\sum_{j,k=1}^{\infty}\dot{Q}_{R,j}\dot{Q}_{R,k}g_{R,kj}&=&\sum_{j>k}^{\infty}\dot{Q}_{R,j}\dot{Q}_{R,k}g_{R,kj}+\sum_{j<k}^{\infty
}\dot{Q}_{R,j}\dot{Q}_{R,k}g_{R,kj}                            \nonumber \\
&=&\sum_{j>k}^{\infty}\dot{Q}_{R,j}\dot{Q}_{R,k}g_{R,kj}+\sum_{r<k}^{\infty}\dot{Q}_{R,r}\dot{Q}_{R,k}g_{R,kr}    \nonumber \\
&=&\sum_{j>k}^{\infty}\dot{Q}_{R,j}\dot{Q}_{R,k}g_{R,kj}+\sum_{k<j}^{\infty}\dot{Q}_{R,k}\dot{Q}_{R,j}g_{R,jk}    \nonumber \\
&=&\sum_{j>k}^{\infty}\dot{Q}_{R,j}\dot{Q}_{R,k}(g_{R,kj}+g_{R,jk})    \nonumber \\
&=&0.
\end{eqnarray}
Here, we used the relation $g_{R,kj}+g_{R,jk}=0$ when $j\neq k$.

The proof of Eq.~(\ref{zhnxuy:1b}) can be given as follows,
\begin{eqnarray}
&&\sum_{j,k=1}^{\infty}\sum_{l,s=1}^{\infty}Q_{R,l}Q_{R,k}g_{R,ls}g_{R,js}g_{R,kj}  \nonumber \\
&=&\sum_{j,s=1}^{\infty}\sum_{l>k}^{\infty}Q_{R,l}Q_{R,k}g_{R,ls}g_{R,js}g_{R,kj}+\sum_{j,s=1}^{\infty}\sum_{l<k}^{\infty
}Q_{R,l}Q_{R,k}g_{R,ls}g_{R,js}g_{R,kj}+\sum_{j,s=1}^{\infty}\sum_{k=1}^{\infty}Q_{R,k}Q_{R,k}g_{R,ks}g_{R,js}g_{R,kj}  \nonumber \\
&=&\sum_{j,s=1}^{\infty}\sum_{l>k}^{\infty}Q_{R,l}Q_{R,k}g_{R,ls}g_{R,js}g_{R,kj}+\sum_{j,s=1}^{\infty
}\sum_{r<k}^{\infty}Q_{R,r}Q_{R,k}g_{R,rs}g_{R,js}g_{R,kj}  \nonumber \\
&&+\sum_{j>s}^{\infty}\sum_{k=1}^{\infty}Q_{R,k}Q_{R,k}g_{R,ks}g_{R,js}g_{R,kj}+\sum_{j<s}^{\infty
}\sum_{k=1}^{\infty}Q_{R,k}Q_{R,k}g_{R,ks}g_{R,js}g_{R,kj}  \nonumber \\
&=&\sum_{j,s=1}^{\infty}\sum_{l>k}^{\infty}Q_{R,l}Q_{R,k}g_{R,ls}g_{R,js}g_{R,kj}+\sum_{j,s=1}^{\infty
}\sum_{k<l}^{\infty}Q_{R,k}Q_{R,l}g_{R,ks}g_{R,js}g_{R,lj}  \nonumber \\
&&+\sum_{j>s}^{\infty}\sum_{k=1}^{\infty}Q_{R,k}Q_{R,k}g_{R,ks}g_{R,js}g_{R,kj}+\sum_{s<j}^{\infty
}\sum_{k=1}^{\infty}Q_{R,k}Q_{R,k}g_{R,kj}g_{R,sj}g_{R,ks}  \nonumber \\
&=&\sum_{j,s=1}^{\infty}\sum_{l>k}^{\infty}Q_{R,l}Q_{R,k}g_{R,ls}g_{R,js}g_{R,kj}+\sum_{s,j=1}^{\infty
}\sum_{k<l}^{\infty}Q_{R,k}Q_{R,l}g_{R,kj}g_{R,sj}g_{R,ls}+\sum_{j>s}^{\infty}\sum_{k=1}^{\infty}Q_{R,k}Q_{R,k}g_{R,ks}g_{R,kj}(g_{R,js}+g_{R,sj}) \nonumber \\
&=&\sum_{j,s=1}^{\infty}\sum_{l>k}^{\infty}Q_{R,k}Q_{R,l}g_{R,kj}g_{R,ls}(g_{R,js}+g_{R,sj})  \nonumber \\
&=&0.
\end{eqnarray}
In the derivation of the last step in the above equation, we also used the relation given in Eqs.~(\ref{parametggt3}).

\subsection{C. The quantized Hamiltonian of the three-cavity optomechanical system}

Next, we derive the Hamiltonian of the three-cavity system. By the Legendre transformation, the classical Hamiltonian of the system can be written as
\begin{equation}
H=p_{1}\dot{q}_{1}+p_{2}\dot{q}_{2}+\sum_{k}P_{L,k}\dot{Q}_{L,k}+\sum_{k}P_{k}\dot{Q}_{k}+\sum_{k}P_{R,k}\dot{Q}_{R,k}-L,  \label{legclssc3}
\end{equation}
where $P_{L,k}$, $P_{k}$, $P_{R,k}$, $p_{1}$, and $p_{2}$ are, respectively, the canonical momentum of the $k$th mode of the first, second, and third cavity field, the first moving mirror, and second moving mirror, defined by,
\begin{subequations}
\label{camont3}
    \begin{align}
P_{L,k}&=\frac{\partial L}{\partial\dot{Q}_{L,k}}=\dot{Q}_{L,k}+\frac{\dot{q}_{2}}{L_{2}+q_{2}}\sum_{j=1}^{\infty}Q_{L,j}g_{L,jk},     \\
P_{k}&=\frac{\partial L}{\partial\dot{Q}_{k}}=\dot{Q}_{k}+\sum_{n=1,2}\sum_{j=1}^{\infty}\frac{g_{jk}^{(n)}\dot{q}_{n}}{q_{1}-q_{2}}Q_{j},  \\
P_{R,k}&=\frac{\partial L}{\partial\dot{Q}_{R,k}}=\dot{Q}_{R,k}+\frac{\dot{q}_{1}}{L_{1}-q_{1}}\sum_{j=1}^{\infty }Q_{R,j}g_{R,jk},   \\
p_{1}&=\frac{\partial L}{\partial\dot{q}_{1}}=\frac{1}{q_{1}-q_{2}}\sum_{j,k=1}^{\infty}g_{jk}^{(1)}Q_{j}P_{k}+\frac{1}{L_{1}
-q_{1}}\sum_{j,k=1}^{\infty}g_{R,jk}Q_{R,j}P_{R,k}+m_{1}\dot{q}_{1},      \\
p_{2}&=\frac{\partial L}{\partial\dot{q}_{2}}=\frac{1}{q_{1}-q_{2}}\sum_{j,k=1}^{\infty}g_{jk}^{(2)}Q_{j}P_{k}+\frac{1}{L_{2}
+q_{2}}\sum_{k,j=1}^{\infty}g_{L,jk}Q_{L,j}P_{L,k}+m_{2}\dot{q}_{2}.
\end{align}
\end{subequations}
Based on Eqs.~(\ref{legclssc3}) and (\ref{camont3}), we can obtain the Hamiltonian of the three-cavity system as
\begin{eqnarray}
H &=&\frac{1}{2}\sum_{k=1}^{\infty}(P_{k}^{2}+\omega_{k}^{2}Q_{k}^{2}) +\frac{1}{2}\sum_{\alpha=L,R}\sum_{k=1}^{\infty
}(P_{\alpha ,k}^{2}+\omega_{\alpha ,k}^{2}Q_{\alpha ,k}^{2})+V(q_{1}) +V(q_{2})                                         \nonumber \\
&&+\frac{1}{2m_{1}}\left(p_{1}-\frac{1}{q_{1}-q_{2}}\sum_{j,k=1}^{\infty
}g_{jk}^{(1)}Q_{j}P_{k}-\frac{1}{L_{1}-q_{1}}\sum_{j,k=1}^{\infty}g_{R,jk}Q_{R,j}P_{R,k}\right)^{2}                      \nonumber \\
&&+\frac{1}{2m_{2}}\left(p_{2}-\frac{1}{q_{1}-q_{2}}\sum_{j,k=1}^{\infty}g_{jk}^{(2)}Q_{j}P_{k}-\frac{1}{L_{2}+q_{2}}\sum_{k,j=1}^{\infty
}g_{L,jk}Q_{L,j}P_{L,k}\right)^{2}.  \label{Hamiltoninaclssc3}
\end{eqnarray}
Following the canonical quantization procedure, the variables $p_{1}$, $p_{2}$, $q_{1}$, $q_{2}$, $Q_{L,k}$, $Q_{k}$, $Q_{R,k}$, $P_{L,k}$, $P_{k}$, and $P_{R,k}$ should be replaced by the corresponding operators $\hat{p}_{1}$, $\hat{p}_{2}$, $\hat{q}_{1}$, $\hat{q}_{2}$, $\hat{Q}_{L,k}$, $\hat{Q}_{k}$, $\hat{Q}_{R,k}$, $\hat{P}_{L,k}$, $\hat{P}_{k}$, and $\hat{P}_{R,k}$, which obey the commutation relations
\begin{subequations}
    \begin{align}
\lbrack \hat{q}_{1},\hat{p}_{1}] &=[\hat{q}_{2},\hat{p}_{2}]=i\hbar,  \\
\lbrack \hat{Q}_{j},\hat{P}_{k}] &=i\hbar\delta_{jk},  \\
\lbrack \hat{Q}_{L,j},\hat{P}_{L,k}] &=i\hbar\delta_{jk},  \\
\lbrack \hat{Q}_{R,j},\hat{P}_{R,k}] &=i\hbar\delta_{jk}.
\end{align}
\end{subequations}
The quantized Hamiltonian of the three-cavity system can then be written as
\begin{eqnarray}
\hat{H} &=&\frac{1}{2}\sum_{k}(\hat{P}_{k}^{2}+\omega_{k}^{2}\hat{Q}_{k}^{2}) +\frac{1}{2}\sum_{\alpha=L,R}\sum_{k}(\hat{P}_{\alpha
,k}^{2}+\omega_{\alpha,k}^{2}\hat{Q}_{\alpha,k}^{2}) +V(\hat{q}_{1}) +V(\hat{q}_{2})                                                \nonumber \\
&&+\frac{1}{2m_{1}}(\hat{p}_{1}+\hat{\Pi}_{1})^{2}+\frac{1}{2m_{2}}(\hat{p}_{2}+\hat{\Pi}_{2})^{2},  \label{quanthmaintc3}
\end{eqnarray}
where the operators $\hat{\Pi}_{1}$ and $\hat{\Pi}_{2}$ describe, respectively, the difference between the mechanical and canonical momentum operators of the first and second moving mirrors,
\begin{subequations}
\label{momnt3}
    \begin{align}
\hat{\Pi}_{1} &=\frac{1}{q_{1}-q_{2}}\sum_{j,k=1}^{\infty}g_{kj}^{(1)}\hat{%
Q}_{j}\hat{P}_{k}+\frac{1}{L_{1}-q_{1}}\sum_{j,k=1}^{\infty}g_{R,kj}\hat{Q}_{R,j}\hat{P}_{R,k},  \\
\hat{\Pi}_{2} &=\frac{1}{q_{1}-q_{2}}\sum_{j,k=1}^{\infty}g_{kj}^{(2)}\hat{%
Q}_{j}\hat{P}_{k}+\frac{1}{L_{2}+q_{2}}\sum_{k,j=1}^{\infty}g_{L,kj}\hat{Q}_{L,j}\hat{P}_{L,k}.
\end{align}
\end{subequations}
Based on Eqs.~(\ref{momnt3}), we can see that $\hat{\Pi}_{l}$ will be reduced to $\Gamma_{l}$ given in Eq.~(\ref{diffemomen}) under the limitations $L_{1}-l_{0}\rightarrow \infty$ and $L_{2}\rightarrow \infty$. Then the Hamiltonian in Eq.~(\ref{quanthmaintc3}) can be reduced to Eq.~(\ref{quanliHamli}) up to the free Hamiltonians of the two external cavities. Therefore, we can ignore the electromagnetic fields outside the cavity when we study the phonon transfer between the two moving end mirrors.

\vspace{8mm} \small{$^*$Contact author: jqliao@hunnu.edu.cn}

\end{document}